\pdfoutput=1
\documentclass[aps,prd,onecolumn,groupedaddress,nofootinbib,secnumarabic,eqsecnum]{revtex4}

\usepackage{epsfig}
\usepackage{subfigure}
\usepackage{graphicx}
\usepackage{amsmath}
\usepackage{multirow}

\def\ie{{\it i.e.}}
\def\eg{{\it e.g.}}
\def\d{{\rm d}}

\def\beq{\begin{equation}}
\def\eeq{\end{equation}}
\def\bea{\begin{eqnarray}}
\def\eea{\end{eqnarray}}

\def\nn{\nonumber}

\def\lag{{\cal L}}
\def\q{{\tilde q}}
\def\Q{{\tilde Q}}
\def\u{{\tilde u}}
\def\sd{{\tilde d}}
\def\g{{\tilde g}}
\def\B{{\tilde B}}
\def\W{{\tilde W}}
\def\H{{\tilde H}}
\def\X{{\tilde \chi}}
\def\hc{{\rm h.c.}}
\def\fp{{f^\prime}}
\def\qp{{q^\prime}}
\def\gp{{g^\prime}}
\def\gw{{g_W}}
\def\cw{{c_W}}
\def\sw{{s_W}}
\def\mw{{m_W}}
\def\mz{{m_Z}}
\def\bsp#1\esp{\begin{split}#1\end{split}}
\def\bpm{\begin{pmatrix}}
\def\epm{\end{pmatrix}}

\begin{document}
\eprint{IPHC-PHENO-11-05, LAPTH-048/11, MS-TP-11-03}

\title{Phenomenology of anomaly-mediated supersymmetry breaking scenarios with non-minimal flavour violation}

\author{Benjamin Fuks}
\email[]{benjamin.fuks@iphc.cnrs.fr}
\affiliation{Institut Pluridisciplinaire Hubert Curien/D\'epartement Recherches
   Subatomiques\\
Universit\'e de Strasbourg/CNRS-IN2P3 \\
23 Rue du Loess, F-67037 Strasbourg, France}

\author{Bj\"orn Herrmann}
\email[]{herrmann@in2p3.fr}
\affiliation{LAPTh, Universit\'e de Savoie, CNRS, B.P.\ 110, F-74941 Annecy-le-Vieux, France}

\author{Michael Klasen}
\email[]{michael.klasen@uni-muenster.de}
\affiliation{Institut f\"ur Theoretische Physik,
 Universit\"at M\"unster,
 Wilhelm-Klemm-Stra{\ss}e 9, D-48149 M\"unster, Germany}

\date{\today}

\begin{abstract}

In minimal anomaly-mediated supersymmetry breaking models,
tachyonic sleptons are avoided by introducing a common scalar mass
similar to the one introduced in minimal supergravity. This may
lead to non-minimal flavour-violating interactions, \eg, in the squark sector.
In this paper, we analyze the viable anomaly-mediated supersymmetry breaking 
parameter space in the
light of the latest limits on low-energy observables and LHC searches,
complete our analytical calculations of flavour-violating
supersymmetric particle production at hadron colliders with those
related to gluino production, and study the phenomenological
consequences of non-minimal flavour violation in anomaly-mediated supersymmetry
breaking scenarios at the LHC. Related cosmological aspects are also briefly
discussed.

\end{abstract}

\maketitle

\section{Introduction \label{sec:intro}}
Many alternatives to the Standard Model (SM) of particle physics have been proposed
over the last thirty years. Among these, supersymmetry (SUSY), and particularly its
minimal version dubbed the Minimal Supersymmetric Standard Model (MSSM)
\cite{Nilles:1983ge, Haber:1984rc}, is one
of the most popular SM extensions. It consists in a symmetry linking fields with
opposite statistics, matching thus a bosonic (fermionic) superpartner with each
fermionic (bosonic) SM degree of freedom. As a consequence, it predicts the
stabilization of the
gap between the electroweak and the Planck scale, gauge coupling unification at
high energies, and a lightest
supersymmetric particle, which is weakly interacting and stable and thus a good
dark matter candidate. Since the
superpartners of the SM particles have not yet been observed and in order to
remain a viable solution to the hierarchy problem, SUSY must be softly broken at
low energy, which makes the SUSY particles massive, with a mass lying in the TeV
range. Therefore, SUSY searches at present hadron colliders, such as the
Tevatron at Fermilab or the LHC at CERN, are important topics of the current
experimental high-energy physics program.

Within the Standard Model, flavour violation in the quark sector arises only
through the rotation of the up- and down-type quark interaction eigenstates
into the basis of physical mass eigenstates. 
Four bi-unitary matrices are required to diagonalize the
quark Yukawa matrices,
which renders the charged-current interactions proportional to the unitary CKM matrix
$V_{\rm CKM}$. 
In the super-CKM basis \cite{Hall:1985dx}, the squark interaction eigenstates
undergo the same rotations as their quark counterparts, so that their
charged-current interactions are also proportional to the CKM matrix. However,
in non-minimal flavour-violating supersymmetric theories, the quark and squark
fields can be misaligned due to additional sources of flavour violation which
are related to the breaking of supersymmetry. As a
consequence, this leads to flavour-violating (non-diagonal) entries in the
squark mass matrices. 

In recent works, we have analysed the cases where such soft terms
appear in (non-minimal) supergravity and gauge-mediated SUSY breaking scenarios
\cite{NMFV_mSUGRA, NMFV_GMSB}. In the first case, supersymmetry is broken in a
hidden sector and transmitted to the visible sector of squarks, sleptons,
gauginos and gluinos through gravitational interactions. Soft masses for
sfermions are induced by direct K\"ahler interactions, which can in general be
flavour non-diagonal \cite{Chamseddine:1982jx, Barbieri:1982eh, Nilles:1983ge}.
In the second case, the breaking of supersymmetry is mediated to the
visible sector via gauge interactions with messenger fields in a
flavour-conserving
fashion \cite{Dine:1993yw, Dine:1994vc, Dine:1995ag, Giudice:1998bp}. However,
it has recently been shown that non-minimal versions of the gauge-mediated
SUSY-breaking mechanism can yield important flavour-violation in the squark and
slepton
sectors \cite{Giudice:1998bp, Tobe:2003nx, Dubovsky:1998nr}. 

When SUSY is broken in a hidden sector, the soft masses also receive
contributions from
quantum effects due to the superconformal anomaly \cite{Randall:1998uk,
Giudice:1998xp, Gherghetta:1999sw, Pomarol:1999ie}. In this work, we therefore 
consider
this so-called anomaly-mediated SUSY-breaking (AMSB) scenario, where
those anomaly-mediated effects are large compared to all other sources of
SUSY-breaking, which are subdominant. We extend our previous work on
flavour violation \cite{NMFV_mSUGRA, NMFV_GMSB} by investigating possible
non-minimal flavour
violation within the AMSB context.

This paper is organized as follows: In Sec.\ \ref{sec:model}, we define
anomaly-mediated supersymmetry breaking scenarios and show how
non-minimal flavour violation can appear. In Sec.\ \ref{sec:scan} we impose
current experimental constraints on the flavour-violating AMSB scenario and
perform scans of the parameter space. In addition, experimentally allowed
benchmark points are defined. Cross sections for 
the production of at least one gluino are analytically and numerically
computed in Sec. \ref{sec:xsec}.  We dedicate Sec.\ \ref{sec:cosmo} to an
analysis of the possible cosmological constraints related to the presence of
cold dark matter in our Universe. Our conclusions are presented in Sec.\
\ref{sec:conclusion}.

\section{Anomaly mediation and flavour violation in the squark sector \label{sec:model}}

In AMSB scenarios, the soft terms are related to the anomalous dimensions of
the different fields and have the feature to be renormalization-group invariant
\cite{Jack:1999aj, ArkaniHamed:2000xj}.
As a consequence, they are fully determined  by the known low-energy gauge and
Yukawa couplings and an overall mass scale $m_{\rm aux}$, 
the vacuum expectation value of the
scalar auxiliary field of the gravitation supermultiplet. This scale is 
expected to be of the order of the gravitino mass $m_{3/2}$, and we assume in the
following, to simplify, $m_{3/2} = m_{\rm aux}$. Consequently, the model
is highly predictive, with fixed mass ratios and distinctive signatures
\cite{Gherghetta:1999sw, Feng:1999fu, Feng:1999hg, Rattazzi:1999qg,
Barr:2002ex}. Among all the
predictions, one finds, however, tachyonic sleptons. This problem must be cured in
order to have a phenomenologically viable model. Several solutions have been
proposed
\cite{Randall:1998uk, Pomarol:1999ie, Chacko:1999am, Katz:1999uw, Jack:2000cd,
Jack:2003qg, Murakami:2003pb, Kitano:2004zd, Ibe:2004gh, Hodgson:2005en,
Jones:2006re, Hodgson:2007kq}, and we adopt here the phenomenological
approach of assuming non-negligible contributions to the scalar
soft masses, induced, {\it e.g.}, by supergravity, which makes their square
positive at the weak scale. However, in this case, 
solving the tachyonic sfermion mass problem can also introduce non-minimal
flavour violation in the theory, through possible non-diagonal flavour-violating
soft mass terms.

The squark mass matrices are written, in the super-CKM basis, as
\beq M_{\tilde{q}}^2 = \left(
  \begin{array}{ccc|ccc} 
    M^2_{L_{q_1}} & \Delta^{q_1 q_2}_{LL} & \Delta^{q_1 q_3}_{LL} & 
      m_{q_1} X_{q_1} & \Delta^{q_1 q_2}_{LR} & \Delta^{q_1 q_3}_{LR} \\
    \Delta^{q_1 q_2\ast}_{LL} & M^2_{L_{q_2}} & \Delta^{q_2 q_3}_{LL} & 
      \Delta^{q_1 q_2\ast}_{RL} & m_{q_2} X_{q_2} & \Delta^{q_2 q_3}_{LR} \\ 
    \Delta^{q_1 q_3\ast}_{LL} & \Delta^{q_2 q_3\ast}_{LL} & M^2_{L_{q_3}} & 
      \Delta^{q_1 q_3\ast}_{RL} & \Delta^{q_2 q_3\ast}_{RL} & m_{q_3} X_{q_3} \\ 
    \hline 
    m_{q_1} X_{q_1}^\ast & \Delta^{q_1 q_2}_{RL} & \Delta^{q_1 q_3}_{RL} &
      M^2_{R_{q_1}} & \Delta^{q_1 q_2}_{RR} & \Delta^{q_1 q_3}_{RR} \\
    \Delta^{q_1 q_2\ast}_{LR}& m_{q_2} X_{q_2}^\ast &  \Delta^{q_2 q_3}_{RL} & 
      \Delta^{q_1 q_2\ast}_{RR} & M^2_{R_{q_2}} & \Delta^{q_2 q_3}_{RR} \\ 
    \Delta^{q_1 q_3\ast}_{LR}& \Delta^{q_2 q_3\ast}_{LR} & m_{q_3} X_{q_3}^\ast&
      \Delta^{q_1 q_3\ast}_{RR} & \Delta^{q_2 q_3\ast}_{RR} & M^2_{R_{q_3}} 
  \end{array} \right) \ , 
\eeq
where the flavour-diagonal elements are given by 
\beq\bsp
   M_{L_{q_i}}^2 =&\ M_{\tilde Q_i}^2 + m_{q_i}^2 + \cos 2\beta \mz^2 (T_q^3
    - e_q s_W^2) \ , \\
   M_{R_{q_i}}^2 =&\ M_{\tilde U_i}^2 + m_{q_i}^2 + \cos 2\beta \mz^2 e_q \sw^2
     \quad \text{for up-type squarks} \ , \\
   M_{R_{q_i}}^2 =&\ M_{\tilde D_i}^2 + m_{q_i}^2 + \cos 2\beta \mz^2 e_q \sw^2
     \quad \text{for down-type squarks} \ ,\\ 
   X_{q_i} =&\ A_{q_i}^\ast - \mu \left\{ 
     \begin{array}{l}
        \cot\beta\quad \text{for up-type squarks} \ , \\
        \tan\beta\quad \text{for down-type squarks} \ .
     \end{array}\right. 
\esp\eeq
The weak isospin quantum numbers are $T_q^3 = \pm1/2$ for
left-handed up-type and down-type (s)quarks, their fractional electromagnetic
charge is denoted by $e_q$, and $m_{q_i}$ is the mass of the quark $q_i$, $i$
being the flavour index, \ie, $d_1=d$, $d_2=s$, $d_3=b$, $u_1=u$, $u_2=c$, and
$u_3=t$. In addition, $m_Z$ is the $Z$-boson mass, and $\sw$ is the sine of the
electroweak mixing angle. The soft supersymmetry-breaking mass terms are
$M_{\tilde Q_i}$ and $M_{\{\tilde U_i,\tilde D_i\}}$ for the left-handed and
right-handed squarks, while the quantities $A_{q_i}$ are the trilinear
couplings between the Higgs bosons and the scalar SUSY particles. In the Higgs sector,
$\mu$ denotes the off-diagonal superpotential Higgs mass parameter, and 
$\tan\beta = v_u / v_d$ is the ratio of vacuum expectation values
of the two Higgs doublets. 
In our phenomenological approach, the off-diagonal parameters $\Delta^{q
\qp}_{ab}$ are arbitrary and can be normalized to the diagonal entries according
to \cite{Gabbiani:1996hi}
\beq
   \Delta_{ab}^{q_{i} q_{j}} = \lambda^{q_{i} q_{j}}_{ab} M_{q_i} M_{q_j} \ . 
\eeq 
Additional sources of quark flavour violation are then parameterized through the 
21 dimensionless (possibly complex) new variables $\lambda^{q_{i} q_{j}}_{ab}$,
since due to $SU(2)$ gauge invariance, the $\Delta_{LL}^{q\qp}$ elements of
the up- and down-type squark squared mass matrices are related to each other,
\beq\label{eq:SU2L}
   M_{\tilde{u},LL}^2 = V_{\rm CKM}  M_{\tilde{d},LL}^2 V_{\rm CKM}^\dag \ ,
\eeq 
as are the associated $\lambda$-parameters. This equation shows that
both squark mass matrices cannot be simultaneously diagonal (without neglecting the
CKM matrix). The diagonalization of the mass matrices $M_{\tilde{u}}^2$ and
$M_{\tilde{d}}^2$ requires the introduction of two $6 \times 6$ matrices $R^u$ 
and $R^d$,
\beq 
  {\rm diag}\, (m_{\tilde u_1}^2, \ldots, m_{\tilde u_6}^2) = 
    R^u M_{\tilde{u}}^2 R^{u\dag}
  \qquad \text{and} \qquad 
  {\rm diag}\, (m_{\tilde d_1}^2, \ldots, m_{\tilde d_6}^2) = 
    R^d M_{\tilde{d}}^2 R^{d\dag} \ , 
\eeq 
where by convention the masses are ordered increasingly, $m_{\tilde q_1} <
\ldots < m_{\tilde q_6}$. These mixing matrices relate the physical mass eigenbasis
to the interaction eigenbasis through 
\beq 
  (\tilde{u}_1, \tilde{u}_2, \tilde{u}_3, \tilde{u}_4, \tilde{u}_5,
    \tilde{u}_6)^t = R^u (\tilde{u}_L, \tilde{c}_L, \tilde{t}_L, \tilde{u}_R,
    \tilde{c}_R, \tilde{t}_R)^t
  \quad \text{and} \quad 
  (\tilde{d}_1, \tilde{d}_2, \tilde{d}_3, \tilde{d}_4, \tilde{d}_5,
    \tilde{d}_6)^t = R^d (\tilde{d}_L, \tilde{s}_L, \tilde{b}_L, \tilde{d}_R,
    \tilde{s}_R, \tilde{b}_R)^t \ .
\eeq

Recently, it has been shown that (minimal) AMSB scenarios with an additional $U(1)$
symmetry satisfy the requirements
of the Minimal Flavour Violation principles \cite{Allanach:2009ne}. In this
case, the $\lambda$-parameters are not free and directly dictated from the
flavour structure of the CKM matrix. In our approach, we are going beyond this
scheme, keeping the flavour-violation parameters independent.

\section{Experimental constraints on AMSB models \label{sec:scan}}

\subsection{Constraints \label{sec:constraints}}

In this Section, we discuss the most relevant experimental
measurements that can be used to constrain the parameter space of the MSSM.
Apart from direct searches
for superpartners at collider experiments, and particularly the
recent results of the ATLAS and CMS experiments at the LHC 
\cite{atlas, cms}, numerous low-energy and
electroweak precision measurements \cite{PDG, HFAG} constrain masses and mixings
of the superpartners. They can often impose stronger limits on the
flavour-violating entries introduced in Sec.\ \ref{sec:model}.

Extensive studies of the kaon sector, $B$- and $D$-meson
oscillations, rare decays, and electric dipole moments suggest that only flavour
mixing involving the second and third generations of squarks can be substantial,
and this only in the left-left and right-right chiral sectors, which mix the
superpartners of the left-handed and right-handed quarks
\cite{Hagelin1992, Brax1995, Gabbiani:1996hi, Ciuchini2007}. For this reason, we
restrict ourselves to the simpler
scenario where only the flavour-mixing parameters related to the second and
third generations and non mixing squark chiralities can be non-vanishing,
\beq
	\lambda_{\rm L} ~\equiv~ \lambda_{LL}^{ct}, \qquad
	\lambda_u ~\equiv~ \lambda_{RR}^{ct} \qquad {\rm and} \qquad
	\lambda_d ~\equiv~ \lambda_{RR}^{sb} \ .
\label{eq:lambda}
\eeq
The parameter related to the mixing among second and third
generation down-type squarks $\lambda_{LL}^{sb}$ is not a free parameter of our
simplified model, since it is connected to
$\lambda_L$ through Eq.\ \eqref{eq:SU2L}.

The measurement of the rare $b\to s\gamma$ decay represents one of the most stringent
constraints on these squark mixings. The inclusive branching ratio is determined
to be
\beq
  {\rm BR}(b\to s\gamma) = \big( 3.55 \pm 0.26_{\rm exp} \pm 0.23_{\rm theo}
    \big) \times 10^{-4} 
  \label{eq:bsg}
\eeq
from BABAR, BELLE, and CLEO data \cite{HFAG}. For the 
theoretical error estimate, we refer to the discussion in Refs.\ \cite{HurthPorodBSG,
Misiak2006}. Squark diagrams contribute already at the one-loop level, as do
the SM particles. As a consequence, this measurement can lead to strong constraints
on the squark masses and couplings, especially on the $\lambda_{\rm L}$ parameter of
the left-left chiral sector, since the lightest neutralinos and charginos
appearing in the loops are mostly winos coupling to the left-handed
components of the squark mass-eigenstates. The same arguments apply to
the $b\to s\mu^+\mu^-$ branching fraction, experimentally measured as  
\cite{HFAG} 
\beq
  {\rm BR}(b\to s\mu^+\mu^-) = \big( 2.23 \pm 0.98_{\rm exp} \pm
    0.11_{\rm theo} \big) \times 10^{-6} \ ,
\eeq
the associated theoretical uncertainties being computed in Ref.\
\cite{HuberHurth}, or to the $B_s^0$-meson branching fractions to a muon pair,
recently bounded from above by the CMS and LHCb experiments \cite{CMS5},
\beq\label{eq:bs0}
  {\rm BR}(B_s^0 \to \mu^+\mu^-) < 1.1 \times 10^{-8} \ ,
\eeq
at $95\%$ confidence level, which is, however, still about four or five times
larger than the SM expectation.

Also, the $B_s^0-\bar{B}_s^0$ oscillations, which have been
recently observed, directly probe the mixing between squarks of the second
and third generations. Since NMFV contributions arise at the same loop level as 
the SM ones, this observable can again be sensitive to non-vanishing
$\lambda$-parameters. Hence, the measured mass difference \cite{HFAG}
\beq
 \Delta M_{B_s^0} = \big( 17.78 \pm 0.12_{\rm exp} \pm 3.3_{\rm theo}
   \big)~{\rm ps}^{-1} \ ,
 \label{eq:dMBs}
\eeq
where the theoretical uncertainty of 3.3~ps$^{-1}$ at the 95\% confidence
level has been derived in Ref.\ \cite{BallFleischer}, allows to constrain the
magnitude of the above-mentioned flavour-violating parameters.

Another important consequence of NMFV mixing in the squark sector is a large
splitting between squark mass eigenvalues. This influences the $Z$- and $W$-boson
self-energies at zero momentum, contributing hence to the electroweak
$\rho$-parameter
\beq
  \Delta\rho = \frac{\Sigma_Z(0)}{m_Z^2} - \frac{\Sigma_W(0)}{m_W^2} = \alpha(m_Z) T 
             = \big( 2.36 \pm 8.65 \big) \times 10^{-4} \  ,
\eeq
the experimental value arising from combined fits of the $Z$-boson mass, width,
and pole asymmetry as well as of the masses of the $W$-boson and the top quark 
\cite{PDG}.

Furthermore, recent measurements of the anomalous magnetic moment of
the muon $(g-2)_{\mu}$ indicate a discrepancy of about $3\sigma$ between the
data and the Standard Model predictions \cite{PDG},
\beq
  a_{\mu}^{\rm exp} - a_{\mu}^{\rm SM} = \big( 25.5 \pm 7.98 \big) \times 10^{-10}
	\qquad {\rm with} \qquad a_{\mu} \equiv (g-2)_{\mu}/2 .
\eeq
This gap could be explained by new physics. In the case of supersymmetric scenarios, the
leading contributions, depending on the smuon, sneutrino, chargino and neutralino
masses, have been found to be proportional
to the sign of the $\mu$-parameter \cite{Moroi1995}. Since negative values
would then increase the discrepancy, we limit ourselves to positive values
of $\mu$. Moreover, since squarks contribute only at the two-loop level, the
dependence on flavour violation in the squark sector is expected
to be considerably reduced.
Finally, from direct searches of the Higgs boson, we ask for the mass of the
lightest Higgs-boson to fulfil 
\beq
  111~{\rm GeV} \lesssim m_h \lesssim 130 \text{ GeV }\ .
\label{eq:higgsmass}\eeq 
The lower bound is based on the exclusion limit of $m_h < 114.4$~GeV from LEP
\cite{PDG}, after accounting for a theoretical uncertainty of about $3$ GeV
\cite{SPheno}, whilst the upper bound corresponds to the non-observation of a
Higgs boson by the ATLAS and CMS experiments \cite{Higgs2, Higgs3}. For
a restricted set of universal scalar mass $m_0$, introduced to solve the tachyonic
slepton problem, smaller than a few TeV, the upper bound is almost always
satisfied.
Considering quantum corrections, the Higgs mass $m_h$ depends
on the squark masses, and thus on the flavour-violating entries in the
associated mass matrices.


\subsection{Analysis of the parameter space related to AMSB scenarios with
non-minimal flavour violation}\label{sec:planes}

In order to study the phenomenology of AMSB models including non-minimal
flavour violation, we have scanned over the parameter space of the model,
generalizing the squark mass matrices by including the three flavour-violation
parameters presented in Eq.\ \eqref{eq:lambda}. Our procedure starts at a
high-energy scale, where we define our input parameters as the gravitino mass
$m_{3/2}$, the universal scalar mass $m_0$ introduced to solve the tachyonic
slepton problem, the ratio of the two neutral Higgs
field vacuum expectation values $\tan\beta=v_u/v_d$, and the sign of the Higgs
mixing parameter ${\rm sgn}(\mu)$. The soft supersymmetry-breaking terms at the
electroweak scale are then obtained through renormalization group running using
the {\tt SPheno} package version 3.0 \cite{SPheno}, which solves the
renormalization group equations numerically to two-loop order and extracts the
particle spectrum and mixings at the electroweak scale at the one-loop level for the
matter and gauge sectors and at the two-loop level for the Higgs sector. It also
computes the electroweak precision and low-energy observables presented in Sec.\
\ref{sec:constraints}.

For the numerical values of the Standard Model parameters, we fix the top quark
pole mass to $m_{\rm top}^{\rm pole} = 173.2$ GeV \cite{Lancaster:2011wr}, 
the bottom
quark mass to $m_b(m_b) = 4.2$ GeV and the $Z$-boson mass to $m_Z =
91.1876$ GeV. The Fermi constant has been taken as $G_F = 1.16637 \times
10^{-5}$ GeV$^{-2}$, and the strong and electromagnetic coupling constants at the
$Z$-pole as $\alpha_s(m_Z) = 0.1176$ and $\alpha(m_Z)^{-1} = 127.934$. At this
stage, the only source of flavour violation lies within the CKM-matrix, which is
calculated using the Wolfenstein parametrization. The corresponding four free
parameters are set to $\lambda^{({\rm CKM})}=0.2272$, $A^{({\rm CKM})}=0.818$,
$\bar{\rho}^{({\rm CKM})} = 0.221$, and
$\bar{\eta}^{({\rm CKM})} = 0.34$ \cite{PDG}.

\begin{figure}
\begin{center}
	\includegraphics[scale=0.33]{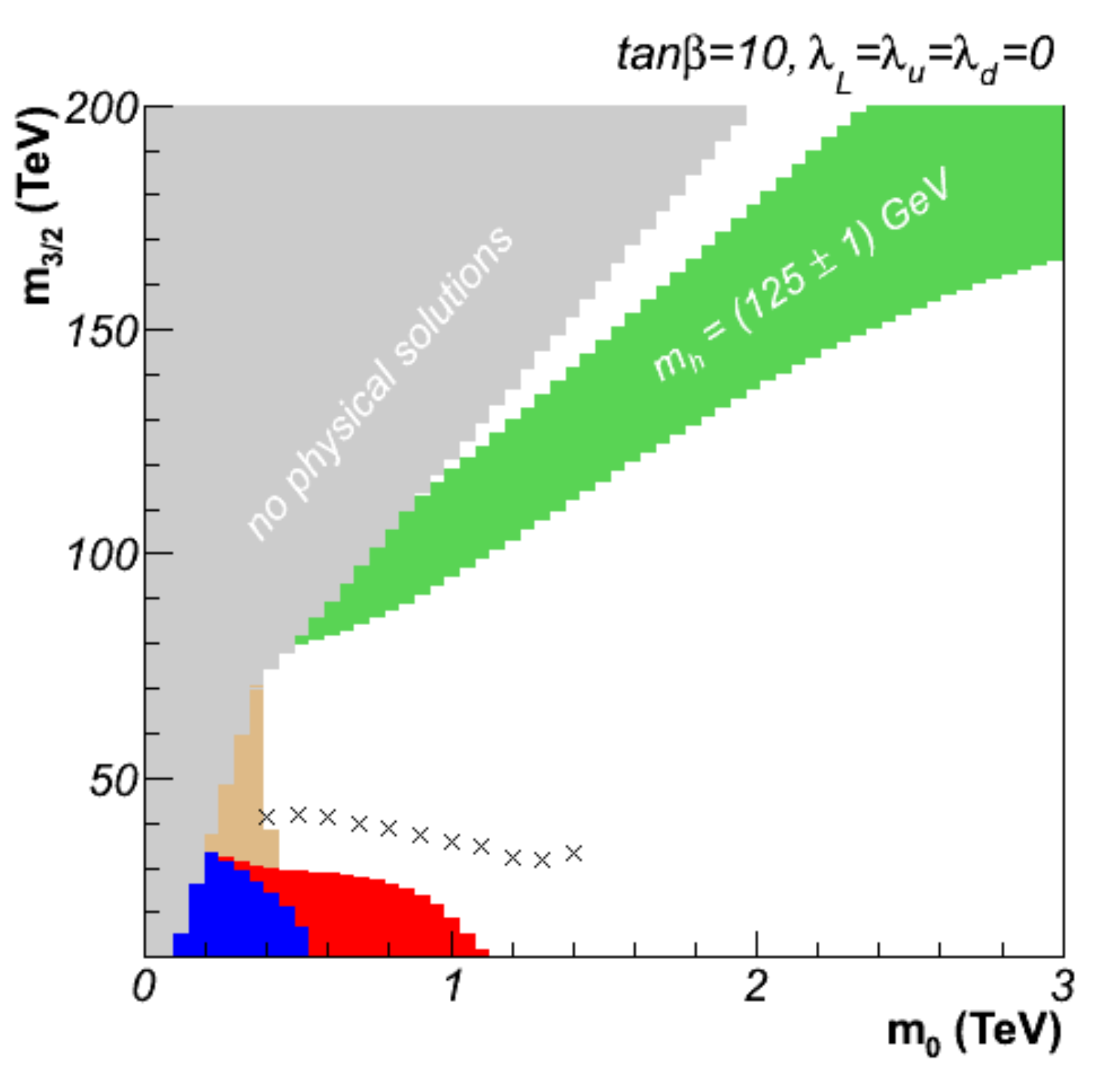}
	\includegraphics[scale=0.33]{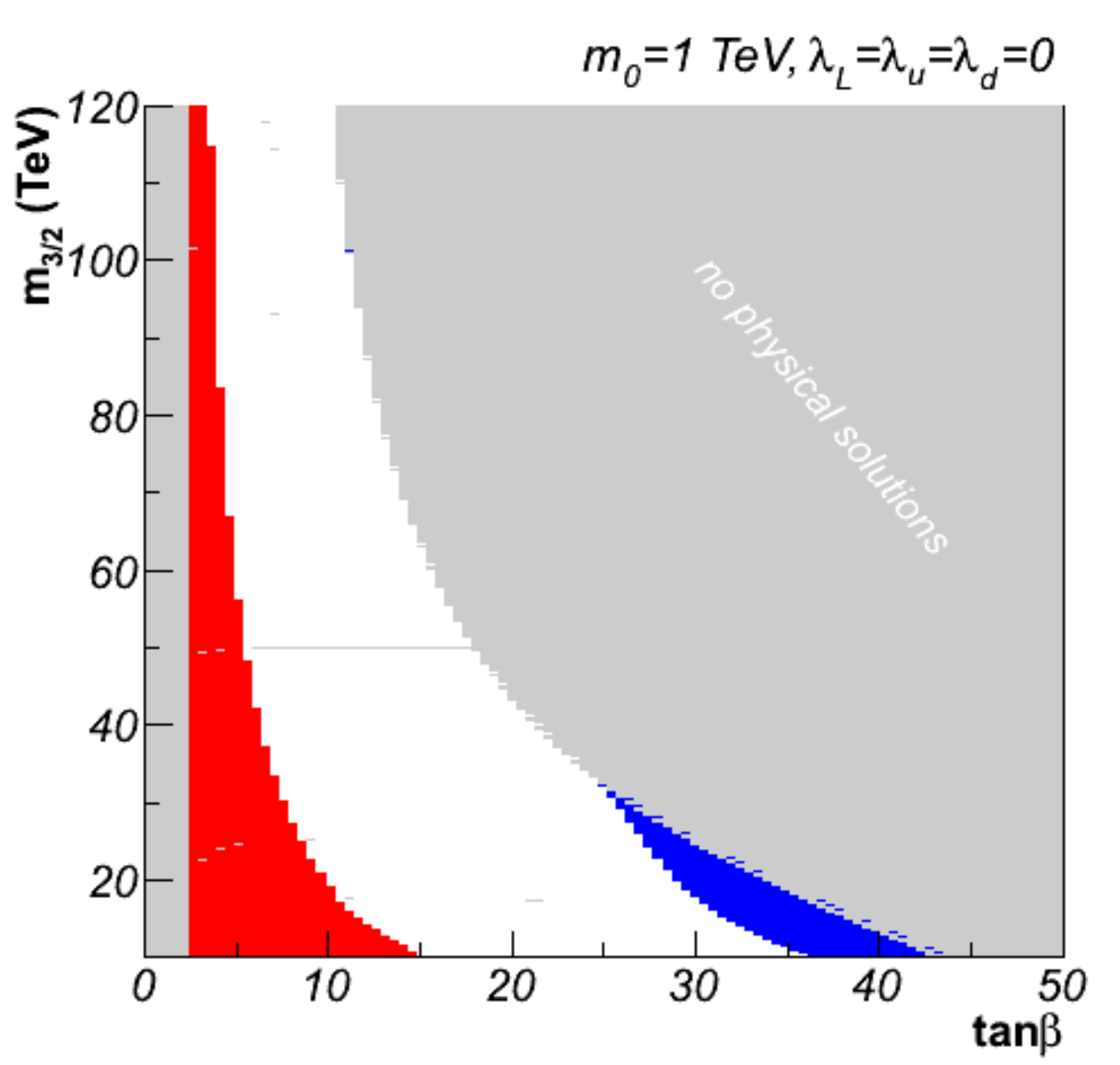}
	\includegraphics[scale=0.33]{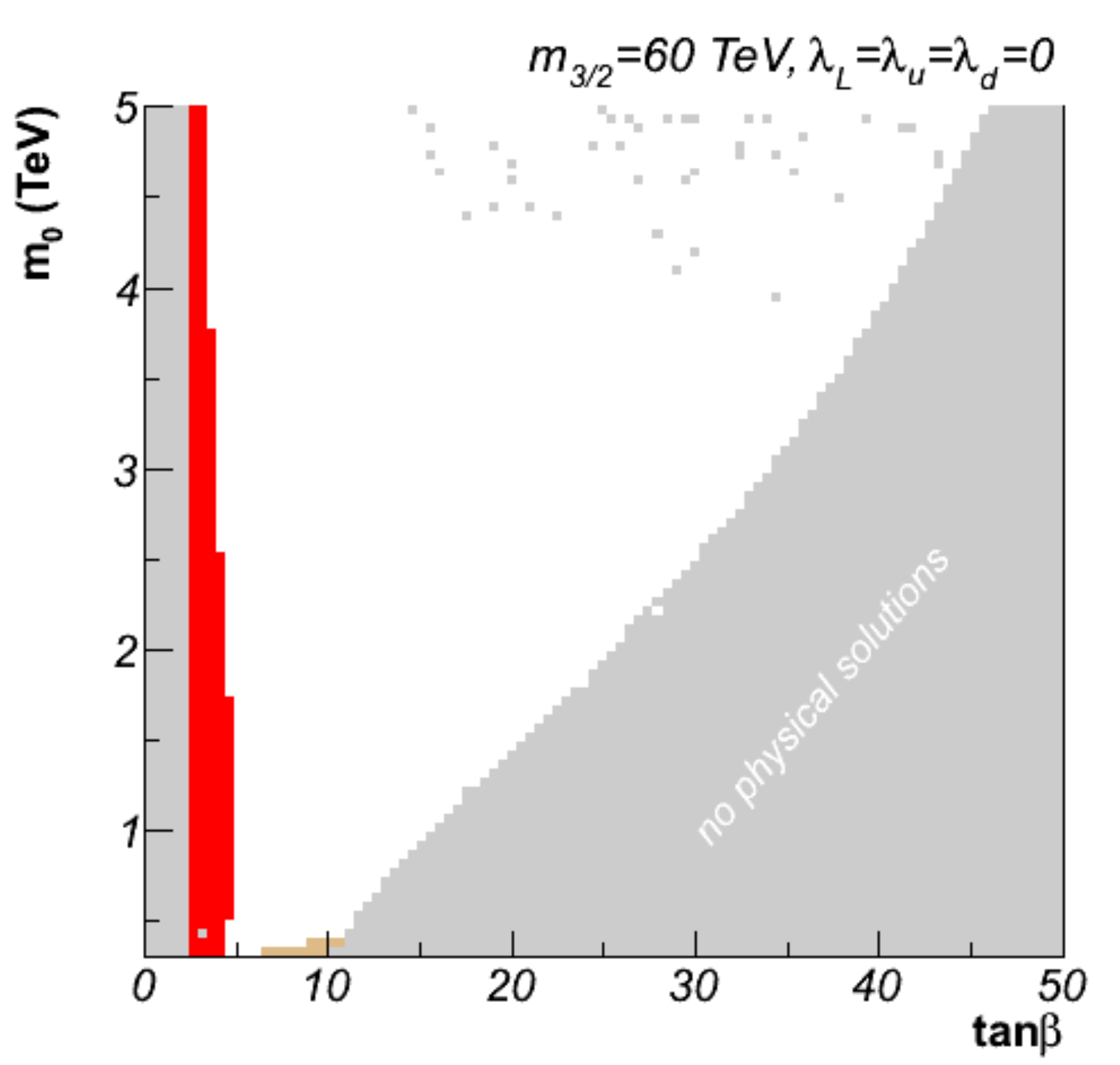}
\end{center}
\caption{Scans of the minimal AMSB parameter space, where all flavour-violating
  parameters, apart from the CKM matrix, vanish, in the $(m_0,m_{3/2})$ plane at fixed $\tan\beta=10$ (left
  panel), in the $(\tan\beta,m_{3/2})$ plane at fixed $m_0=1$ TeV (central
  panel) and in the $(\tan\beta,m_0)$ plane at fixed $m_{3/2}=60$ TeV (right
  panel). We show regions where there is no physical solution to the
  renormalization group equations (grey), or which are excluded by the
  constraints
  related to the $b\to s\gamma$ branching ratio (blue) and the Higgs mass (red).
  The regions, where the agreement between theory and
  experiment for the anomalous magnetic moment of the muon is restored, are
  presented in beige. On the left panel, we also indicate the region where the
  mass of the lightest Higgs boson
  is close to 125 GeV (green) and the exclusion limit obtained (crosses) from the
  reinterpretation of the LHC results on supersymmetric particle searches in the
  context of AMSB scenarios (see Ref.\ \cite{Allanach:2011qr}).}
\label{fig:MFV}
\end{figure}

In Fig.\ \ref{fig:MFV}, we present typical scans of the \textit{minimal} AMSB
parameter space, \ie, when flavour-violation is induced by the CKM-matrix
alone and all $\lambda$-parameters vanish. We show the examples of the 
$(m_0,m_{3/2})$ plane at fixed $\tan\beta=10$ (left panel), the $(\tan\beta,m_{3/2})$
plane at fixed $m_0=1$ TeV (central panel) as well as the $(\tan\beta,m_0)$ plane at
fixed $m_{3/2}=60$ TeV (right panel). All experimental limits described in Sec.\
\ref{sec:constraints} are imposed at the $2\sigma$-level. We observe that the
low-mass regions with a relatively small $m_0$, attractive from a collider point
of view, are strongly disfavoured by both the measurements of the $b\to s\gamma$
branching ratio and the too low predicted mass for the lightest Higgs boson. The
latter also excludes small values of $\tan\beta$ for a large range of $m_0$ and
$m_{3/2}$ masses. In addition, a significant part of the regions where the
predicted value for the $b\to s \gamma$ branching ratio lies outside the $2
\sigma$-range deduced from Eq.\ \eqref{eq:bsg} is also excluded by the constraints
associated to the $b\to s\mu^+\mu^-$ branching ratio. Therefore, these regions
have not been shown on Fig.\ \ref{fig:MFV}, for clarity. Predictions for the
other considered $B$-physics observables, $\Delta M_{B_s^0}$ and the branching
ratio $B_s^0 \to \mu^+ \mu^-$, are mostly 
compatible with data and hence absent from the figure, as is the computed value
for the $\Delta\rho$ parameter. This quantity only restricts the parameter
space at very heavy masses, which lie outside the mass range presented in the
figures.

On the left panel of Fig.\ \ref{fig:MFV}, we also show the exclusion limits on
the parameter space of AMSB scenarios obtained from a reinterpretation of the
results of the direct searches for superparticles at the LHC in the AMSB
context \cite{Allanach:2011qr}, as well as the region where the mass
of the lightest Higgs boson is close to 125 GeV, a value favoured by the recent
observations of the ATLAS and CMS experiments \cite{Higgs2, Higgs3}. However,
in the absence of a confirmed signal for a Higgs boson in that mass range, we do not
consider this last limit for our analysis in the sequel. 

The size of the regions, where the gap between the data and the
predictions for the anomalous magnetic moment of the muon $a_\mu$ is closed, is
relatively small. Moreover, a significant fraction of it is excluded by the $b\to
s\gamma$ constraint, as it has also been found in Refs.\ \cite{Martin:2001st,
Allanach:2009ne}. Hence, a possible explanation of the discrepancy between
theory and data for this observable by contributions related to AMSB scenarios
is rather difficult. However, the dominant supersymmetric contributions to
$a_\mu$ highly depend on the slepton masses, entering into the theoretical
calculation already at the one-loop level. Since in the context of AMSB
scenarios, any prediction associated to the slepton sector is tightly linked to the
employed solution to solve the tachyonic slepton problem, we choose to
relax the constraint associated to the anomalous magnetic moment of the muon.
This is also motivated by the fact that non-minimal flavour violation in the
squark sector, \ie, the scope of this paper, contributes subdominantly, at the
two-loop level, to $a_\mu$, so that it becomes almost independent of the
$\lambda$-parameters as it was already noticed in the discussions of Refs.\
\cite{NMFV_mSUGRA, NMFV_GMSB}.

\begin{figure}
\begin{center}
	\includegraphics[scale=0.33]{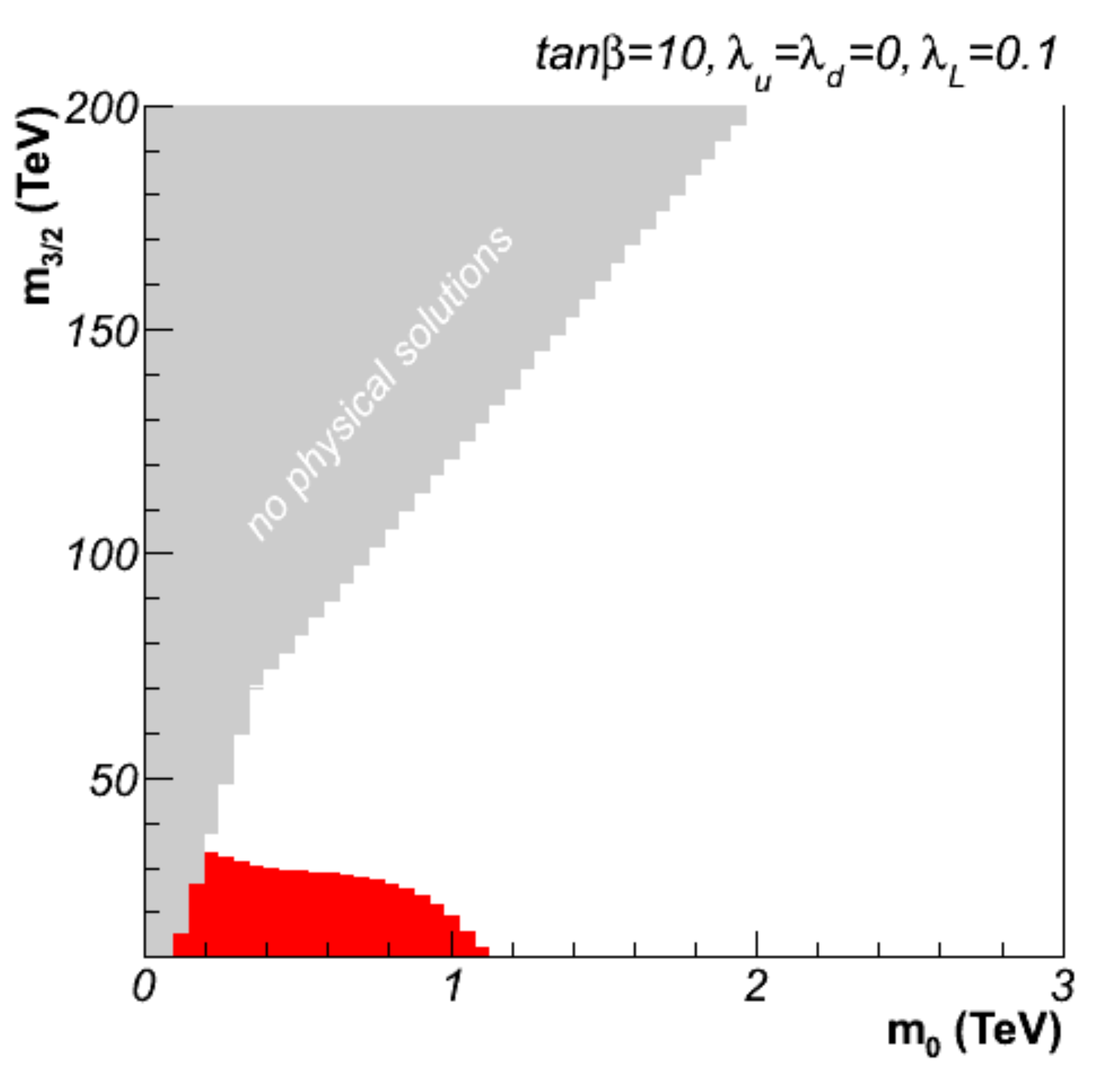}
	\includegraphics[scale=0.33]{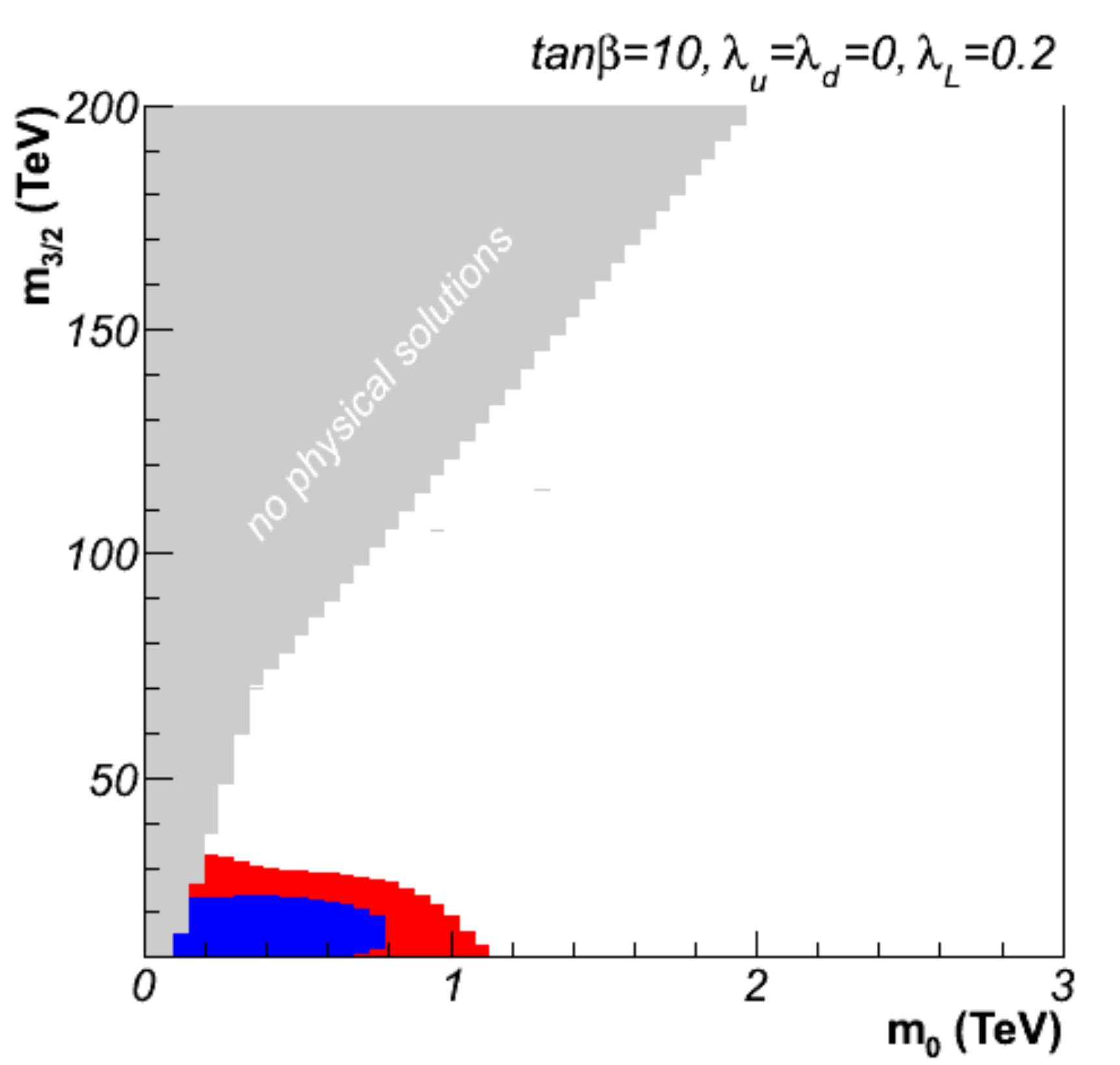}
	\includegraphics[scale=0.33]{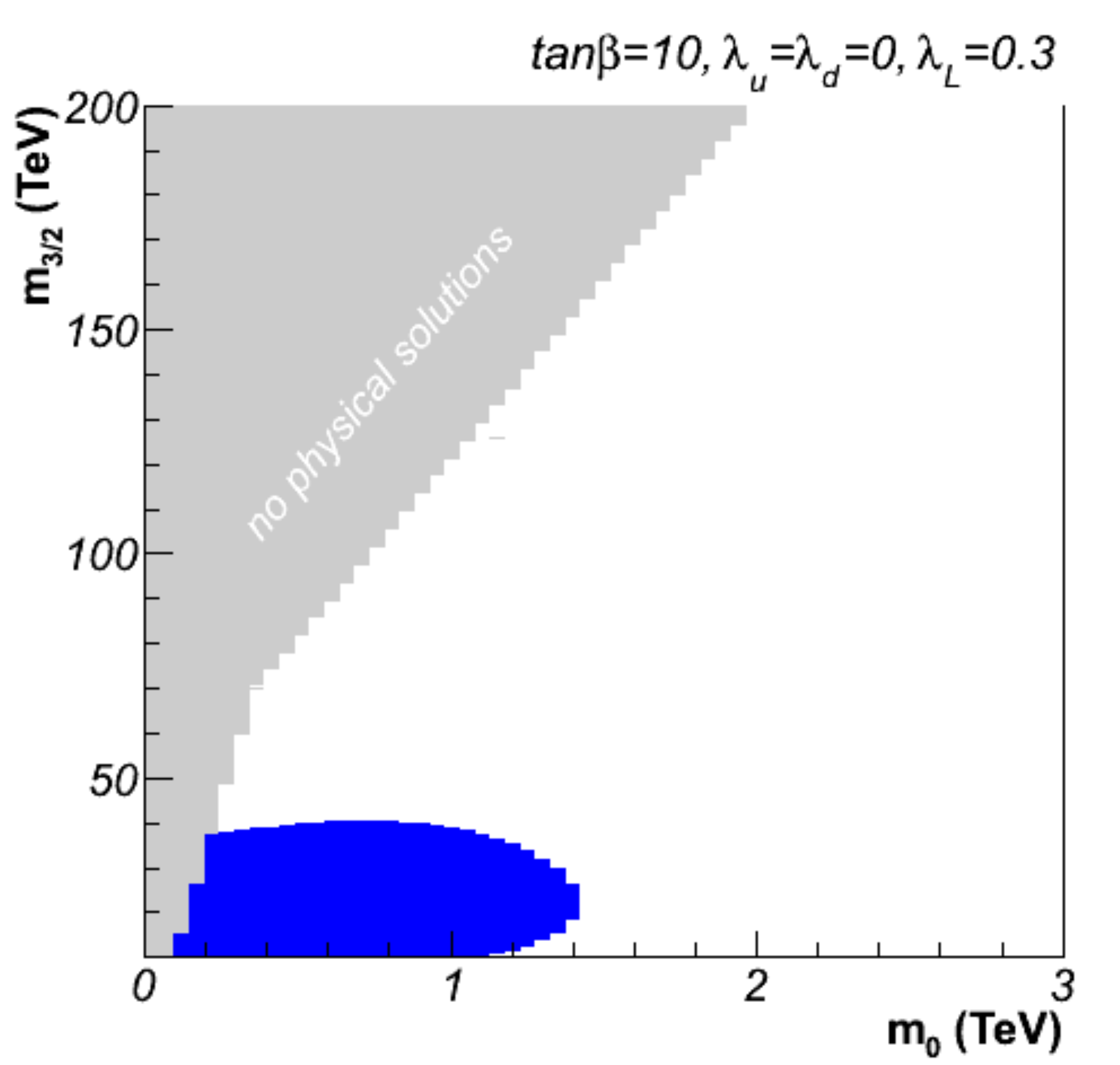}\\
	\includegraphics[scale=0.33]{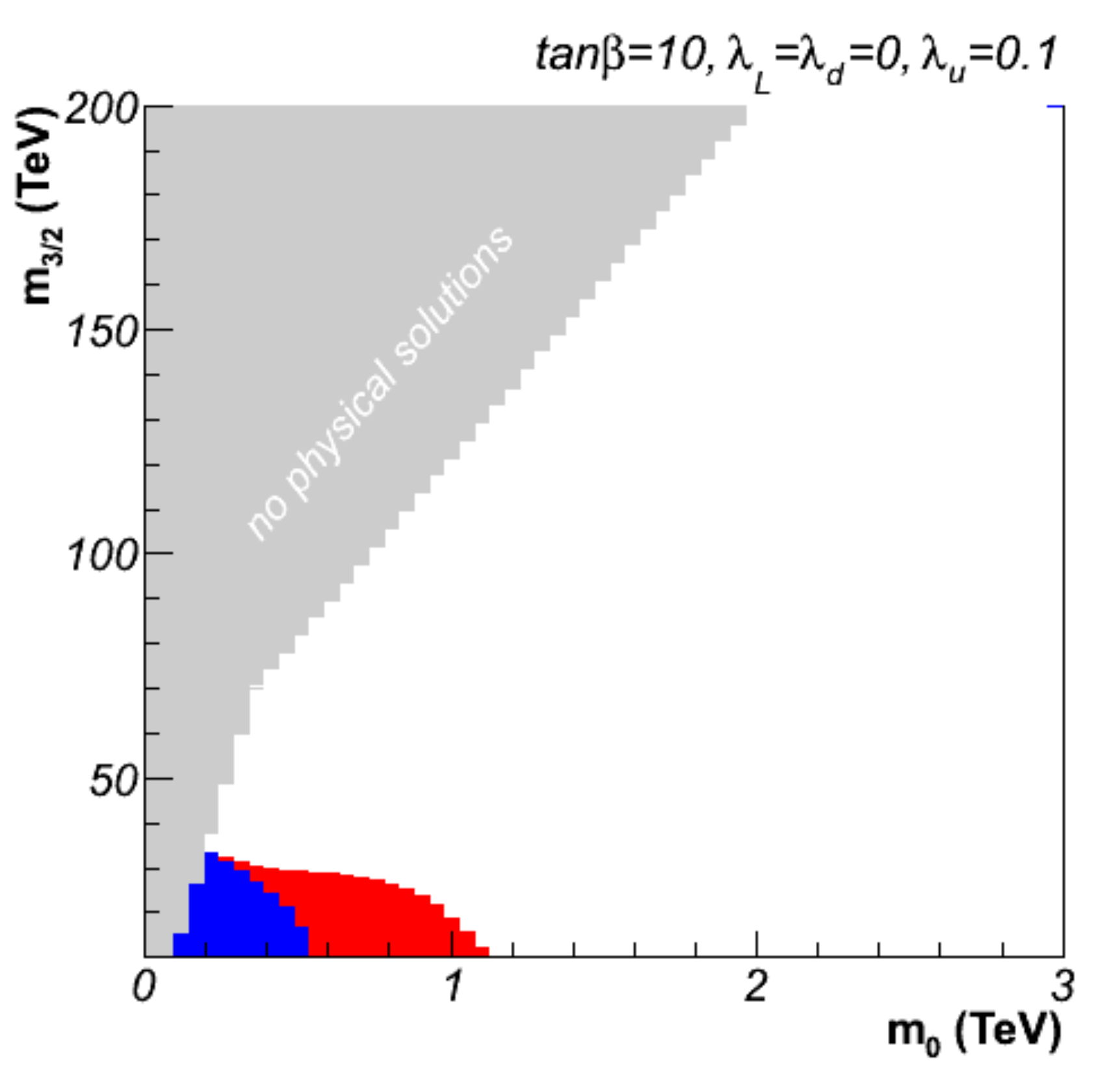}
	\includegraphics[scale=0.33]{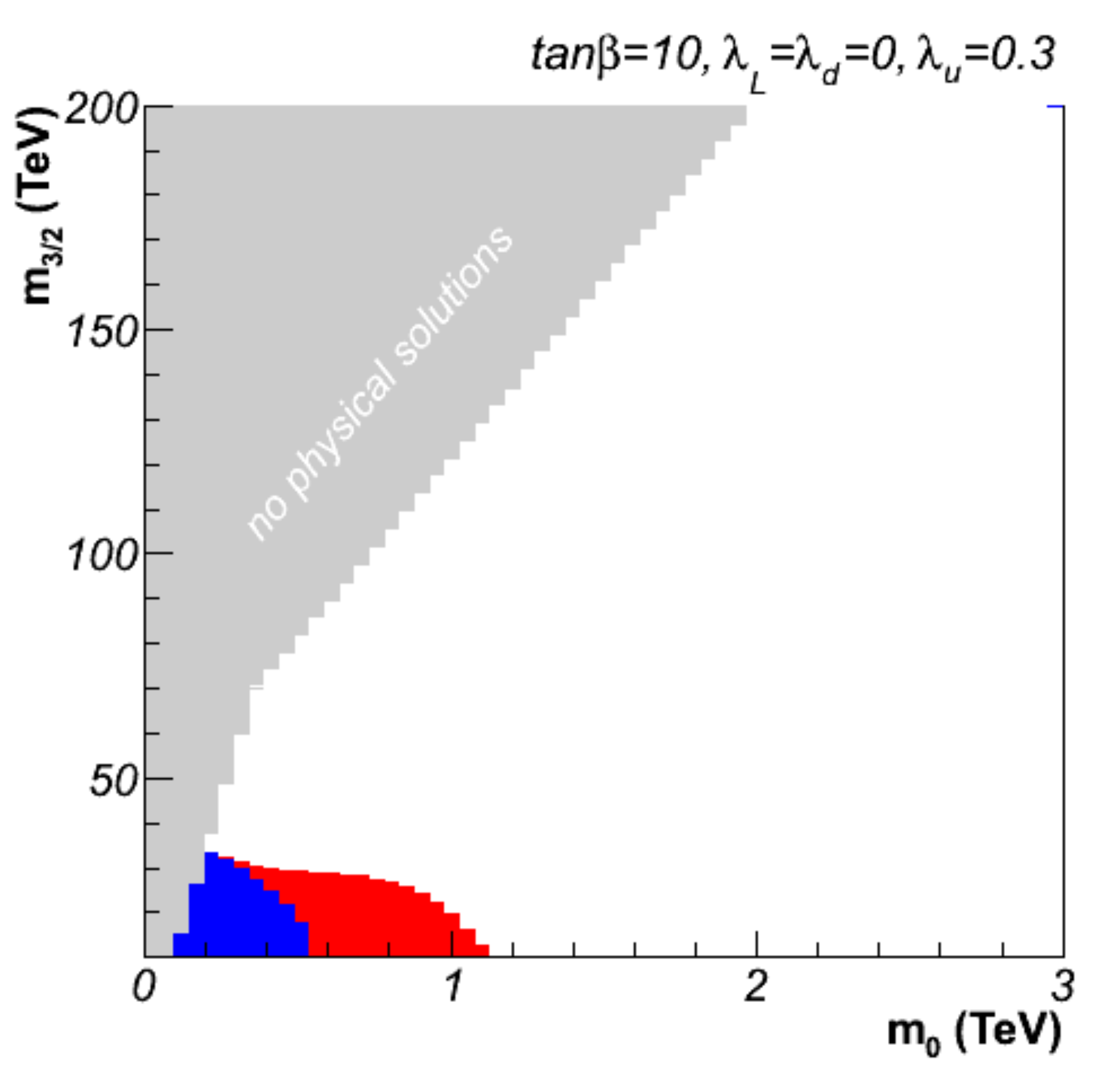}
	\includegraphics[scale=0.33]{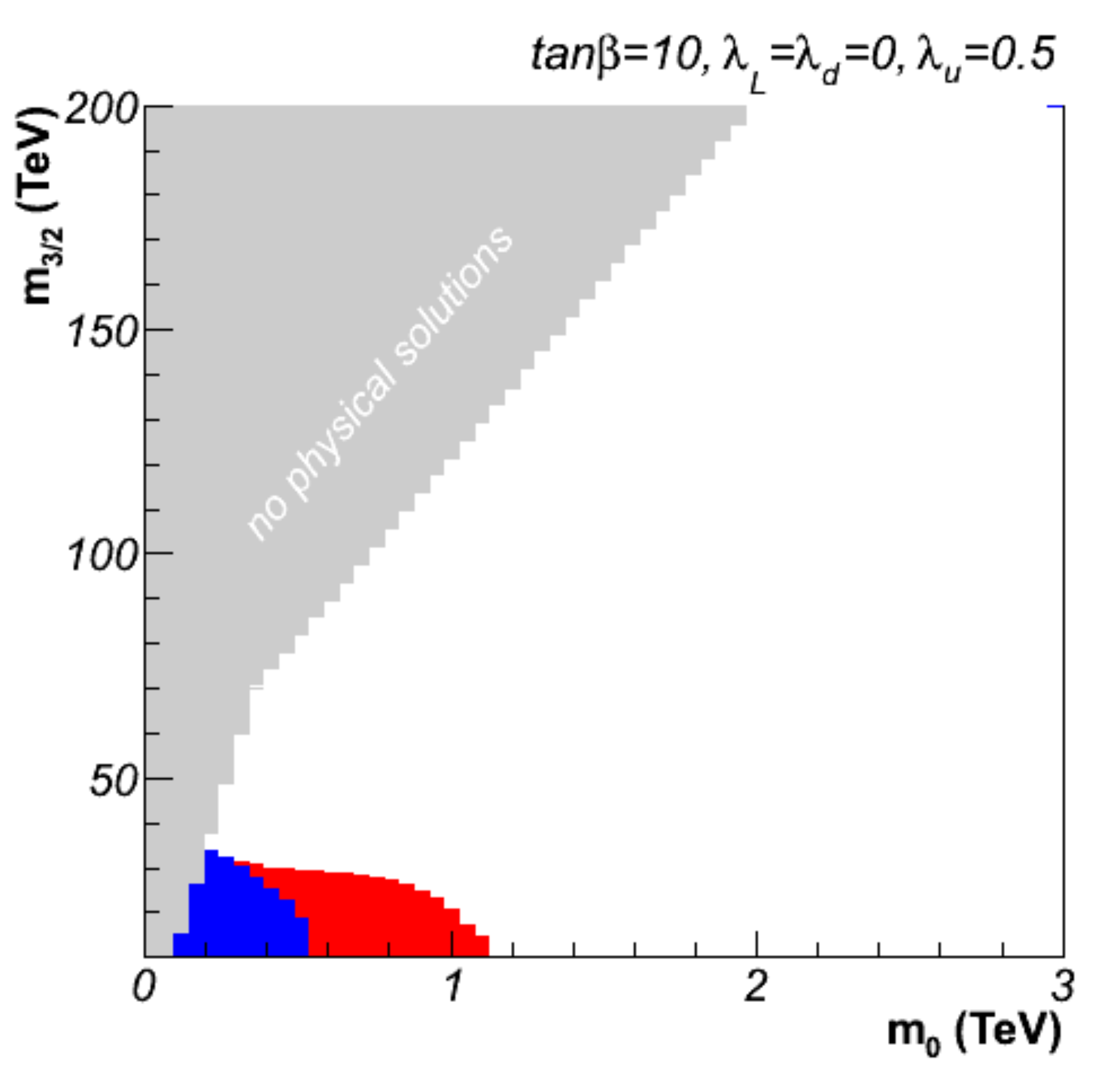}\\
	\includegraphics[scale=0.33]{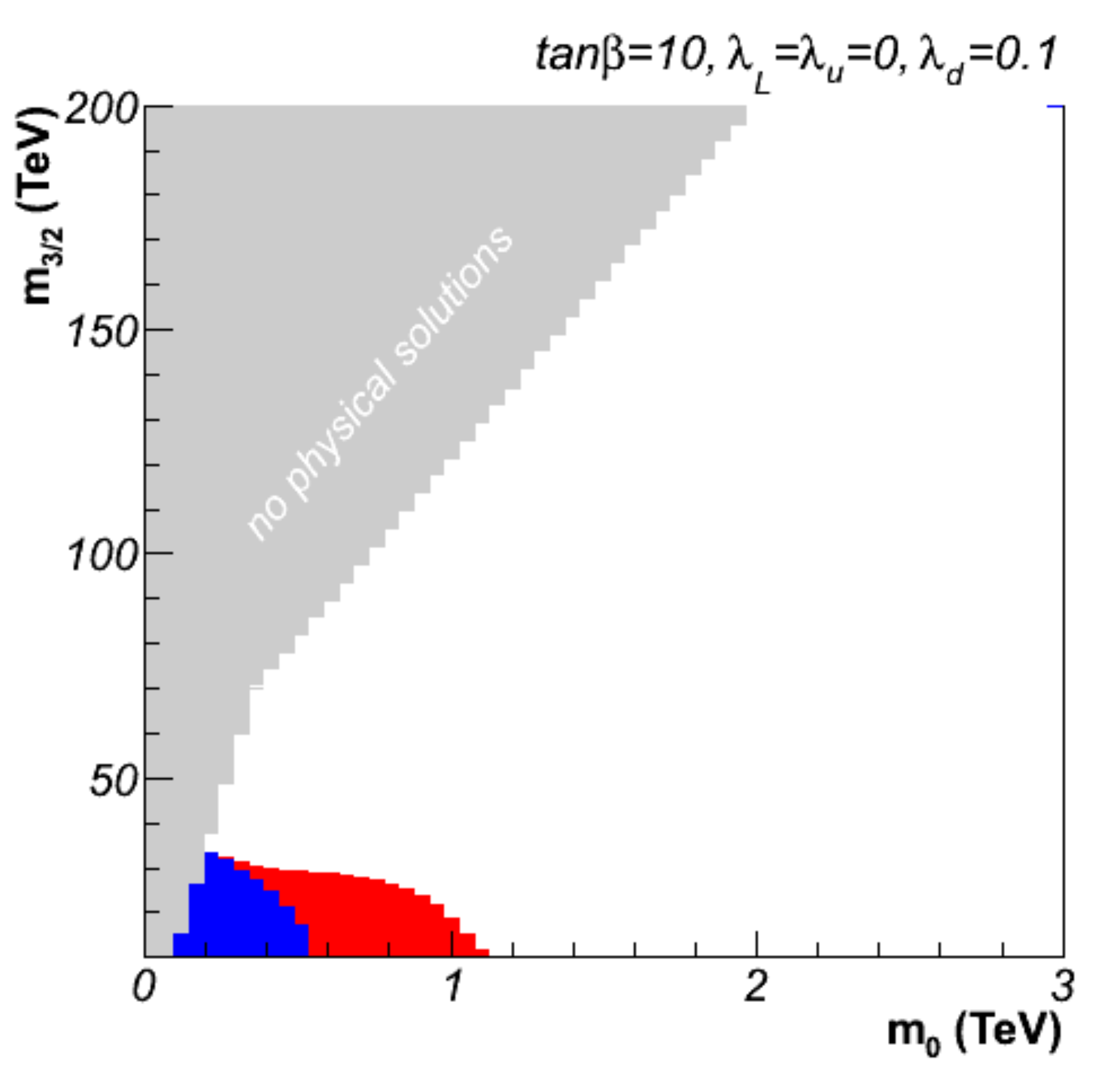}
	\includegraphics[scale=0.33]{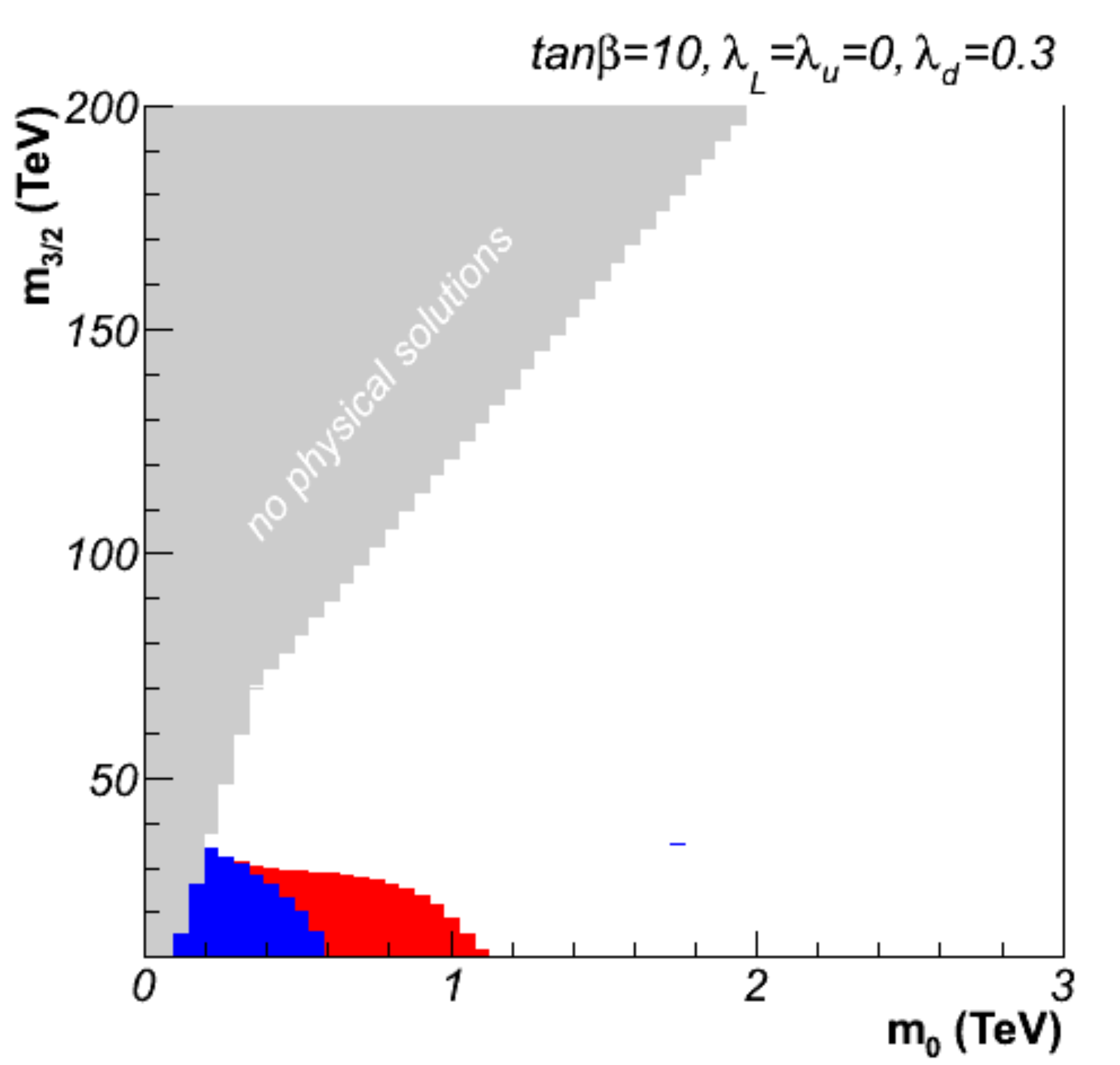}
	\includegraphics[scale=0.33]{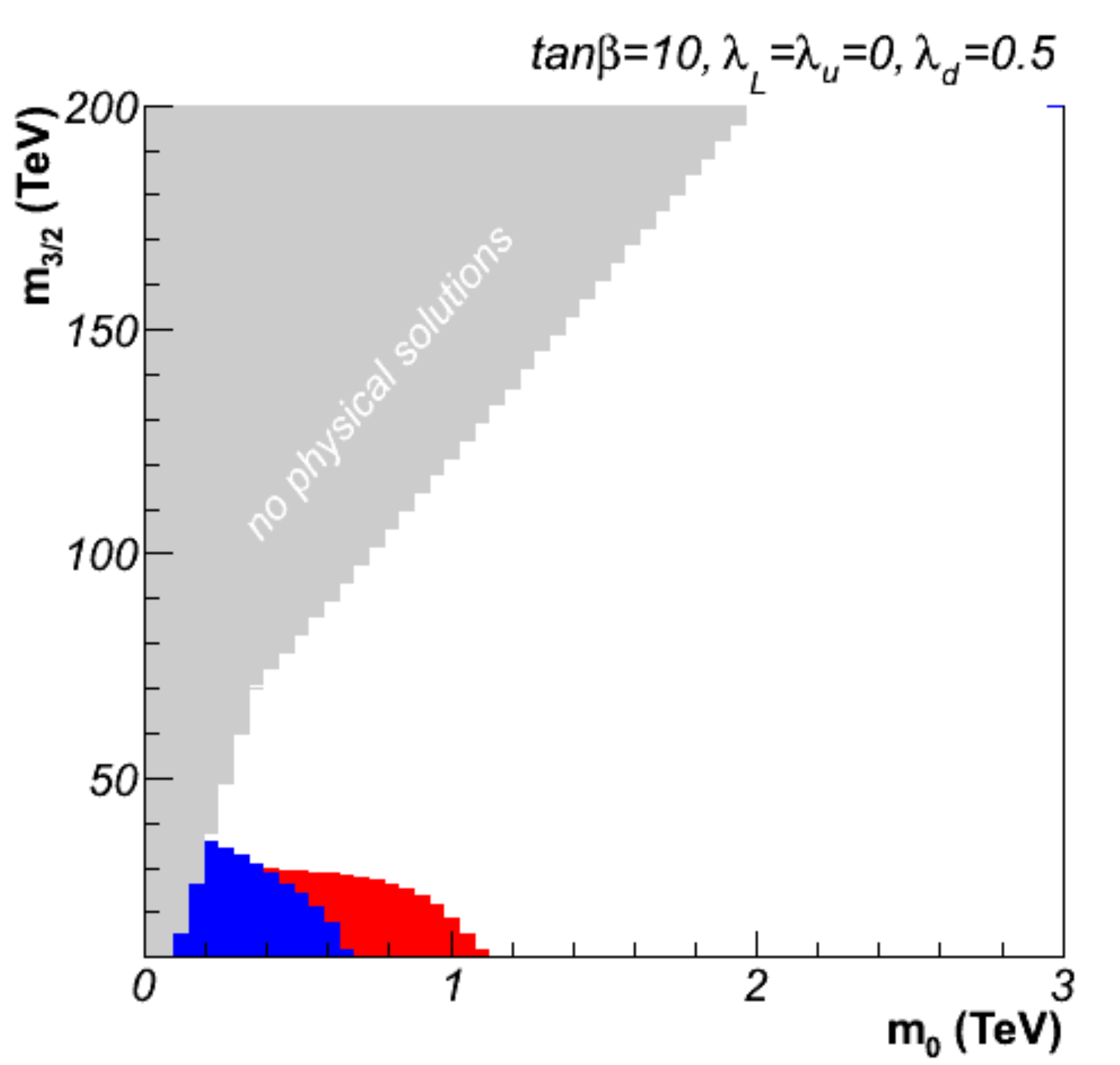}
\end{center}
\caption{Typical scans of NMFV AMSB scenarios in the $(m_0,m_{3/2})$ plane with a
  fixed value of $\tan\beta=10$. The sign of the
  $\mu$-parameter is chosen positive and we present the results for different
  values of the NMFV-parameters $\lambda_{\rm L}$ (upper panels), $\lambda_u$
  (central panels) and $\lambda_d$ (lower panels). We show regions excluded due
  to the absence of physical solutions to the renormalization group equations
  (grey), by the constraints associated to the $b\to s\gamma$ branching ratio
  (blue) and the Higgs mass (red).}
\label{fig:tb10} 
\end{figure}

\begin{figure}
\begin{center}
	\includegraphics[scale=0.33]{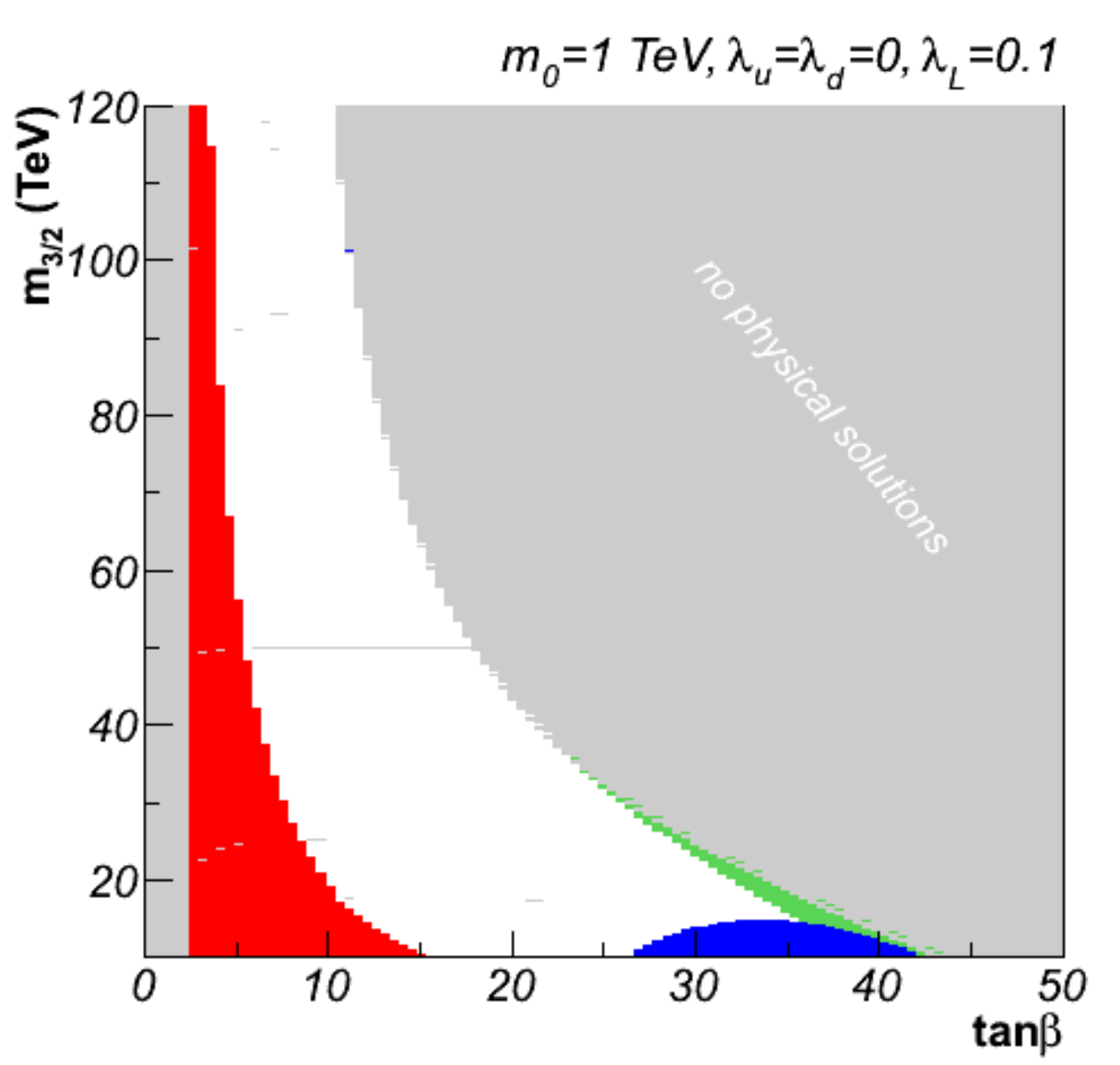}
	\includegraphics[scale=0.33]{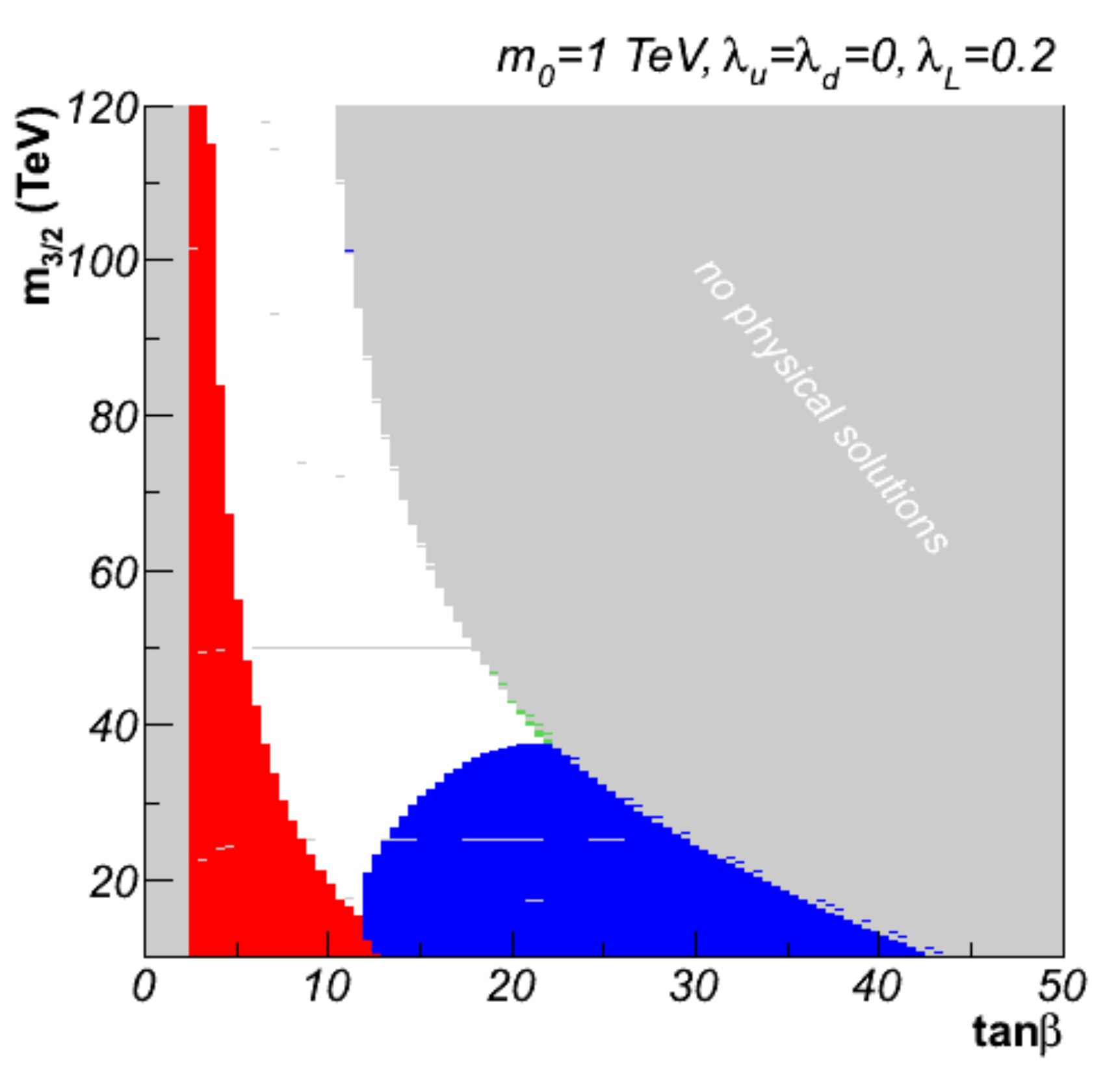}
	\includegraphics[scale=0.33]{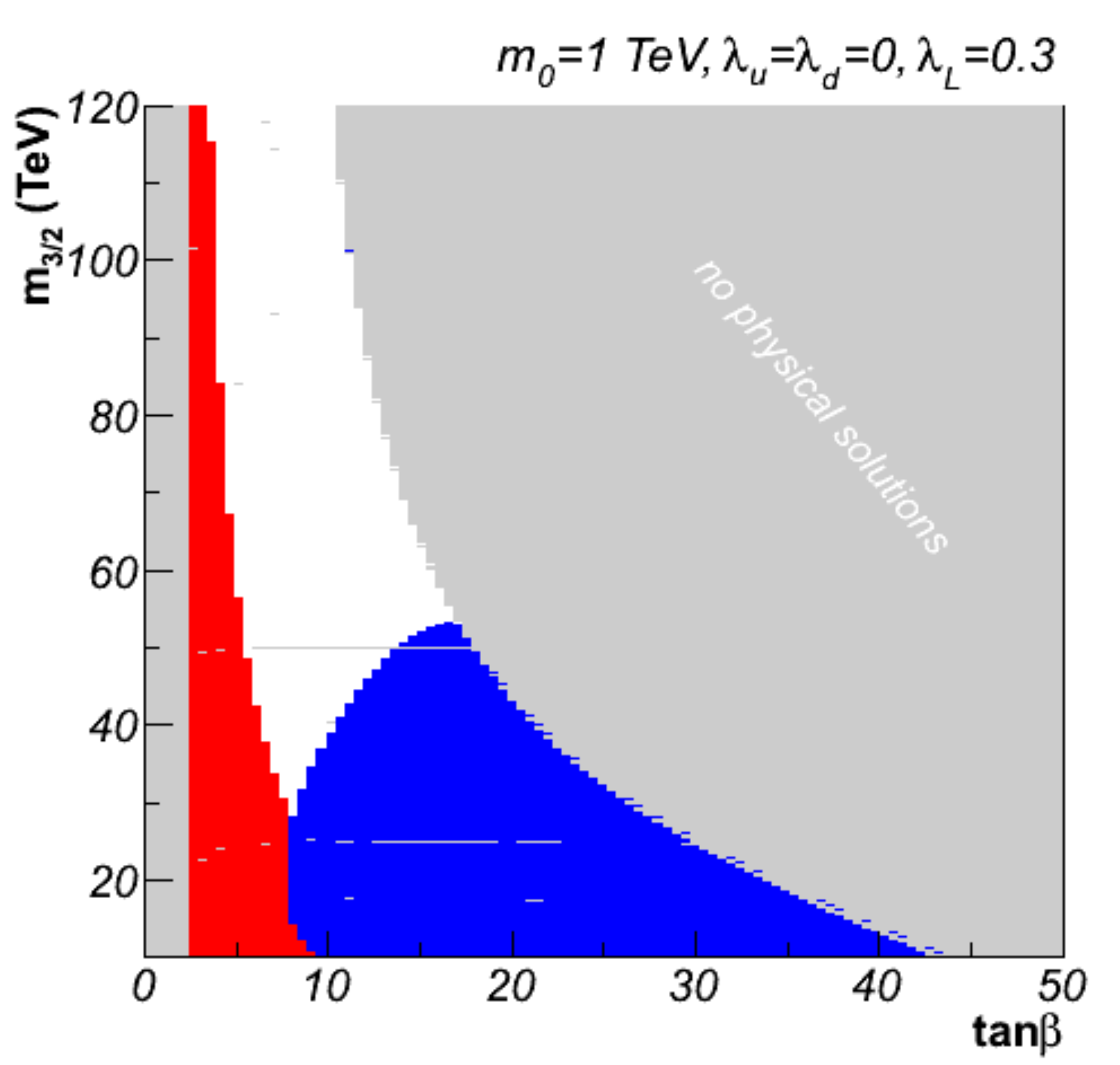}\\
	\includegraphics[scale=0.33]{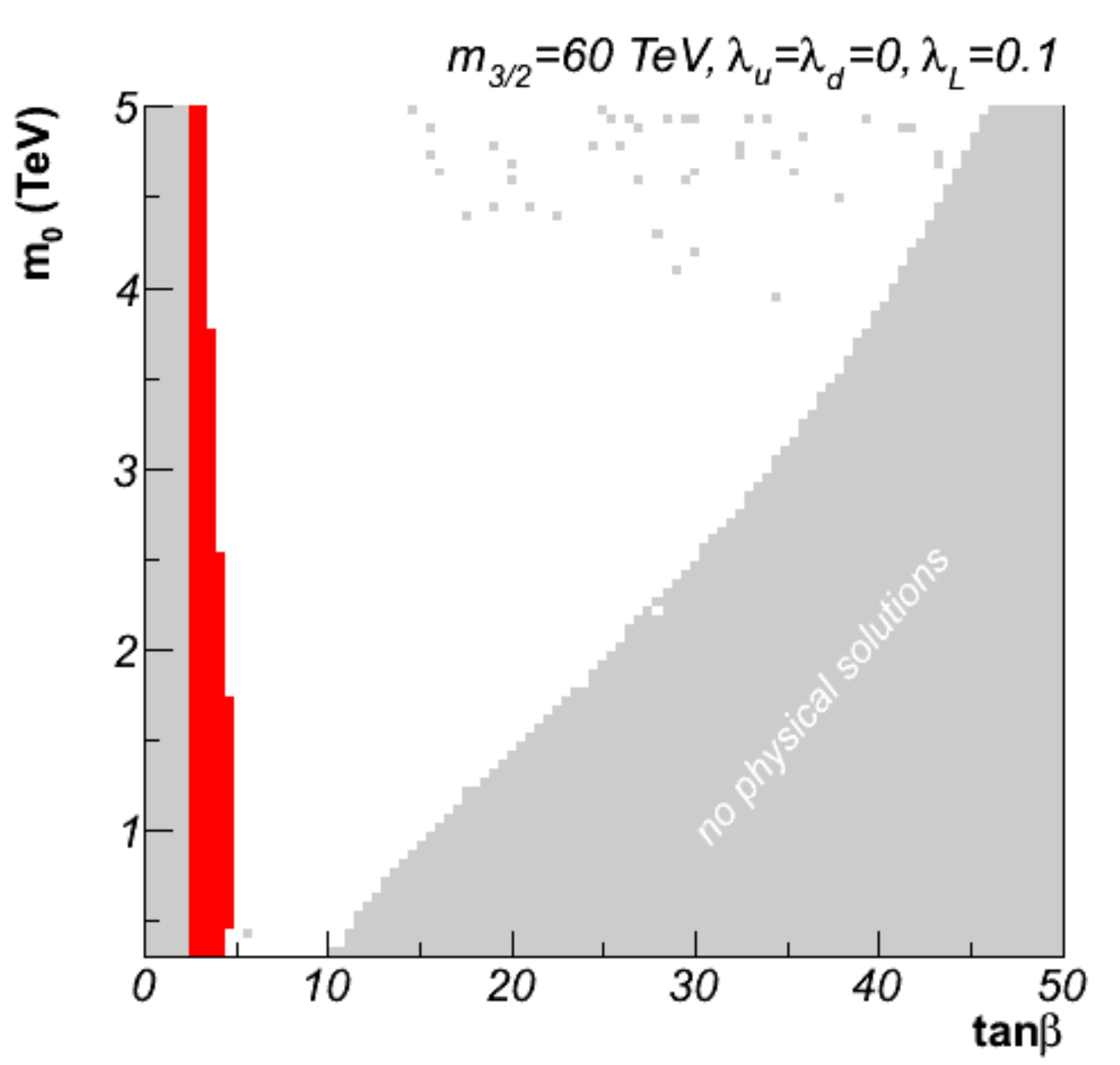}
	\includegraphics[scale=0.33]{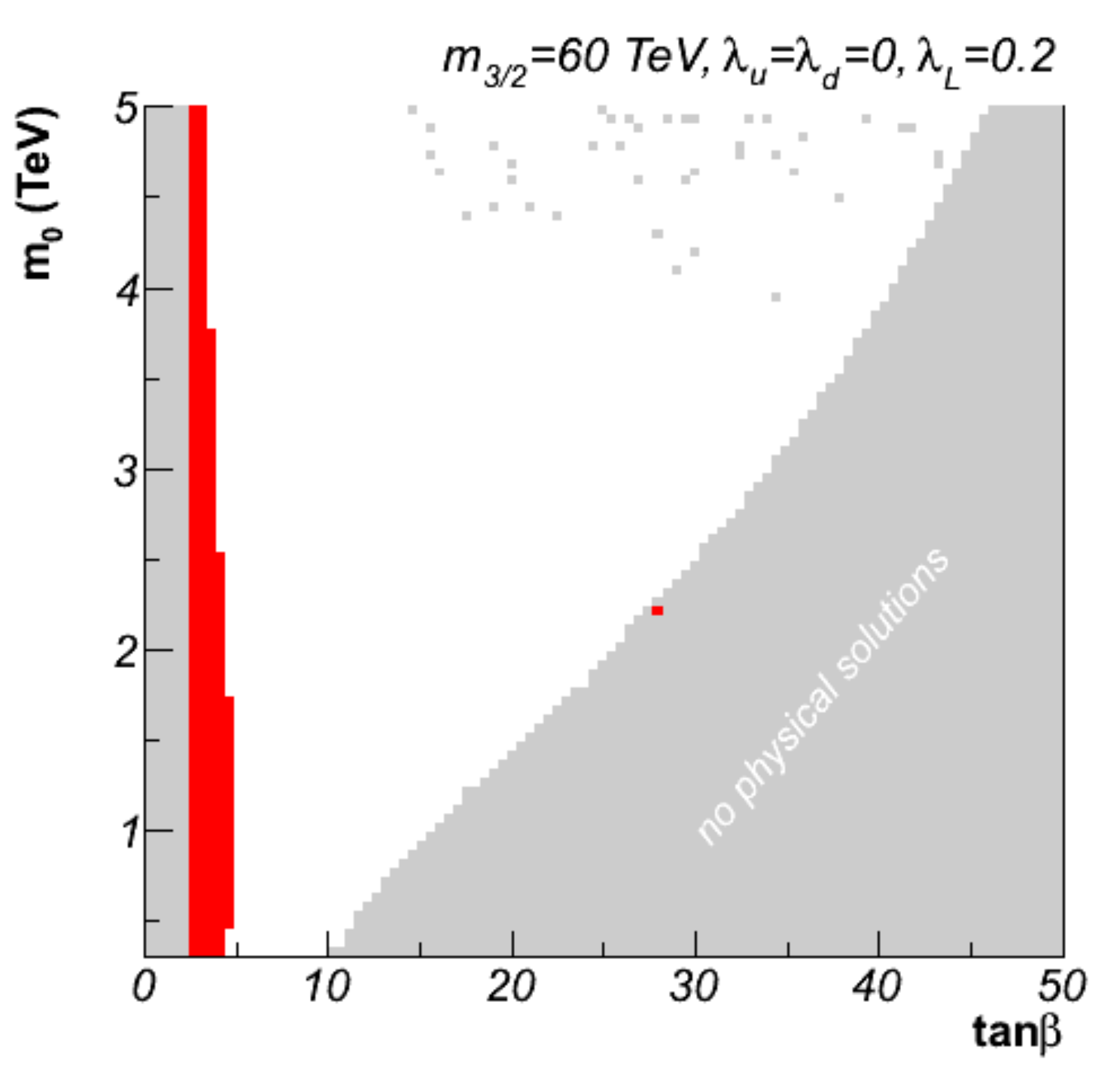}
	\includegraphics[scale=0.33]{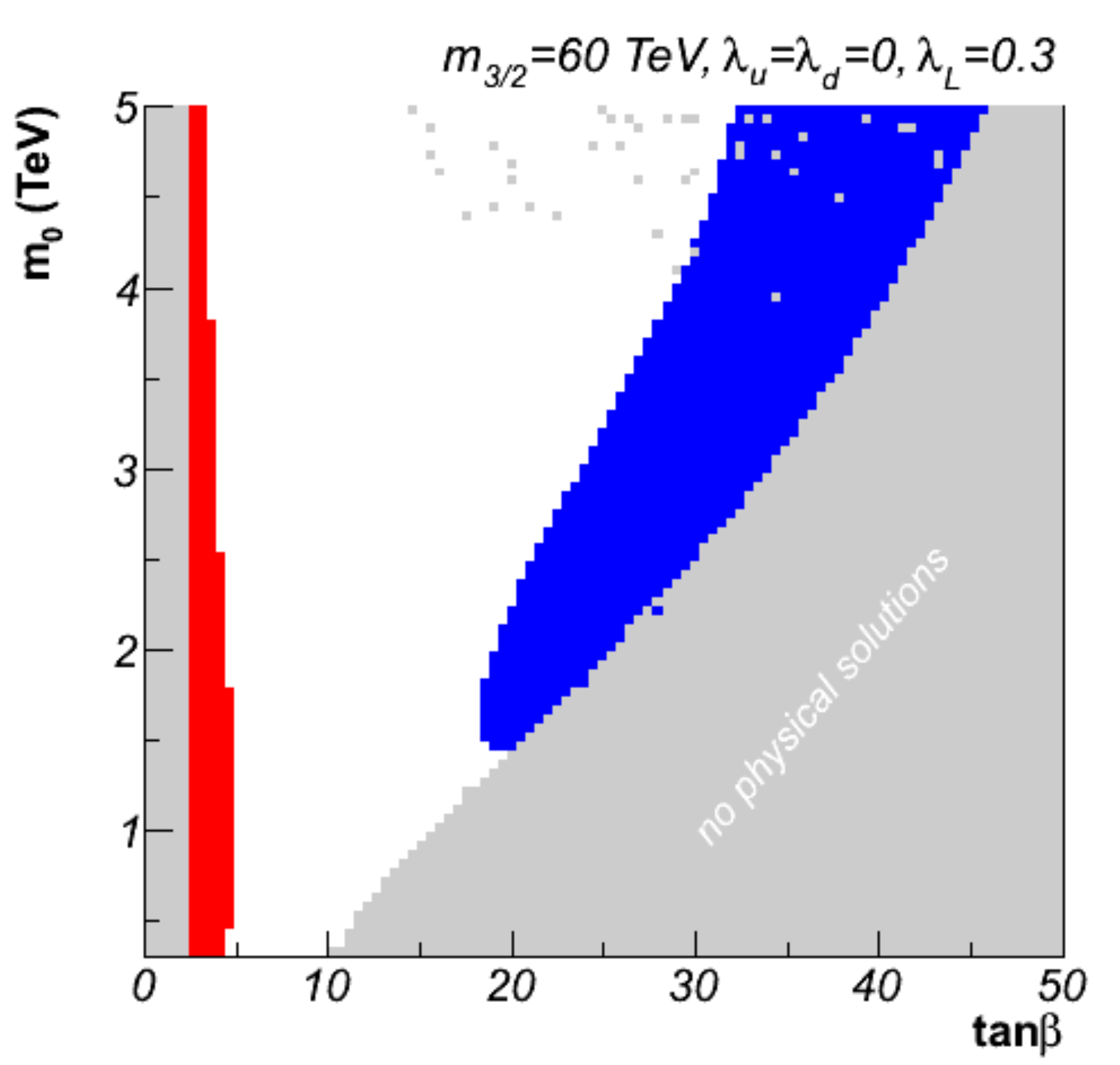}
\end{center}
\caption{Scans of NMFV AMSB scenarios in the $(\tan\beta, m_{3/2})$ plane with a
  fixed value of $m_0=1$ TeV (upper panels) and in the $(\tan\beta, m_0)$ plane
  with a fixed value of $m_{3/2}=60$ TeV (lower panels). The sign of the
  $\mu$-parameter is chosen positive and we present the results for different
  values of the NMFV-parameter $\lambda_L$. We show regions excluded due to the
  absence of physical solutions to the renormalization group equations (grey), by
  the constraints associated to the $b\to s\gamma$ branching ratio (blue), the
  $B_s^0$-meson branching to a muon pair ${\rm BR}(B_s^0 \to \mu^+\mu^-)$
  (green) and the lightest 
  Higgs-boson mass (red).}
\label{fig:m60}
\end{figure}

We now turn to non-minimal flavour violation in AMSB scenarios, where
non-vanishing $\lambda$-parameters can arise, \eg, from non-trivial K\"ahler
interactions as in supergravity. In our phenomenological approach, 
we introduce non-minimal flavour violation at
the weak scale, generalizing the squark mass matrices by introducing the
three parameters defined in Eq.\ \eqref{eq:lambda}. We use again the {\tt
SPheno 3.0} program to diagonalize the squark sector and compute the
flavour and weak observables presented in Sec.\ \ref{sec:constraints}. In Figs.\
\ref{fig:tb10} and \ref{fig:m60}, we depict the impact of the considered NMFV
parameters on these observables, showing the associated constraints on the
non-minimal AMSB parameter space. We impose the above-mentioned limit for the
Higgs-boson mass, \ie, $m_h \gtrsim 111$ GeV (see the discussion in 
Section \ref{sec:constraints}), as well as all the other 
constraints at the 2$\sigma$ confidence level. We present different typical AMSB
planes for different values of the flavour-violating $\lambda$-parameters,
keeping only the most constraining observables for clarity and neglecting the
anomalous magnetic moment of the muon for the reasons mentioned above.

Typical scans in the $(m_0, m_{3/2})$ planes of the AMSB parameter space are
shown in Fig.\ \ref{fig:tb10}. We use a fixed value of $\tan\beta=10$ and choose
a positive sign for the $\mu$-parameter. Comparing with the corresponding
minimal AMSB results (left panel of Fig.\ \ref{fig:MFV}), we find that the most
sensitive observable is the $b\to s\gamma$ branching ratio, which directly
probes
squark mixing in the left-left chiral sector, \ie, the $\lambda_L$-parameter.
This strong dependence is due to squark and neutralino-chargino loops, involving
$SU(2)_L$ interactions between squarks, quarks and neutralinos or charginos,
proportional to the squark mixing matrices $R^u$ and
$R^d$ (see Eq.\ \eqref{eq:couplstr} below). This also explains the less
pronounced sensitivity to NMFV mixing in the right-right chiral sectors
($\lambda_u$ and $\lambda_d$), where
the results do not show any strong dependence on the $\lambda$-parameters. At
intermediate values of $\lambda_{\rm L}\sim 0.1$, the interplay between squark
masses and mixings is such that almost all the parameter space accessible by
renormalization group running remains allowed by the $b\to s
\gamma$ branching ratio compared to the minimal results of Fig.\
\ref{fig:MFV}. Similar conclusions hold for the $b\to s\mu\mu$ branching ratio,
not presented on the figures for clarity. The other considered observables, \eg,
the constraint on the lightest Higgs boson mass, are barely
sensitive to non-minimal flavour violation, as the dependence on the squark
sector is subdominant.  

In Fig.\ \ref{fig:m60}, we investigate the dependence of the
predictions for weak and flavour observables on $\tan\beta$. 
We first fix the universal scalar
mass $m_0$ to 1 TeV and scan over the $(\tan\beta, m_{3/2})$ plane. Secondly,
we fix the gravitino mass $m_{3/2}$ to 60 TeV and scan over the $(\tan\beta, m_0)$
plane. In both cases, the sign of the $\mu$-parameter is again chosen to be
positive. As for the analysis in Fig.\ \ref{fig:tb10}, the most sensitive
observables to non-minimal flavour violation are the $b\to s\gamma$ and $b\to
s\mu\mu$ branching ratios. Since the constraints related to the second decay are
weaker and overlapping with those of the first, they are again omitted from the
figures. The regions of the parameter space excluded by the constraints
associated to the $b\to s \gamma$ branching ratio increase 
with the values of the flavour-violating parameter $\lambda_{\rm
L}$. For small values, squark mixings and mass effects compensate,
which leaves again almost all the parameter space open. In fact, the
exclusion contours related to the $b\to s \gamma$ branching ratio even
vanish, when 
$\lambda_{\rm L}$ increases slightly. Such contours reappear only at
much larger $\lambda_{\rm L}$ values. As a consequence, for
$\tan\beta \gtrsim 20$, possible flavour violating entries in the squark mass
matrices are reduced to be rather modest, \ie, $\lambda_{\rm L} \lesssim 0.15$.
On the other hand, very low values of $\tan\beta$ are excluded by
the bounds on the lightest Higgs boson mass for all values of $\lambda_{\rm L}$.
It is also interesting to remark that the predictions for the ${\rm BR}(B_s^0 \to
\mu^+\mu^-)$ observable lie above the current experimental bounds of Eq.\
\eqref{eq:bs0} in the rather large $\tan\beta$ region, for moderate values of
the mixing parameter of the left-left chiral sector ($\lambda_{\rm L} \in [0.1 -
0.2]$) and for relatively light gravitino and universal scalar masses.

Let us finally note that the results for non-minimal flavour violation in the
right-right chiral sector are unchanged with respect to the minimal case shown
in Fig.\ \ref{fig:MFV}, and we subsequently do not present the corresponding
figures. 

\subsection{Benchmark scenarios for non-minimally flavour violating AMSB scenarios \label{sec:benchmark}}

Inspecting the various AMSB planes presented in Sec.\ \ref{sec:planes}, we
select three benchmark scenarios allowed by the present low-energy and electroweak
precision constraints, which permit a sizeable mixing between second and third
generation squarks and which are collider-friendly in the sense that one
or several superpartners could be produced with a large rate at the LHC.
This means that at least some of the superpartners should not be too heavy.
The chosen benchmark points are presented 
in Tab.\ \ref{tab:points}, together with the SPS9 benchmark point \cite{SPS}
which is shown as a reference. As a generic feature for all the four points, the
gravitino mass is taken as $m_{3/2}=60$ TeV and the sign of the off-diagonal
Higgs mixing parameter is positive, for the reasons discussed in Sec.\
\ref{sec:constraints}. Due to the renormalization group running invariance of the
gaugino mass parameters $M_1$, $M_2$ and $M_3$, the dependence of the masses of
the gluino, the two lightest neutralinos and the lightest chargino, all mainly
gaugino-like, on $m_0$ and $\tan\beta$ is reduced and
subdominant. Therefore, they are roughly identical for all scenarios, with
$m_{\tilde{g}} \sim 1300 - 1400$~GeV, $m_{\tilde{\chi}^0_1} \sim
m_{\tilde{\chi}^{\pm}_1} \sim 175$~GeV, and $m_{\tilde{\chi}^0_2} \sim 550$~GeV.
In contrast, the scalar spectrum is quite different.

\begin{table}
\caption{Three AMSB benchmark scenarios allowing for sizeable flavour-violating
         entries in the squark mass matrices with respect to the second and
         third generation mixing in the left-left and/or the right-right chiral
         squark sectors. The SPS9 scenario is also presented for comparison.
         We indicate the SUSY-breaking parameters
         at the high scale and the resulting masses at low energy (after
         renormalization group running) for the gluino, the lightest squarks and
         sleptons, the two lightest neutralinos, the lightest chargino, and
         the lightest Higgs-boson. The values are presented assuming constrained
         minimal flavour-violation, \ie, when $\lambda_L=\lambda_u=\lambda_d=0$.}
\begin{tabular}{|c||cccc||ccccccccc|}
  \hline
  & &&& &  \multicolumn{9}{c|}{[GeV]} \\
  & $m_{3/2} {\rm [TeV]}$ & $m_0 {\rm [TeV]}$ & $\tan\beta$ & ${\rm sgn}(\mu)$ & $m_{\tilde{g}}$ &
    $m_{\tilde{u}_1}$ & $m_{\tilde{d}_1}$ & $m_{\tilde{\ell}_1}$ &
    $m_{\tilde{\nu}_1}$ & $m_{\tilde{\chi}^0_1}$ & $m_{\tilde{\chi}^0_2}$ &
    $m_{\tilde{\chi}^{\pm}_1}$ & $m_{h^0}$ \\ 
  \hline\hline
  SPS9 & 60   & 0.45 & 10  & + & 1321.5 &  950.0 & 1133.3 &  339.7 & 364.1 &
    173.5 & 542.2 & 173.7 & 118.4  \\
\hline
  I & 60   & 1    & 10  & + & 1352.3 & 1087.6 & 1332.9 &  947.8 & 957.5 & 174.5
    & 547.2 & 174.7 & 116.9  \\
  II& 60   & 2    & 20  & + & 1394.2 & 1469.7 & 1865.4 & 1906.4 & 1940.6 & 176.4
    & 550.8 & 176.6 & 118.7  \\
  III & 60   & 3    & 30  & + & 1417.0 & 1950.7 & 2460.7 & 2743.1 & 2860.6 & 176.9
    & 552.6 & 177.1 & 121.0  \\
  \hline
\end{tabular}
\label{tab:points}
\end{table}

The point SPS9 presents a low value of $\tan\beta=10$ as
well as a relatively low sfermion mass
parameter $m_0=450$ GeV. Therefore, the masses of the 
colour-neutral scalar partners of the SM
fermions remain rather moderate, while the squark masses are comparable to the gluino
mass. The lightest Higgs boson mass,
$m_{h^0}=118.4$ GeV, lies well within the limit of Eq.\ (\ref{eq:higgsmass}). 
Our first benchmark point I differs very little from the SPS9
scenario, with a
moderately larger universal scalar mass $m_0=1$ TeV. As a consequence, the
slepton masses are of about 1 TeV, while the squarks are only slightly heavier.
Even if the lightest Higgs boson mass is now a bit smaller, it lies still well
above the excluded limit. 

The points II and III feature higher values for $\tan\beta=20$ and 30,
respectively. In order to be able to solve the renormalization group equations,
one must then either decrease the value of $m_{3/2}$ or increase the value of
$m_0$, as it can be seen from Fig.\ \ref{fig:MFV}. Since the first possibility does
not allow for large NMFV mixing in the squark sector due to the $b\to s \gamma$
branching ratio constraint, we adopt the second choice and fix the universal
scalar mass to $m_0=2$ TeV and 3 TeV for the scenarios II and III, respectively.
Sfermions are therefore considerably heavier, with masses lying in the $2 - 3$
TeV range. The mass of the lightest Higgs boson lies still
above the constraint of Eq.\ \eqref{eq:higgsmass}. Let us note that for these
two scenarios, the squarks are lighter than the sleptons, which may lead to
non-leptonic supersymmetric cascade decays as a typical collider signature. 

\begin{figure}
\begin{center}
	\includegraphics[width=.31\columnwidth]{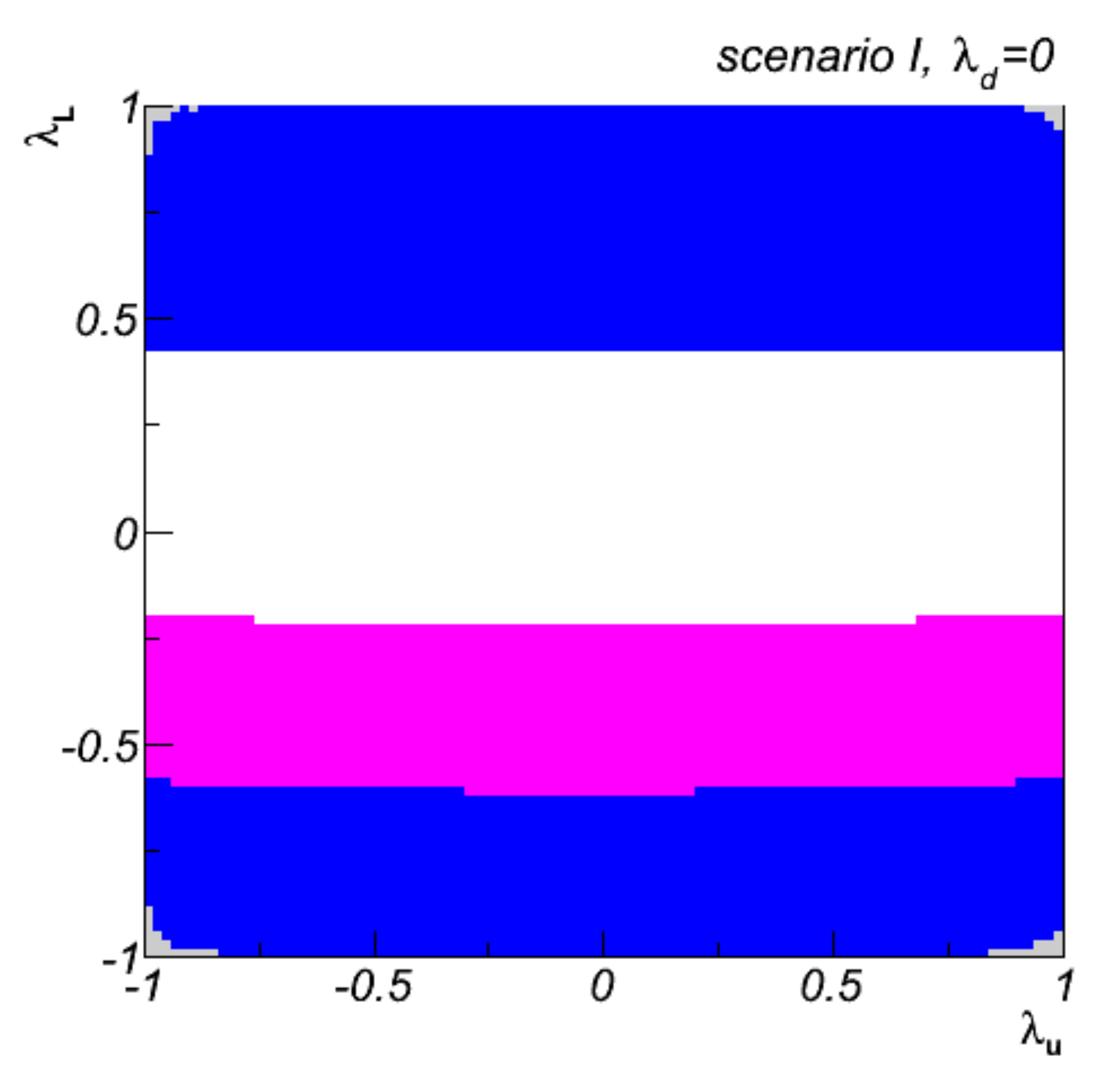} 
	\includegraphics[width=.31\columnwidth]{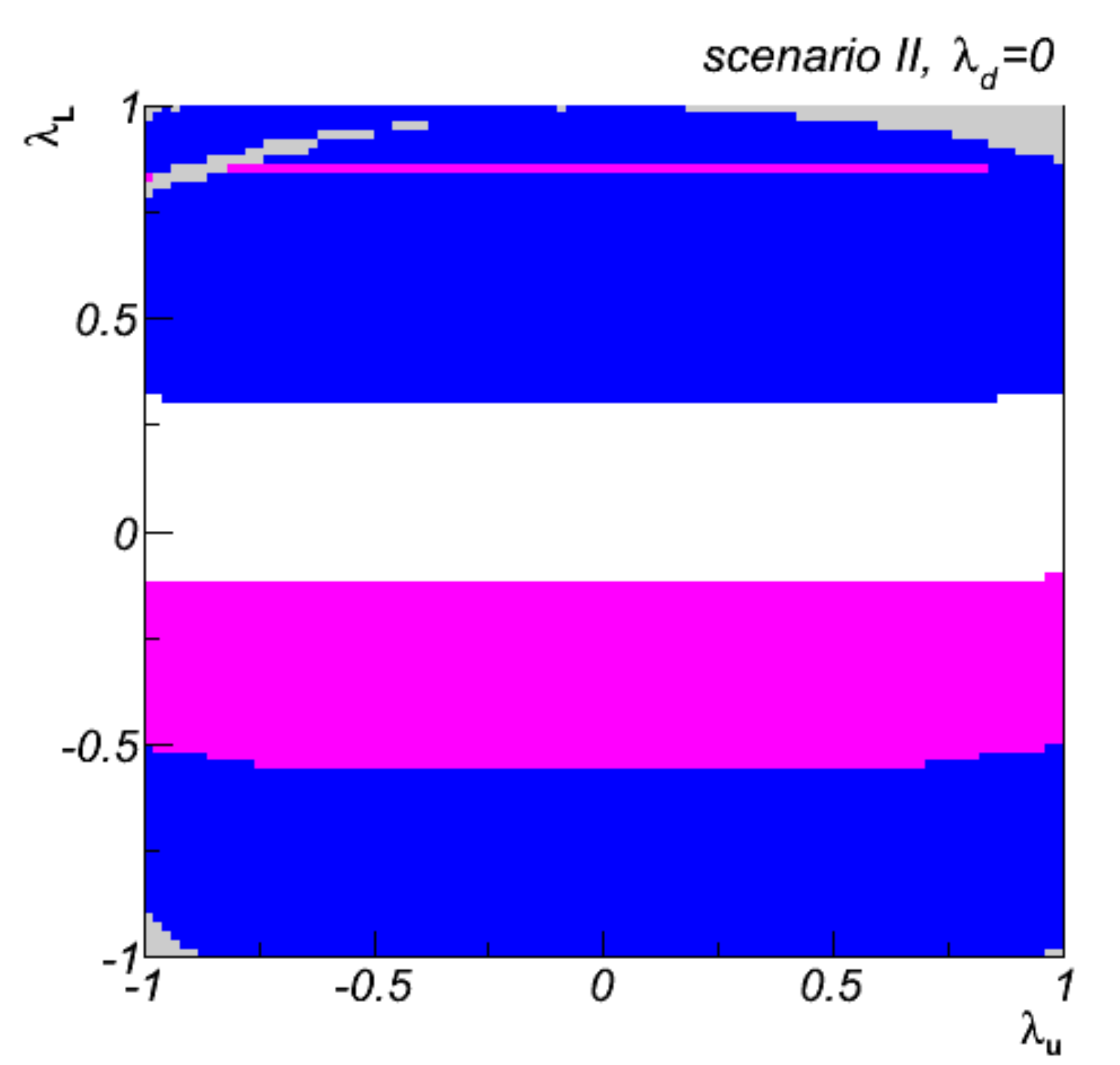} 
	\includegraphics[width=.31\columnwidth]{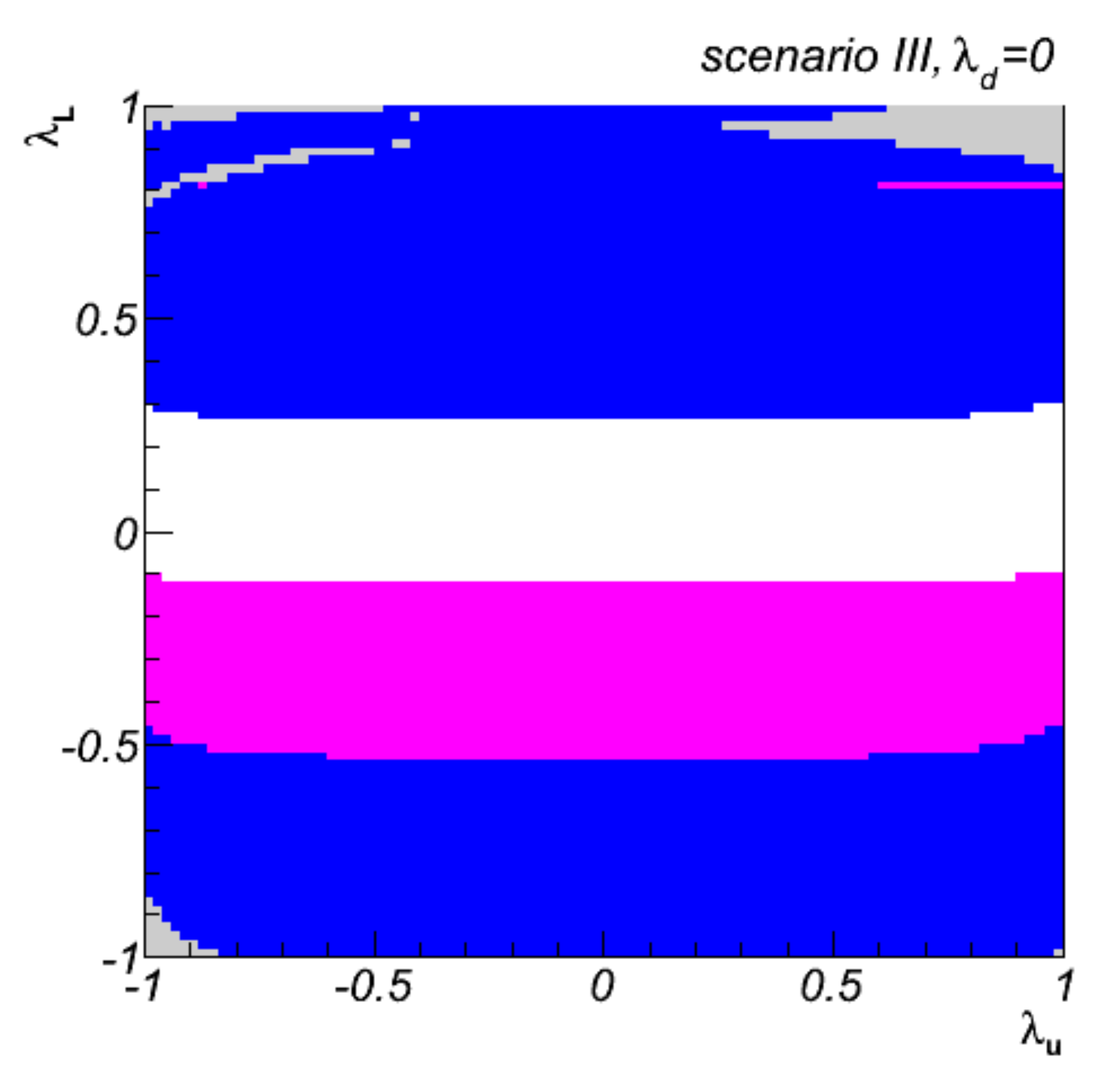} \\
	\includegraphics[width=.31\columnwidth]{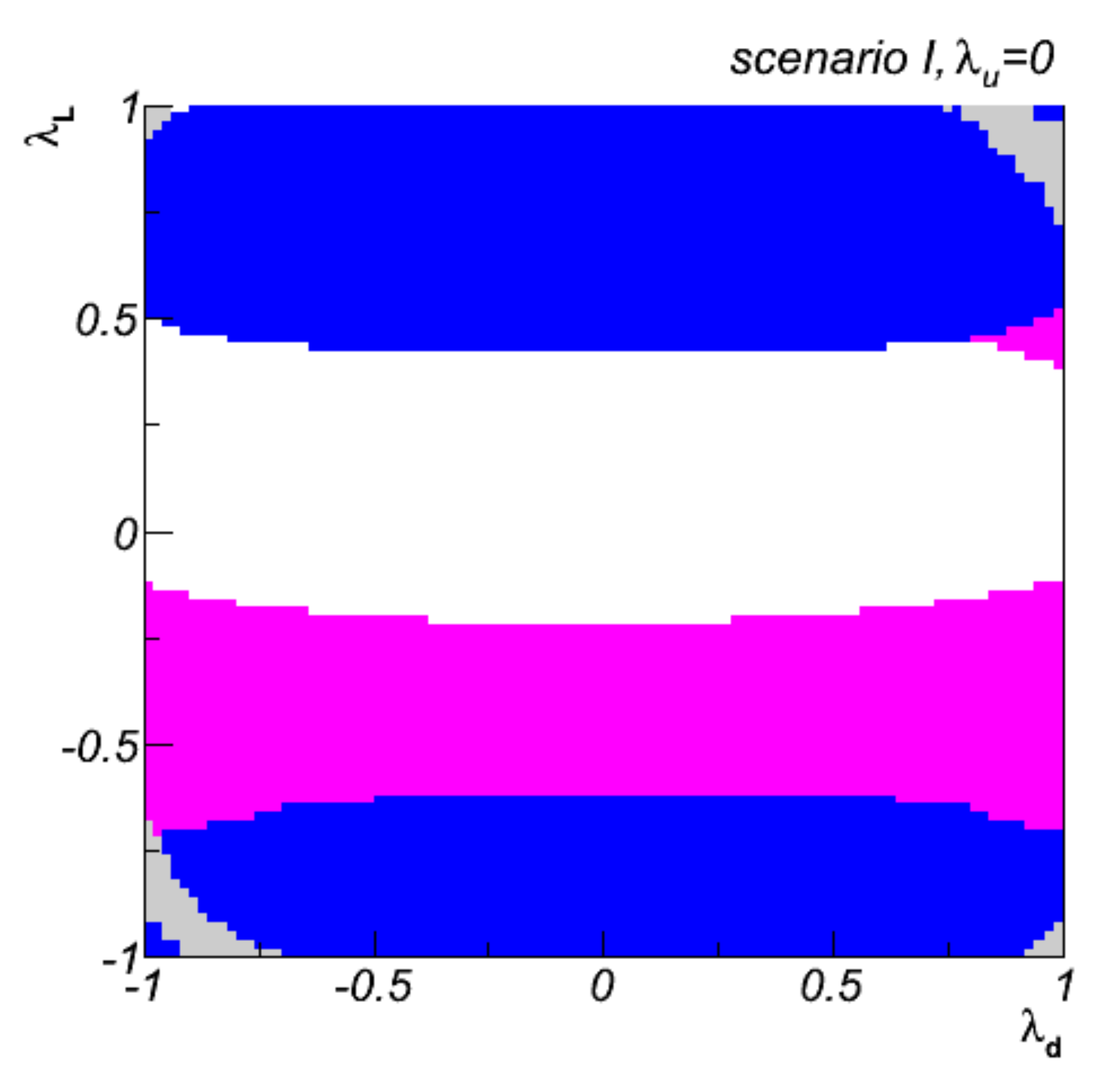} 
	\includegraphics[width=.31\columnwidth]{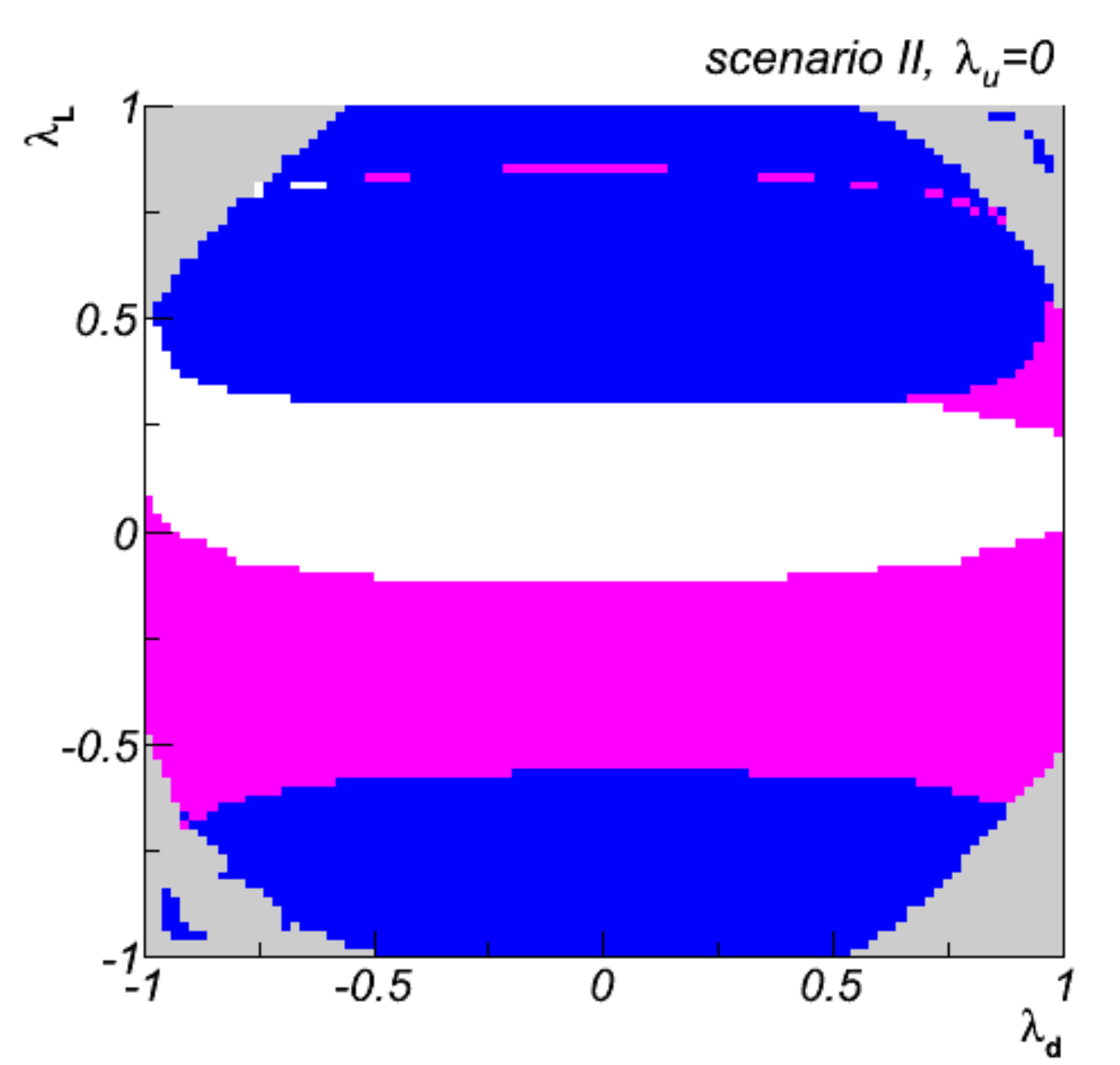} 
	\includegraphics[width=.31\columnwidth]{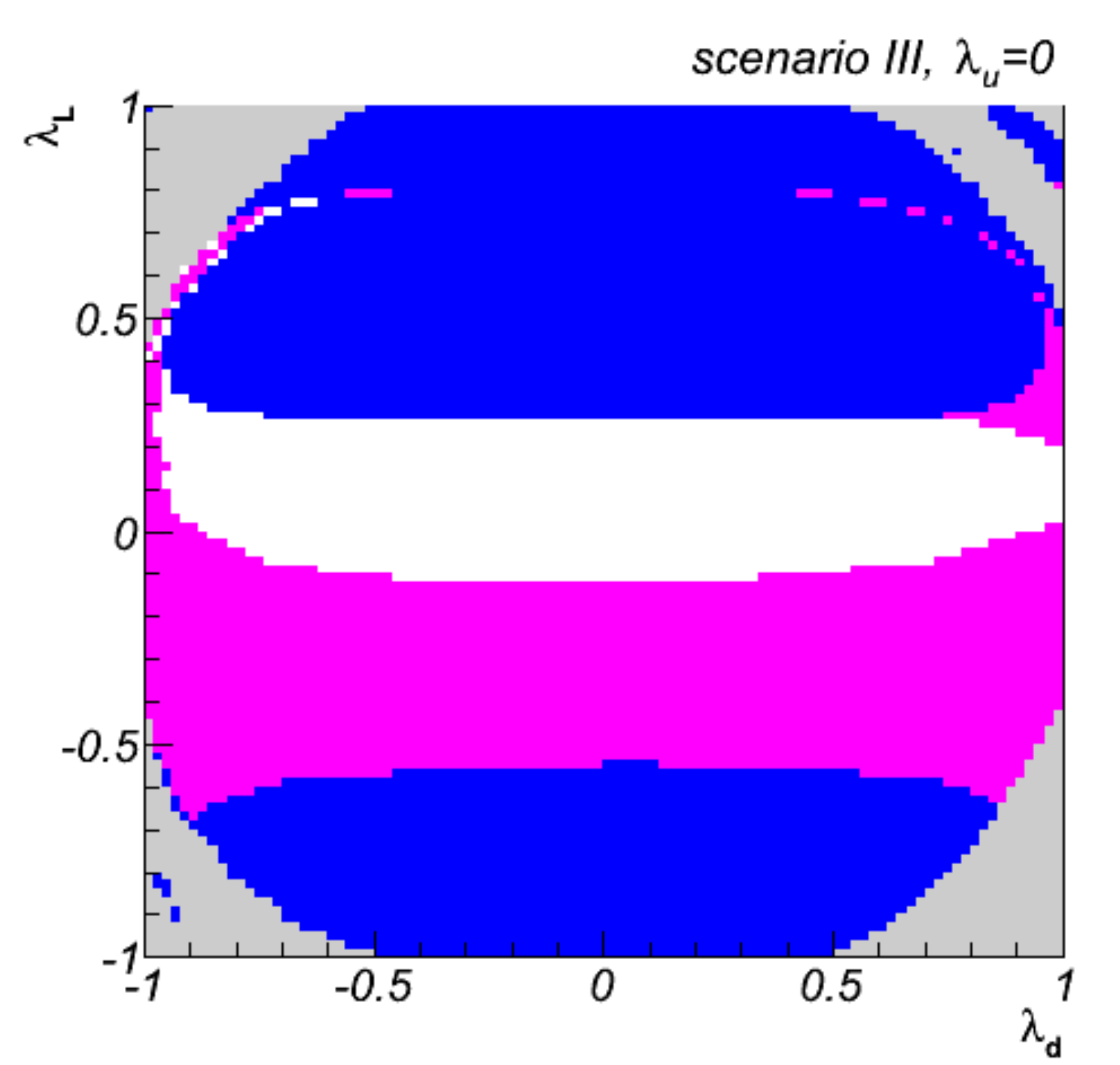} 
\end{center}
\caption{ $(\lambda_{\rm L}, \lambda_u)$ (upper panel) and $(\lambda_{\rm L},
   \lambda_d)$ (lower panel) planes for the three scenarios I (left), II (center)
   and III (right) presented in Tab.\ \ref{tab:points}. We show regions excluded
   due to the absence of physical solutions for the diagonalization of the mass
   matrices at the one-loop level (grey), the constraints associated to the
   $b\to s\gamma$ branching ratio (blue) and the $B$-meson mass difference $\Delta
   M_{B_s^0}$ (purple).}
\label{fig:LLRR}
\end{figure}

Starting from the minimally flavour-violating scenarios I, II, and III, we now
include possible non-minimal flavour-violation in the squark sector, \ie, we
allow for $\lambda_{\rm L}$, $\lambda_u$ and $\lambda_d$ to be
non-vanishing. However, we restrain ourselves to the cases where the
off-diagonal entries in the squark mass matrices do not exceed the diagonal
ones, \ie, $|\lambda| < 1$. We investigate in Fig.\ \ref{fig:LLRR} the
dependence of the predictions for the observables BR($b\to s \gamma$), $\Delta
M_{B_s^0}$, BR($b\to s\mu\mu$) and BR($B_s^0 \to \mu^+\mu^-$) 
on the NMFV-parameters and derive the
allowed ranges for $\lambda_{\rm L}$, $\lambda_u$, and $\lambda_d$ under the assumption of a single
non-vanishing parameter. We present in the $(\lambda_{\rm
L}, \lambda_u)$ (upper panel) and $(\lambda_{\rm L}, \lambda_d)$ (lower panel)
planes the regions excluded after confronting the AMSB predictions to the
observed experimental values. The regions where
the mass matrices, at the one-loop level, cannot be diagonalized are indicated
in grey.

As can be seen, strong constraints are imposed by the decay $b\to s\gamma$ and
the meson mass difference $\Delta M_{B_s^0}$. For all scenarios, while the $b\to
s\gamma$ branching ratio data forbid too large (absolute) values for the
$\lambda_{\rm L}$ parameter, $B_s^0-\bar{B}_s^0$ oscillations almost completely
forbid negative values. Contrary, in the right-right chiral sector non-minimal flavour
violation is left rather unconstrainted, if not too large values of $\lambda_{\rm
L}$ are assumed. Hence, in the case of the point I and assuming a single
dominant $\lambda$-parameter hypothesis, \ie, allowing only one single
non-vanishing off diagonal element in the squark mass matrices, NMFV in the
left-left chiral sector is constrained to $-0.22 \lesssim \lambda_{\rm L}
\lesssim 0.42$ (with the two other parameters being set to
$\lambda_u=\lambda_d=0$).
On the other hand, assuming non-minimal flavour violation only in the
right-right chiral sector ($\lambda_{\rm L}=0$), we observe that both the
$\lambda_u$ and $\lambda_d$ parameters are left almost unconstrained, leading to
possible large flavour violating effects.

Similarly, regarding scenario II and accounting
for all constraints, the left-left chiral sector NMFV parameter is restricted to
$-0.12 \lesssim \lambda_{\rm L} \lesssim 0.3$ in the $\lambda_u=\lambda_d=0$
case, and
$\lambda_u$ is left unconstrained in the $\lambda_{\rm L}=0$ case. However, the
higher value of $\tan\beta=20$ renders the $\Delta M_{B_s^0}$ observable
sensitive to very high (absolute) values of $\lambda_d$, due to enhanced Yukawa
couplings with the
down-type (s)quark sector. This yields the constraint $-0.92\lesssim \lambda_d
\lesssim 0.96$ if the flavour-violation is assumed to be located only in the
down-type squark right-right chiral sector.

This effect on $\Delta M_{B_s^0}$ is a bit more pronounced 
for the third scenario,
with its large value of $\tan\beta = 30$, yielding moderate constraints
for scenarios with non-minimal flavour-violation in the right-right
down-type squark chiral sector ($-0.88\lesssim\lambda_d \lesssim 0.9$). 
For NMFV in the left-left chiral sector, with $\lambda_u=\lambda_d=0$, one
observes that the allowed range for $\lambda_{\rm L}$ is
now severely restricted, as for the second scenario, 
to $-0.12\lesssim \lambda_{\rm L} \lesssim 0.26$. All the
results are summarized in Tab.\ \ref{tab:ranges}.

\begin{table}
\caption{Ranges for the flavour-violating parameters $\lambda_{\rm L}$,
$\lambda_u$ and $\lambda_d$ compatible with the low-energy and electroweak
precision observables in the case of our three reference scenarios of Tab.\
\ref{tab:points}. The limits are given under the assumption of a single NMFV
parameter, \ie, where only one single parameter is allowed to vary, the other
being set to zero. If no value is indicated, the whole explored range of $-1 <
\lambda_{u,d} < 1$ is allowed.}
\begin{tabular}{|c|ccc|}
	\hline
	     & \qquad $\lambda_{L}$ \qquad & \qquad $\lambda_{u}$ \qquad & \qquad $\lambda_{d}$ \qquad \\
	\hline
	   I & $[-0.22, 0.42]$ & --                   & --                  \\
	   II & $[-0.12, 0.30]$ & --                   & $[-0.92, 0.96]$     \\
	   III & $[-0.12, 0.26]$  & --        		 & $[-0.88, 0.90]$      \\
	\hline
\end{tabular}
\label{tab:ranges}
\end{table}

\begin{figure}
\begin{center}
	\includegraphics[scale=0.35]{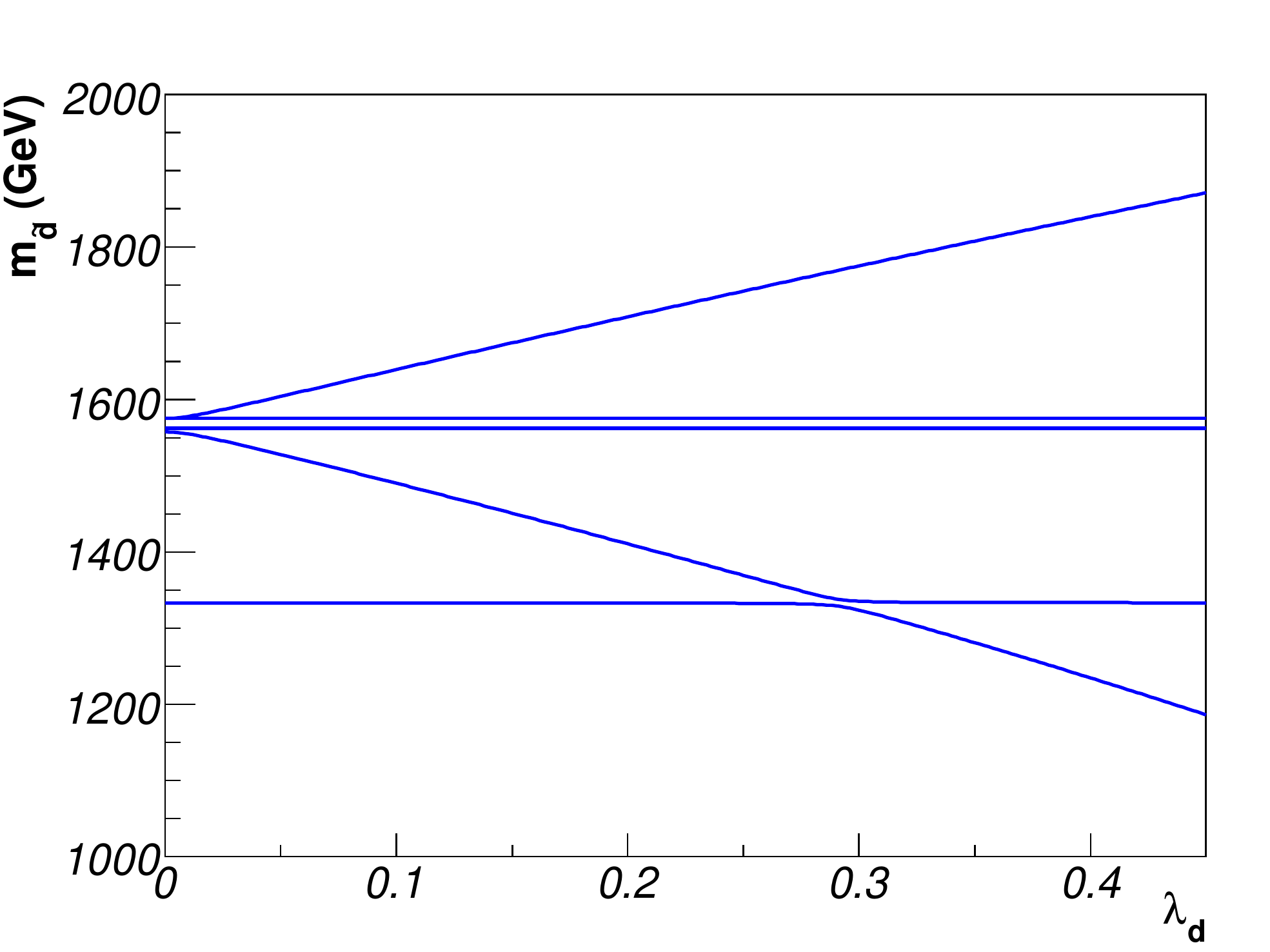} \quad
	\includegraphics[scale=0.35]{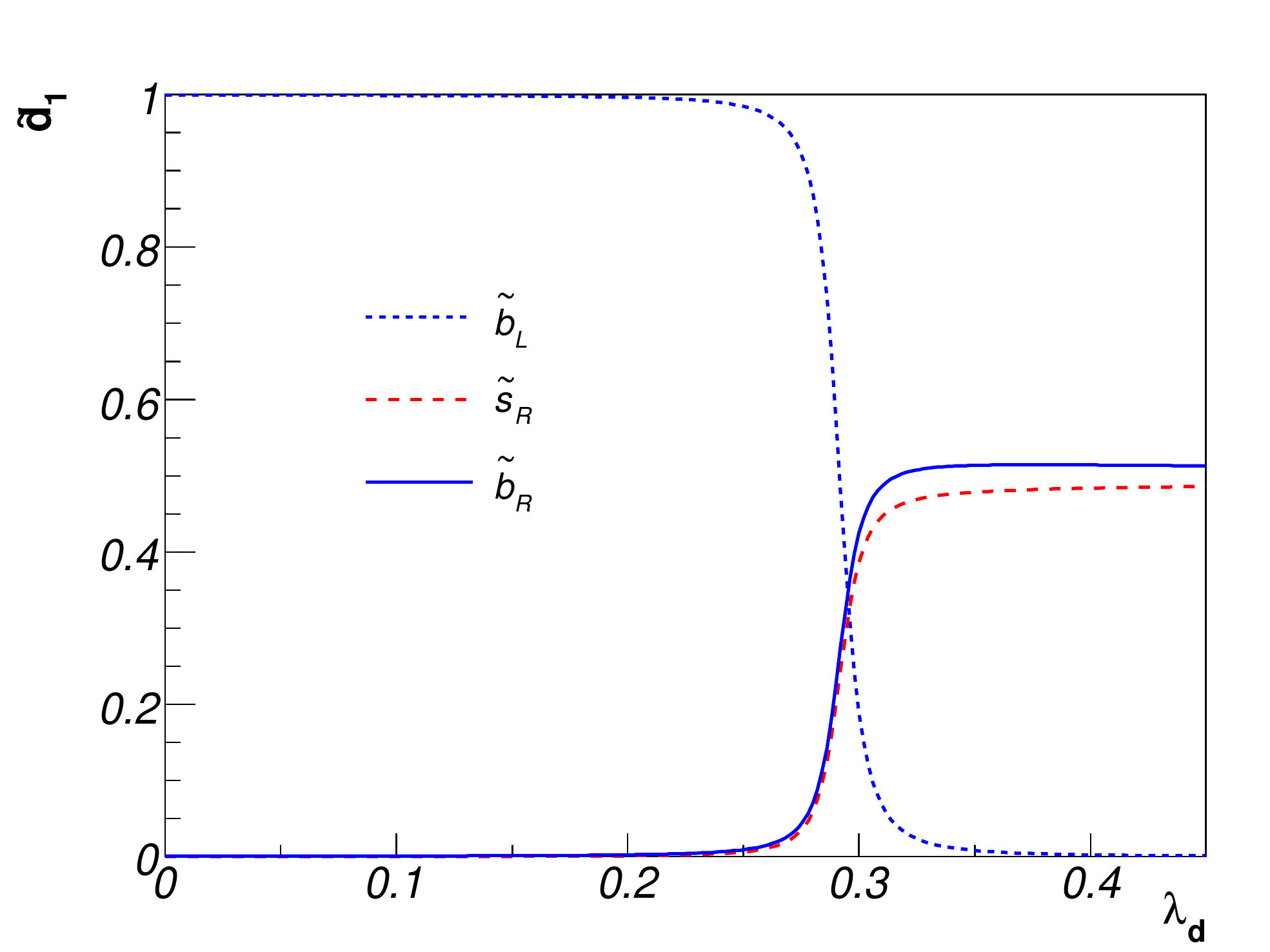} \\
	\includegraphics[scale=0.35]{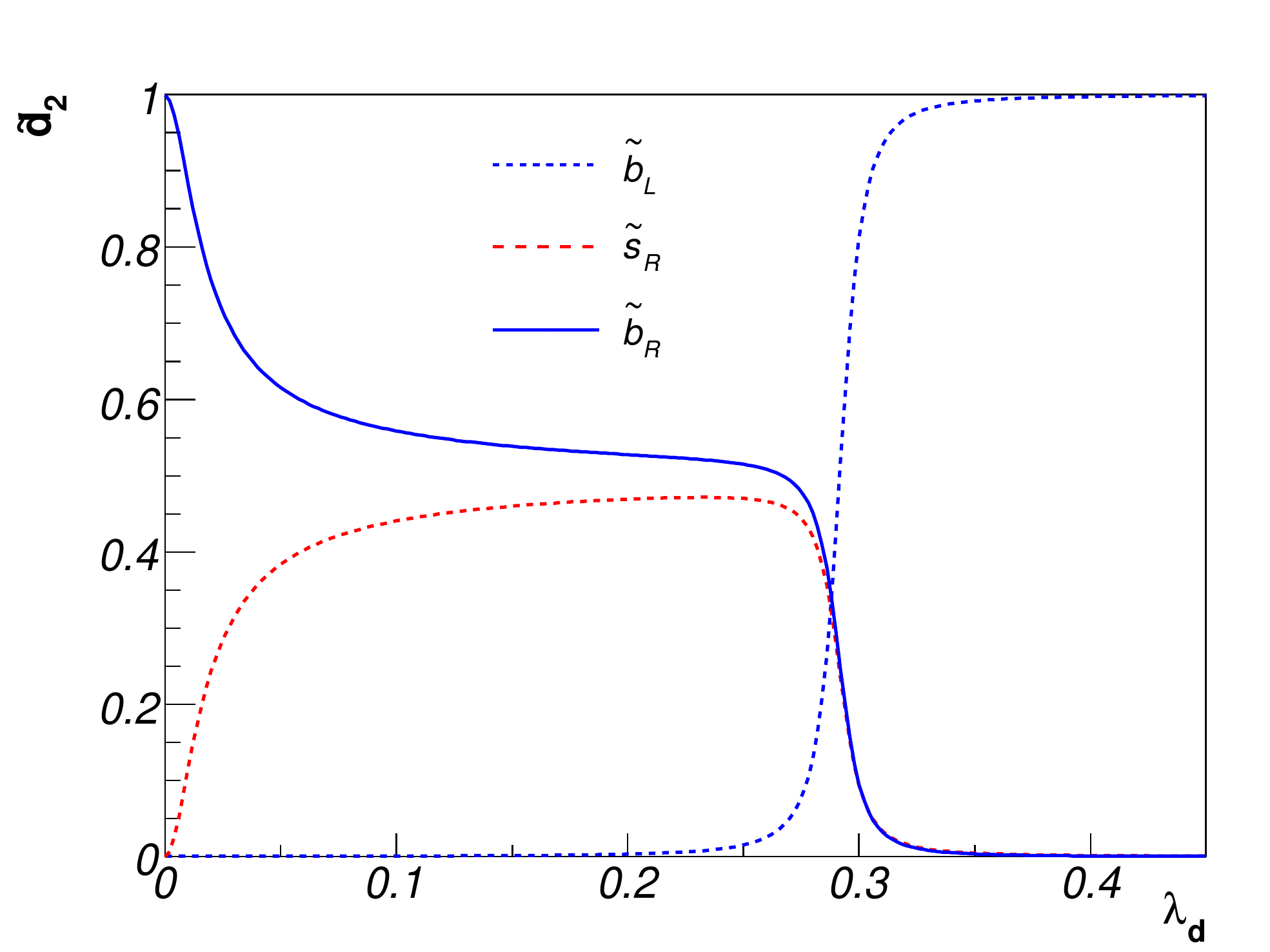} \quad
	\includegraphics[scale=0.35]{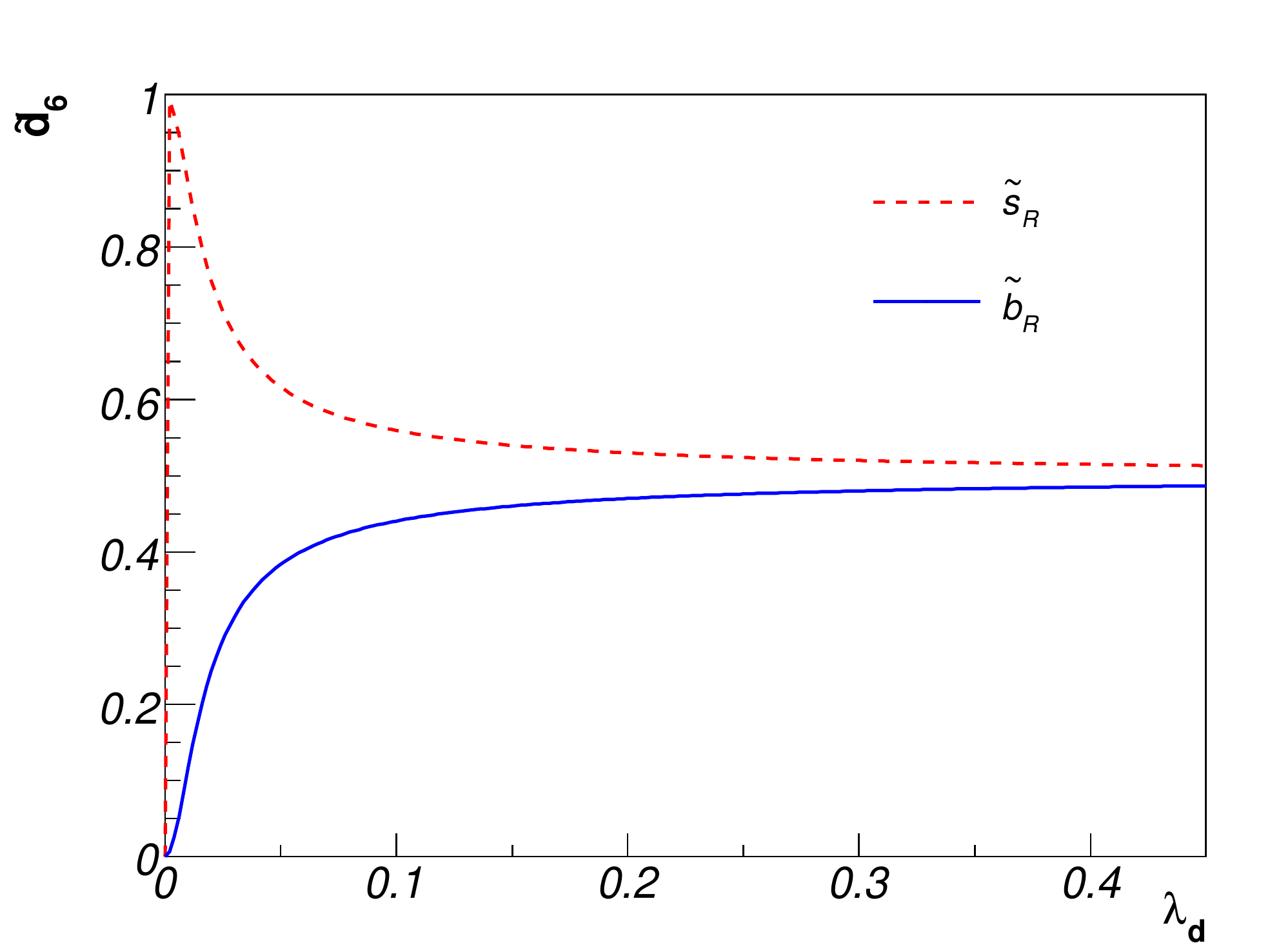} 
\end{center}
\caption{Dependence of masses and flavour decomposition of the lightest, second
lightest, and heaviest down-type squarks on the flavour-violating parameter
$\lambda_{d}$ for the scenario I of Tab.\ \ref{tab:points}.}
\label{fig:squarks1}
\end{figure}

For scenario I, we show in Fig.\ \ref{fig:squarks1} the
dependence of the mass eigenvalues and the flavour decomposition of selected
down-type squarks on the NMFV-parameter $\lambda_d$, as an example, which
induces a
$\tilde{b}_R-\tilde{s}_R$ mixing. With increasing off-diagonal elements in the
squark mass matrix, the resulting splitting of the physical mass eigenvalues
becomes more important. Consequently, the lightest down-type squark
$\tilde{d}_1$ becomes lighter, while the mass of the heaviest down-type state
$\tilde{d}_6$ increases.

\begin{figure}
\begin{center}
	\includegraphics[scale=0.35]{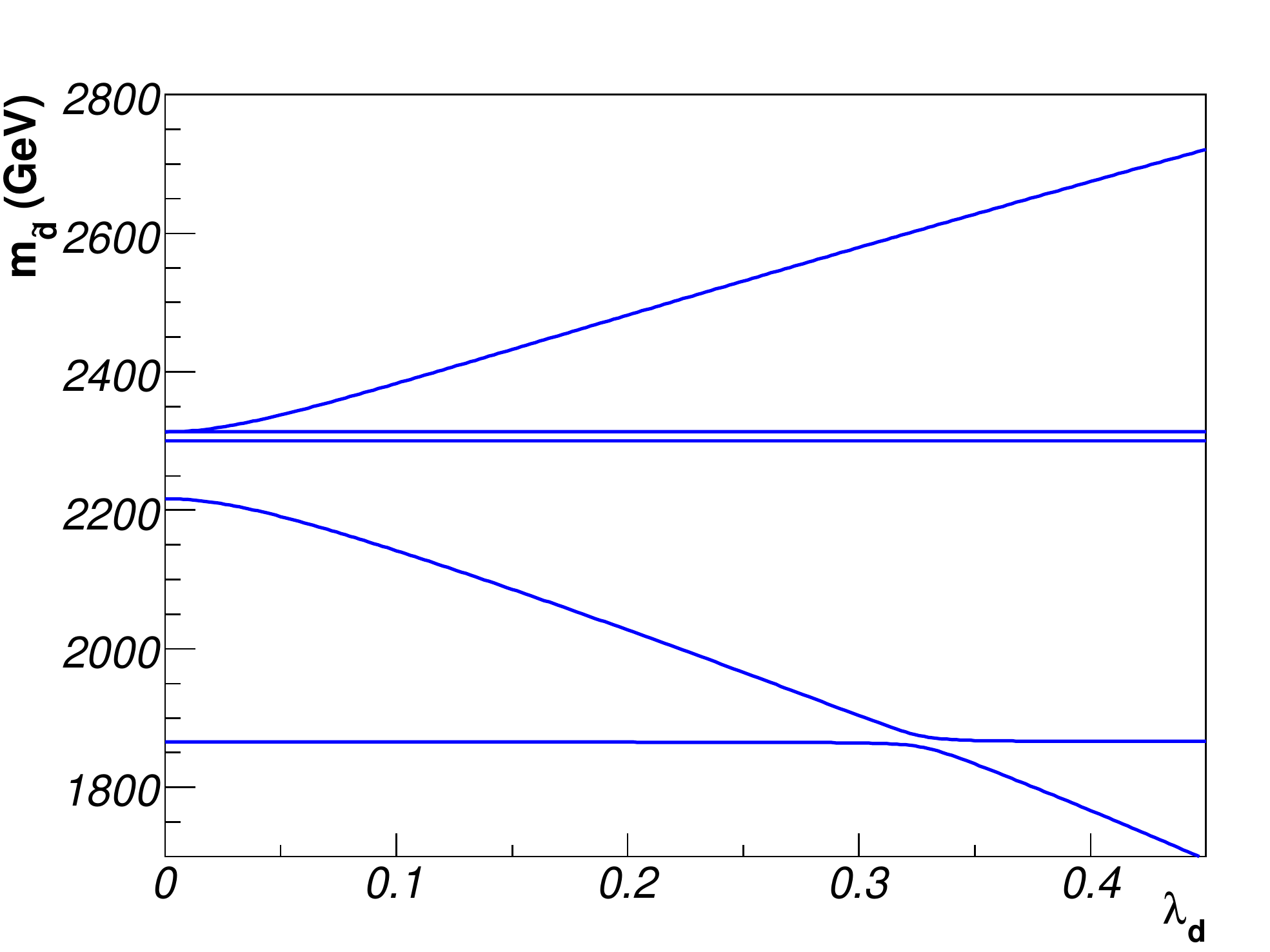} \quad
	\includegraphics[scale=0.35]{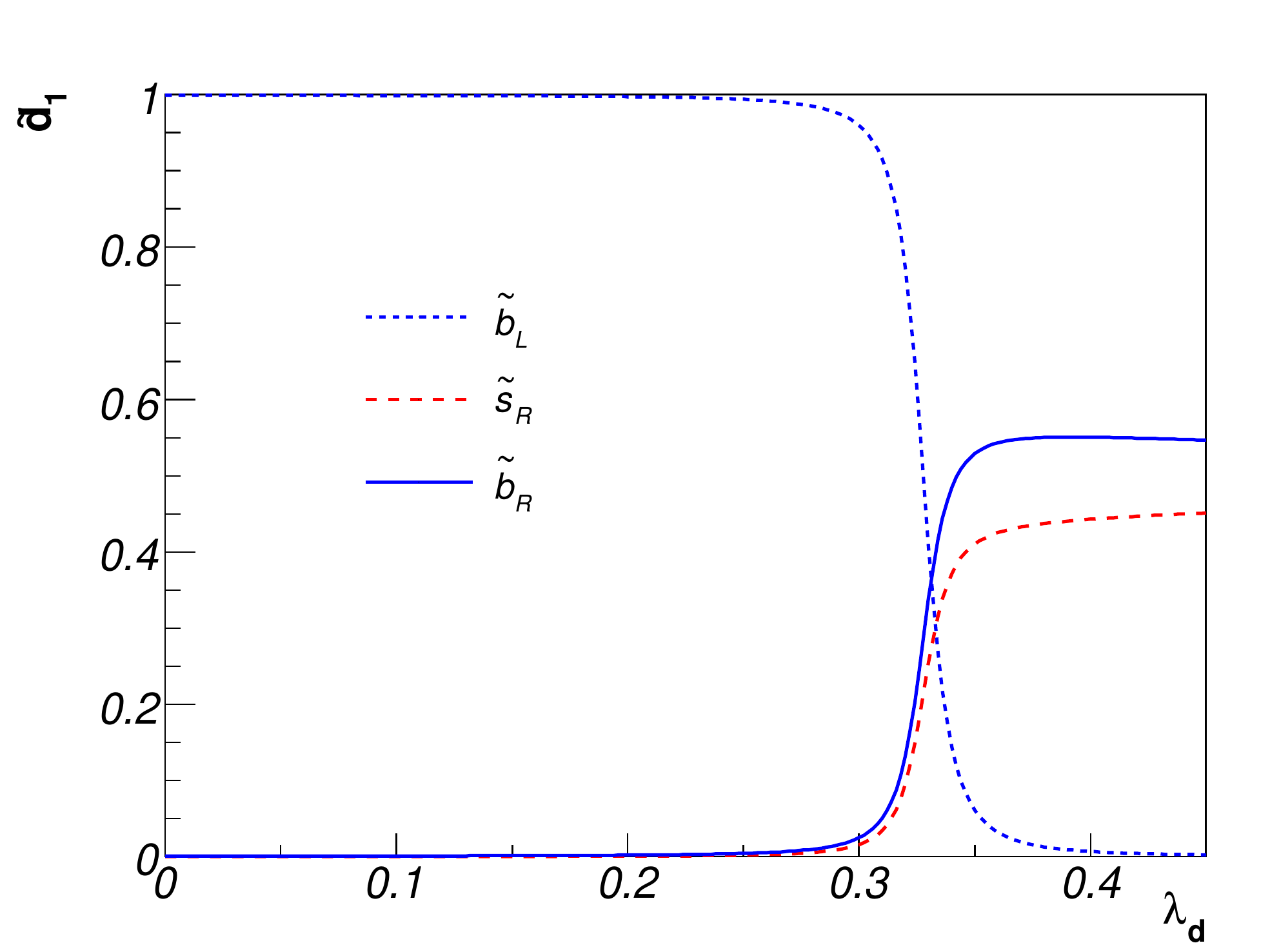} \\
	\includegraphics[scale=0.35]{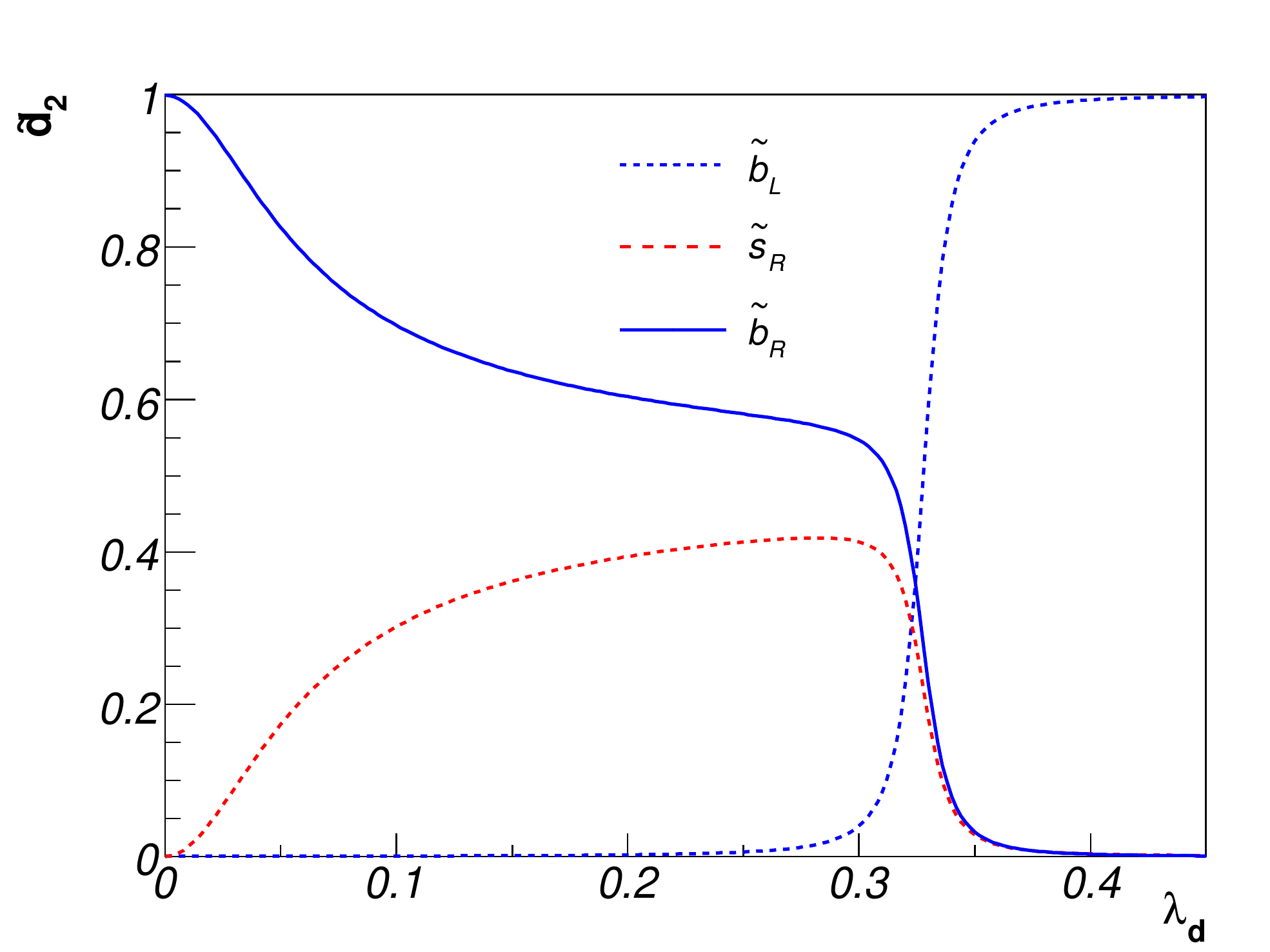} \quad
	\includegraphics[scale=0.35]{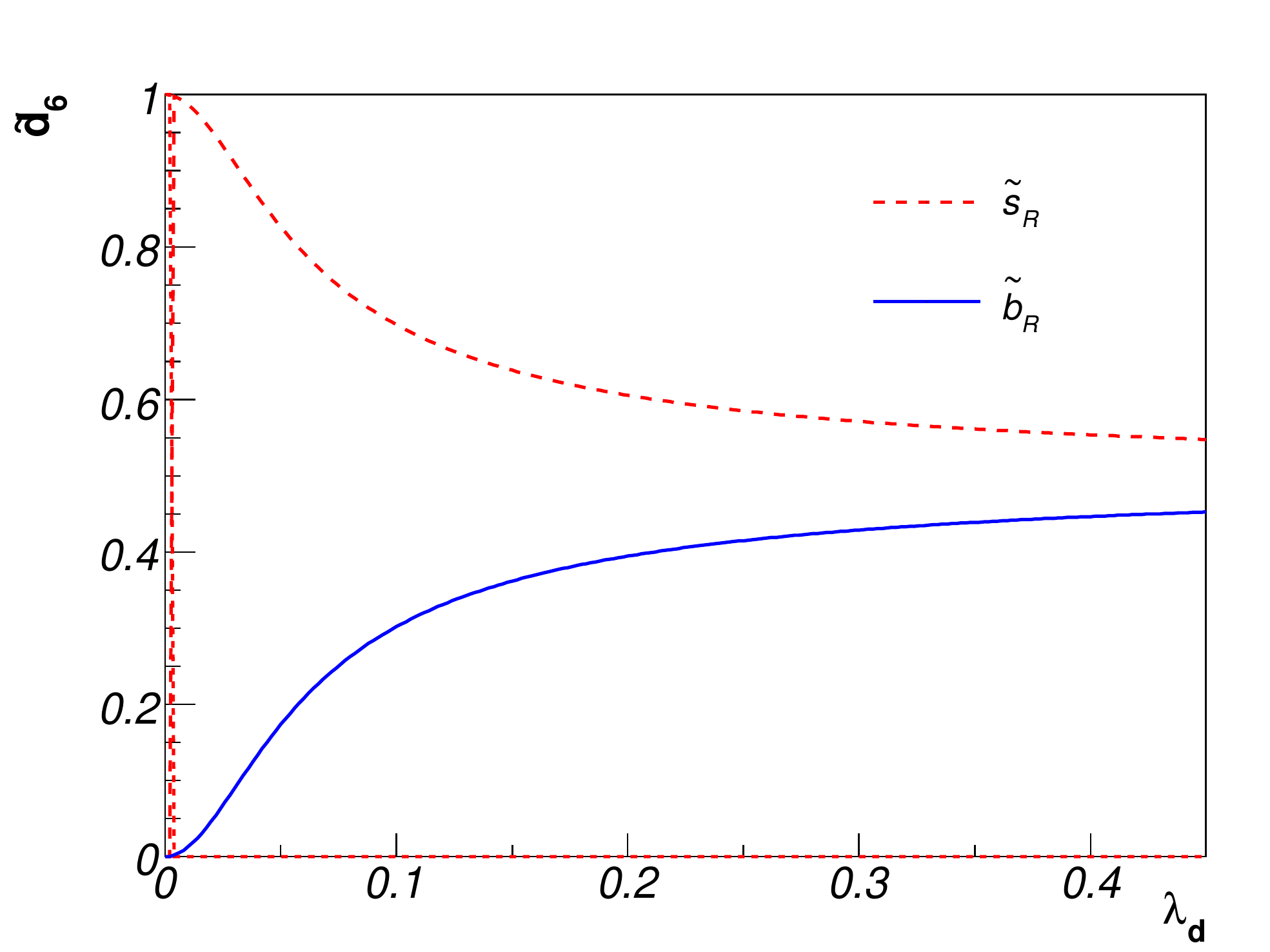} 
\end{center}
\caption{Same as Fig.\ \ref{fig:squarks1} for the scenario II of Tab.\ \ref{tab:points}.}
\label{fig:squarks2}
\end{figure}

\begin{figure}
\begin{center}
	\includegraphics[scale=0.35]{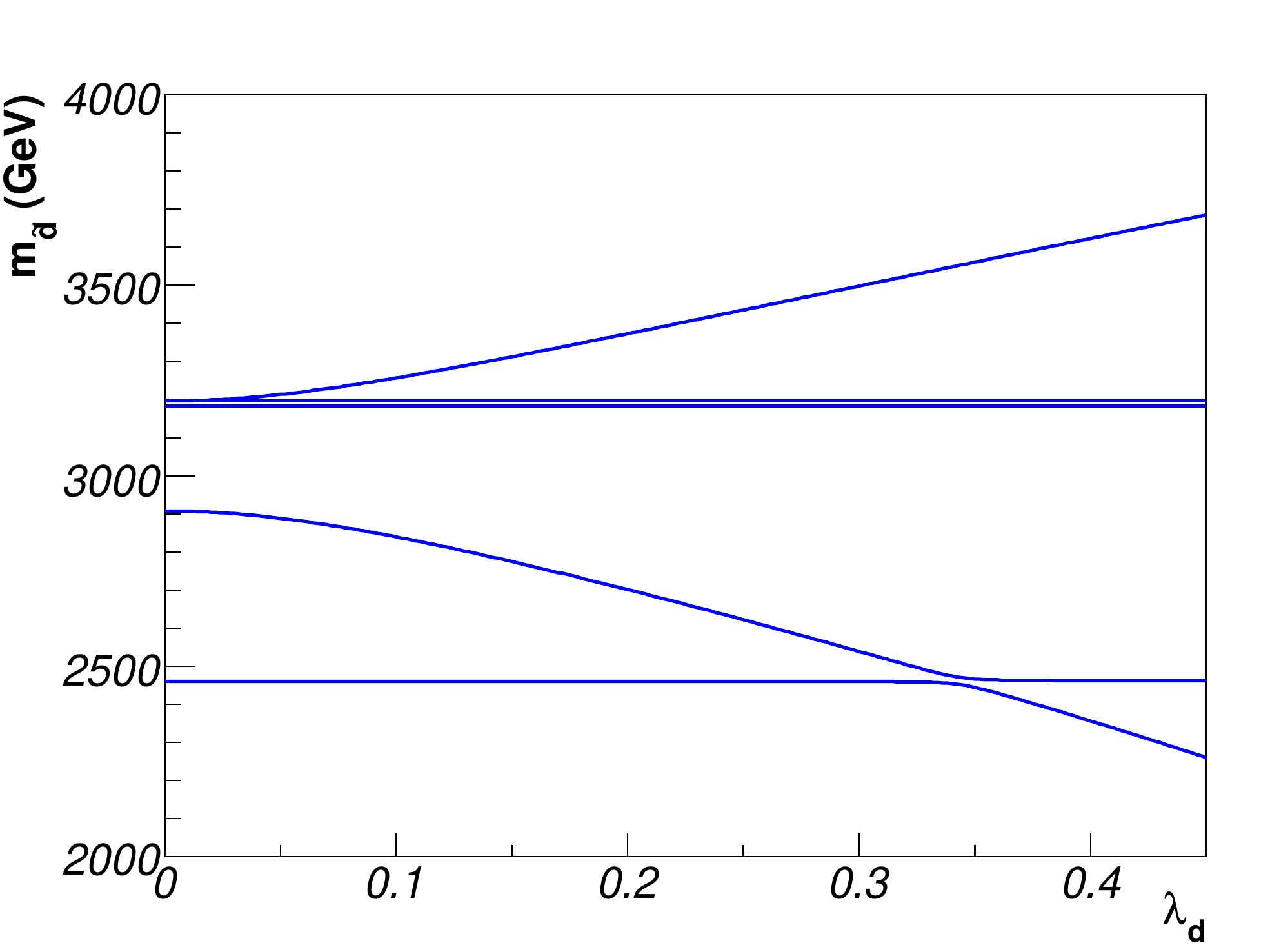} \quad
	\includegraphics[scale=0.35]{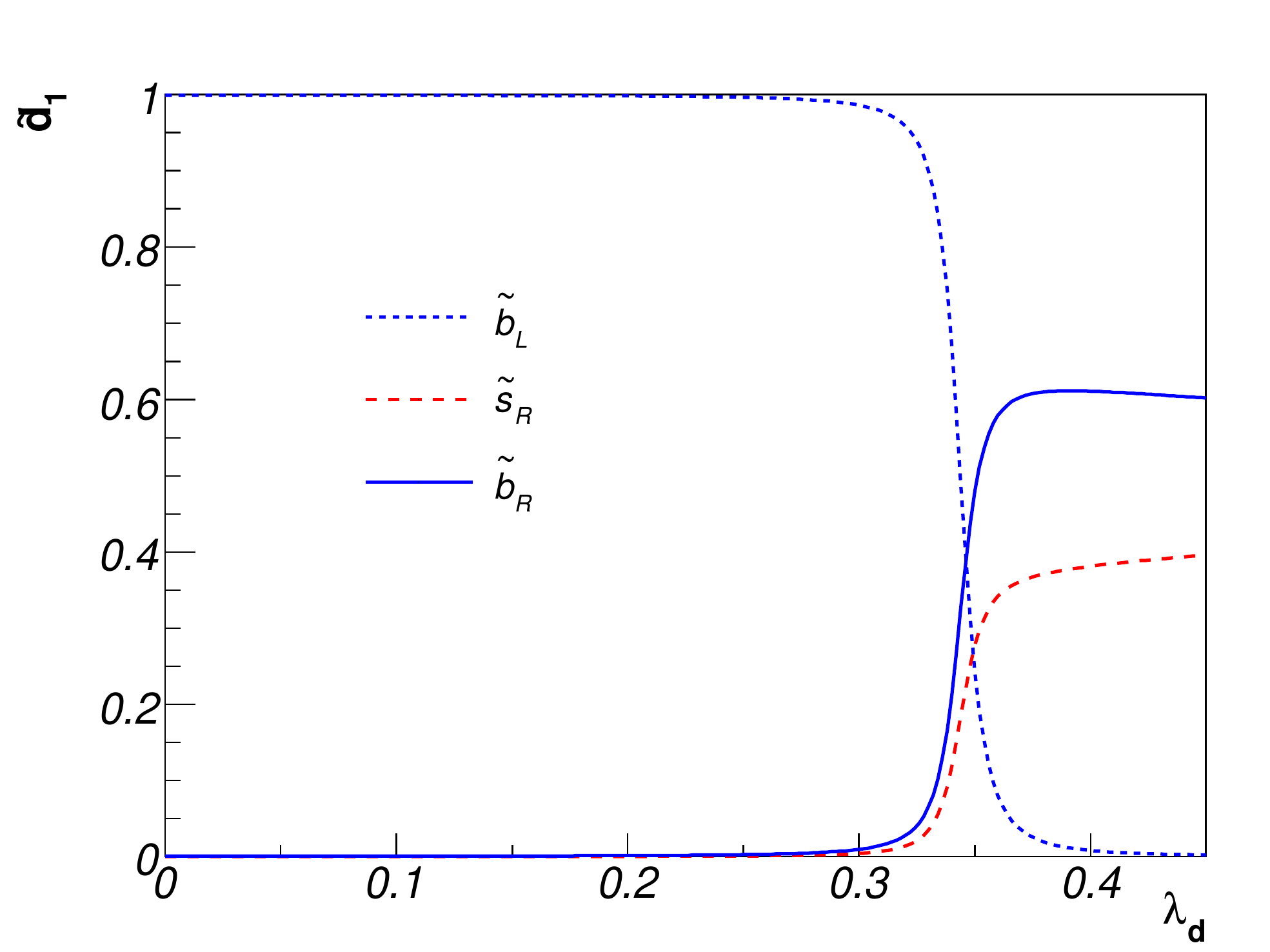} \\
	\includegraphics[scale=0.35]{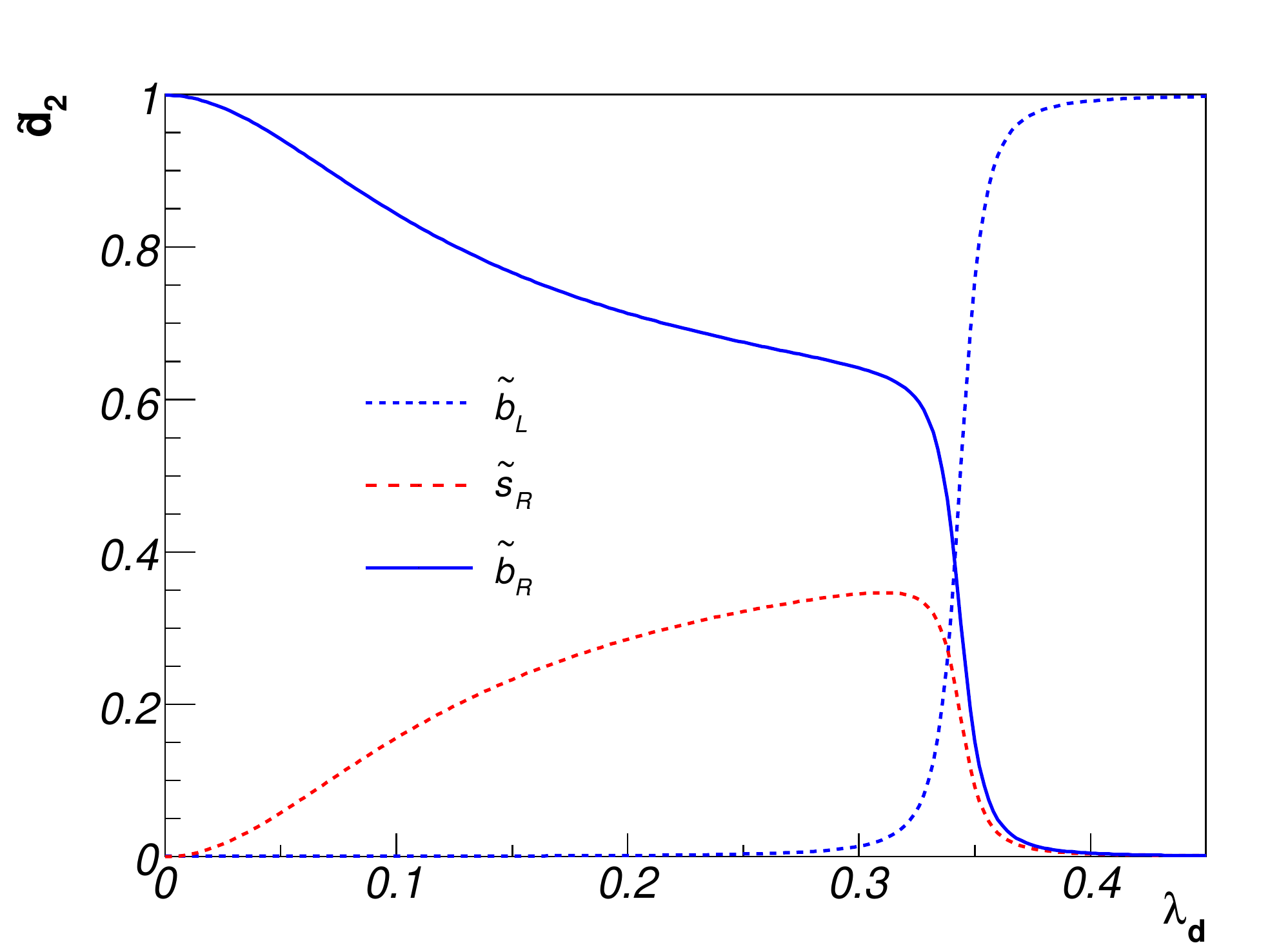} \quad
	\includegraphics[scale=0.35]{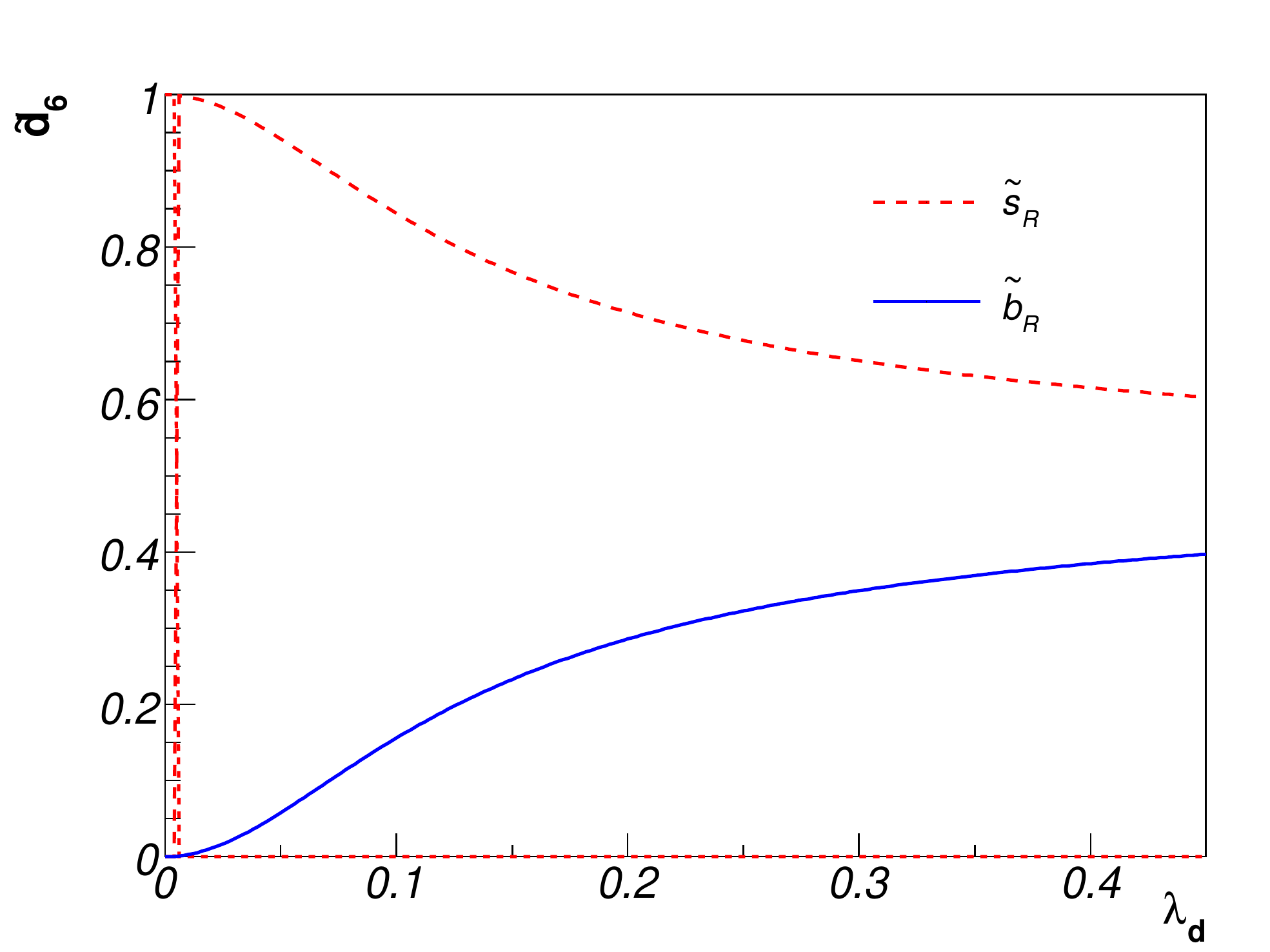} 
\end{center}
\caption{Same as Fig.\ \ref{fig:squarks1} for the scenario III of Tab.\ \ref{tab:points}.}
\label{fig:squarks3}
\end{figure}

The altered mass splitting is accompanied by changes in the flavour
decomposition of the involved squark states. In particular, at points where two
squark mass eigenvalues should cross, the two corresponding eigenstates exchange
their flavour content and undergo a so-called ``avoided crossing''
\cite{NMFV_mSUGRA}. As can be seen in Fig.\ \ref{fig:squarks1}, in the MFV-case
the lightest down-type squark $\tilde{d}_1$ is a pure left-handed sbottom. For
$\lambda_d \gtrsim 0.3$, this state acquires a sizeable admixture of
the $\tilde{s}_R$ and $\tilde{b}_R$ squark, while its $\tilde{b}_L$ content drops
accordingly. In a similar fashion, the next-to-lightest down-type squark 
$\tilde{d}_2$ is purely bottom-like 
at $\lambda_d = 0$ and gets a sizeable strange admixture in the
flavour-violating case. At $\lambda_d \sim 0.3$, an avoided crossing of the
two mass eigenvalues of $\tilde{d}_1$ and $\tilde{d}_2$ is observed and the two
states exchange their flavour content. For $\lambda_d > 0.3$, the lightest
down-type squark is then a mixture of the $\tilde{s}_R$ and $\tilde{b}_R$
eigenstates, whilst the next-to-lightest one becomes a pure $\tilde{b}_L$ state.
Among the other mass eigenstates, only the heaviest down-type squark is fairly
affected by $\lambda_d$ mixings. In the absence of additional flavour violation,
\ie, in the $\lambda_d = 0$ case, this state is purely $\tilde{d}_R$-like
(not shown in the corresponding graph for the sake of visibility). However, for
$\lambda_d \gtrsim 0$, it immediately undergoes an ``avoided crossing'' with
$\tilde{d}_5$ and exchanges its flavour content, being then a pure $\tilde{s}_R$
state. With increasing values of the $\lambda_d$ parameter, this squark
gradually gets a larger and larger $\tilde{b}_R$ component, and maximal mixing is
reached for $\lambda_d
\gtrsim 0.4$.

Avoided crossings and similar mass splittings also occur for the two other
scenarios, as presented in Figs.\ \ref{fig:squarks2} and \ref{fig:squarks3}, as well
as for up-type squarks and the variations of the other flavour-violating
parameters $\lambda_{\rm L}$ and $\lambda_u$ (not shown). 
This behaviour can lead to interesting phenomenological consequences for 
production and decays of squarks and gluinos, as it has been shown in Refs.\ 
\cite{NMFV_mSUGRA, NMFV_GMSB, HurthPorod, QFV_Gluino1, QFV_Squark1, QFV_Squark2,
QFV_Gluino2,Hiller2011}. In particular, the dependence of the gluino production
on the
non-minimal flavour-violation parameters is discussed in the following Section. 

\begin{figure}
\begin{center}
	\includegraphics[scale=0.35]{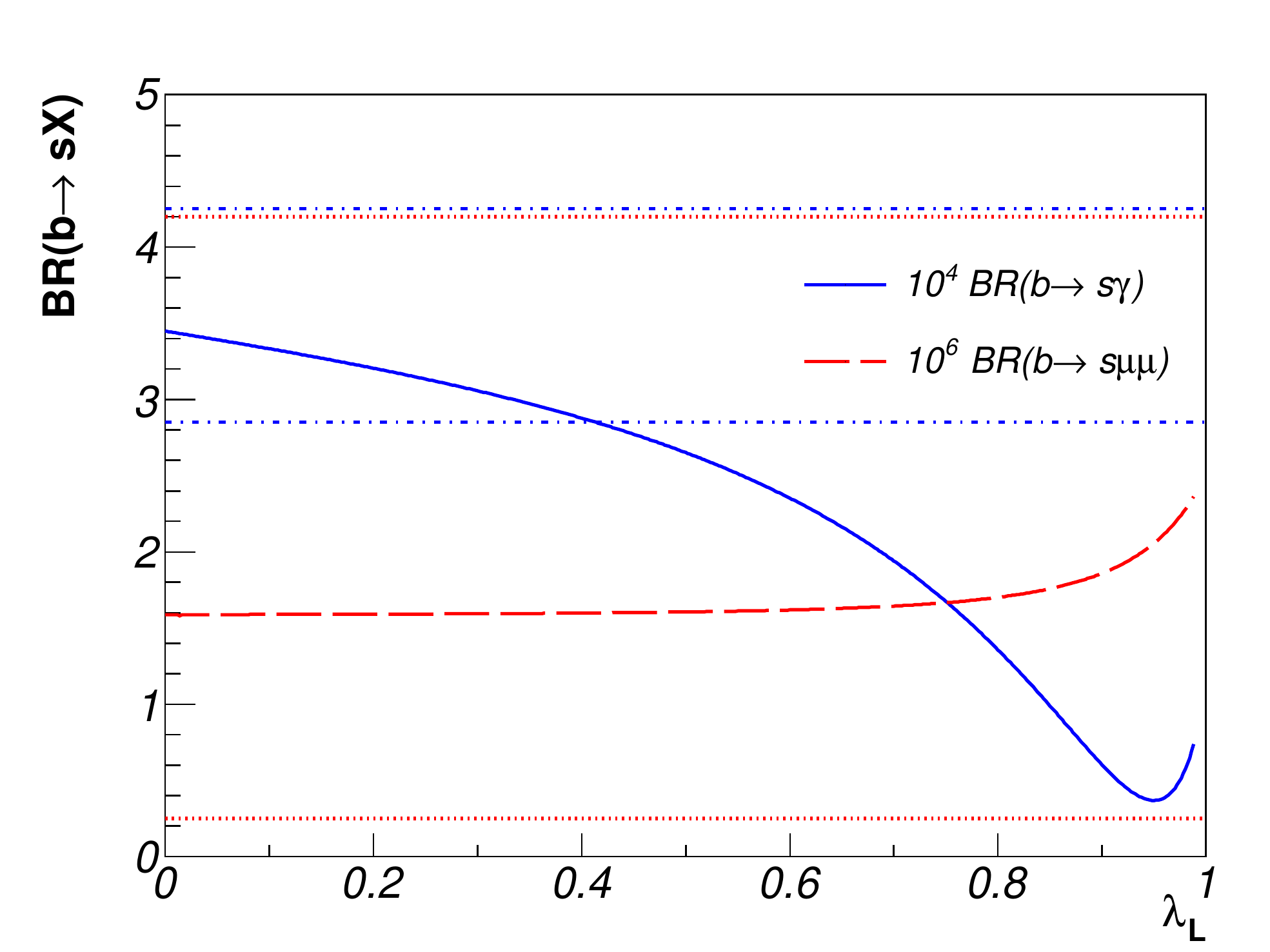} \qquad
	\includegraphics[scale=0.35]{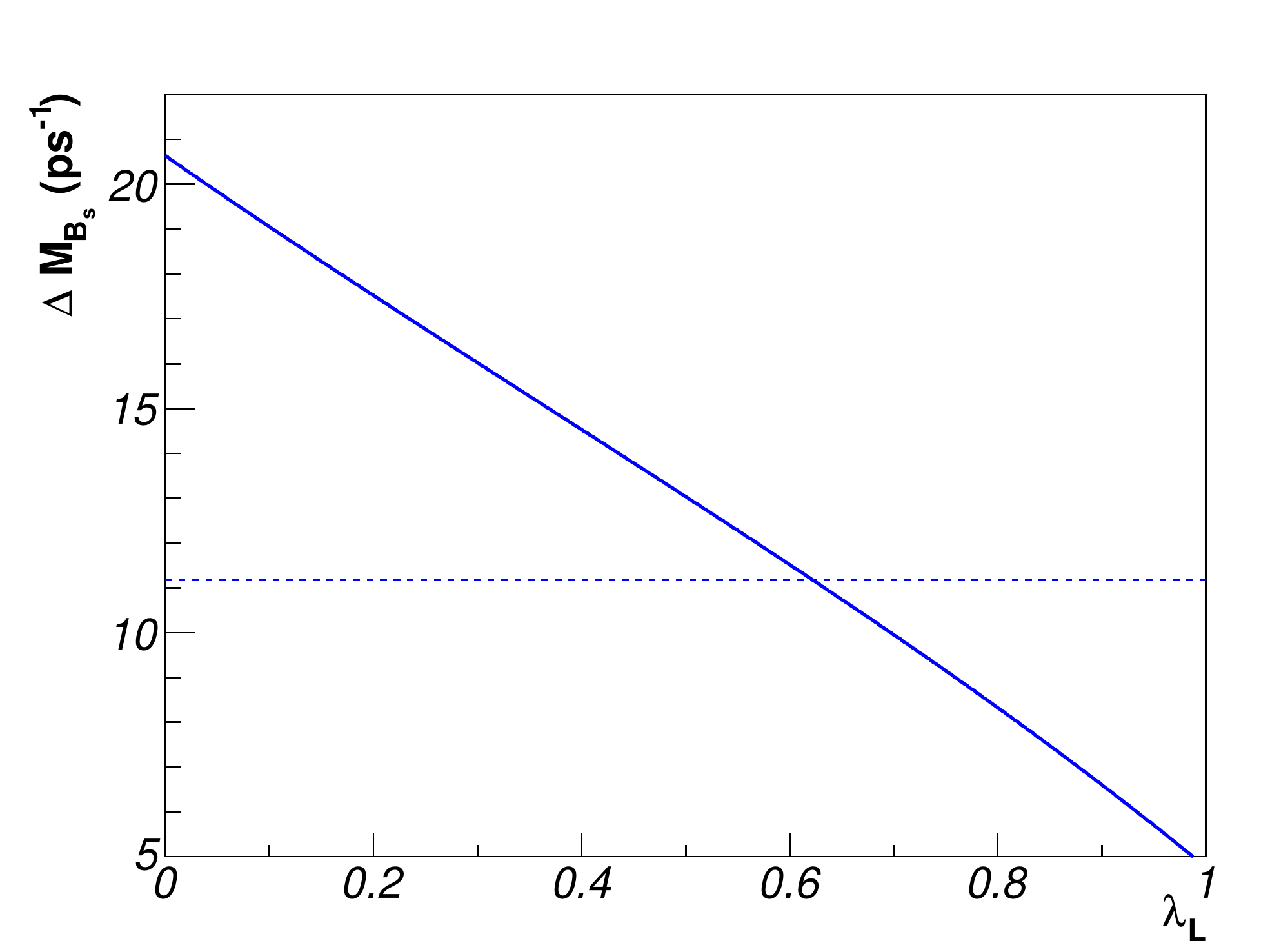} 
\end{center}
\caption{Dependence of BR($b\to s\gamma$), BR($b\to s\mu\mu$), and $\Delta
M_{B_s^0}$ on the flavour-violating parameter $\lambda_L$ for the scenario I of
Tab.\ \ref{tab:points}. We also show horizontal lines corresponding to the
experimental upper and lower limits for BR($b\to s\gamma$) (blue dashed),
BR($b\to s\mu\mu$) (red dotted), and $\Delta M_{B_s^0}$ (blue dashed), as discussed in Sec.\ \ref{sec:constraints}.}
\label{fig:obs1}
\end{figure}

\begin{figure}
\begin{center}
	\includegraphics[scale=0.35]{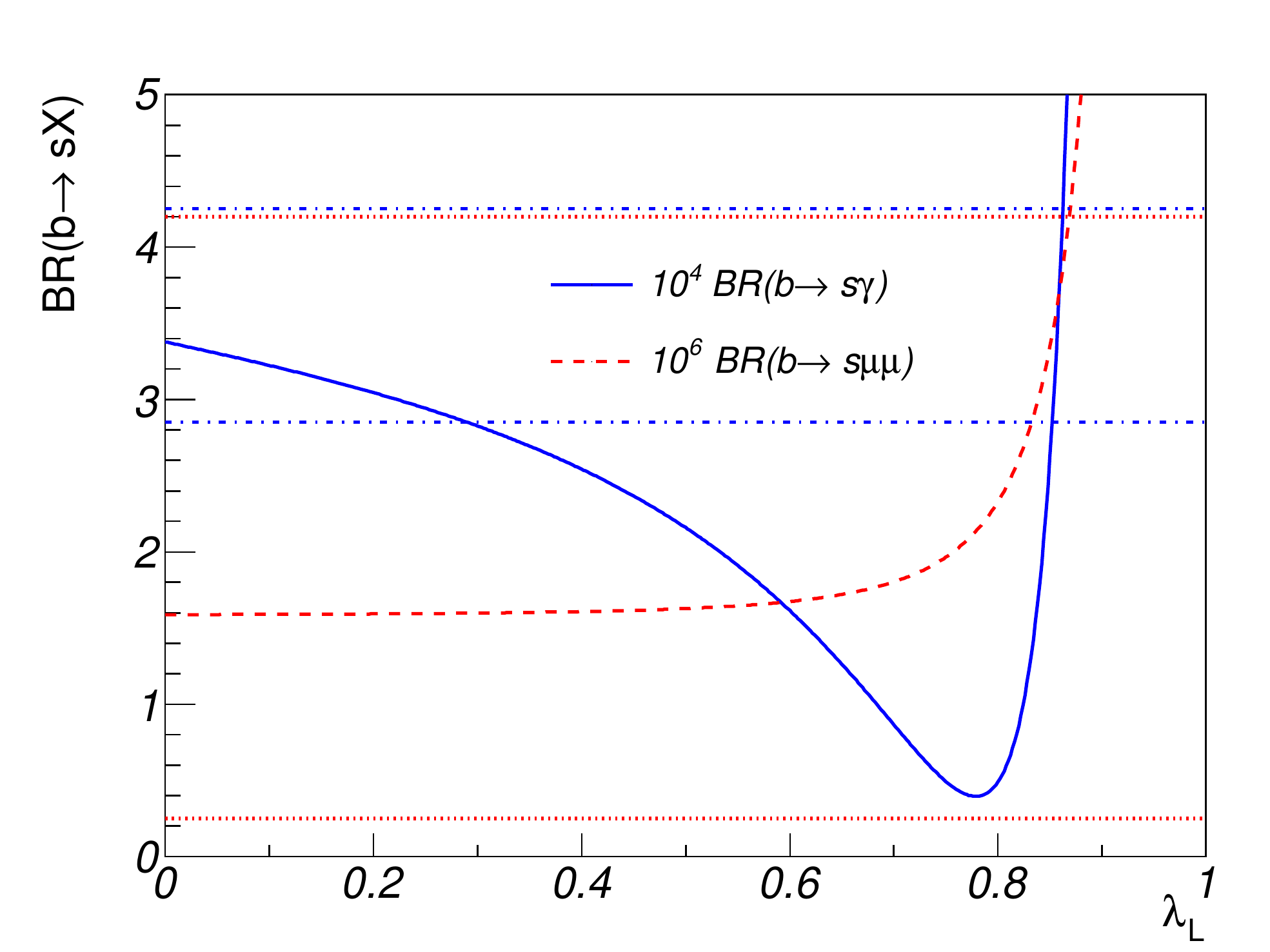} \qquad
	\includegraphics[scale=0.35]{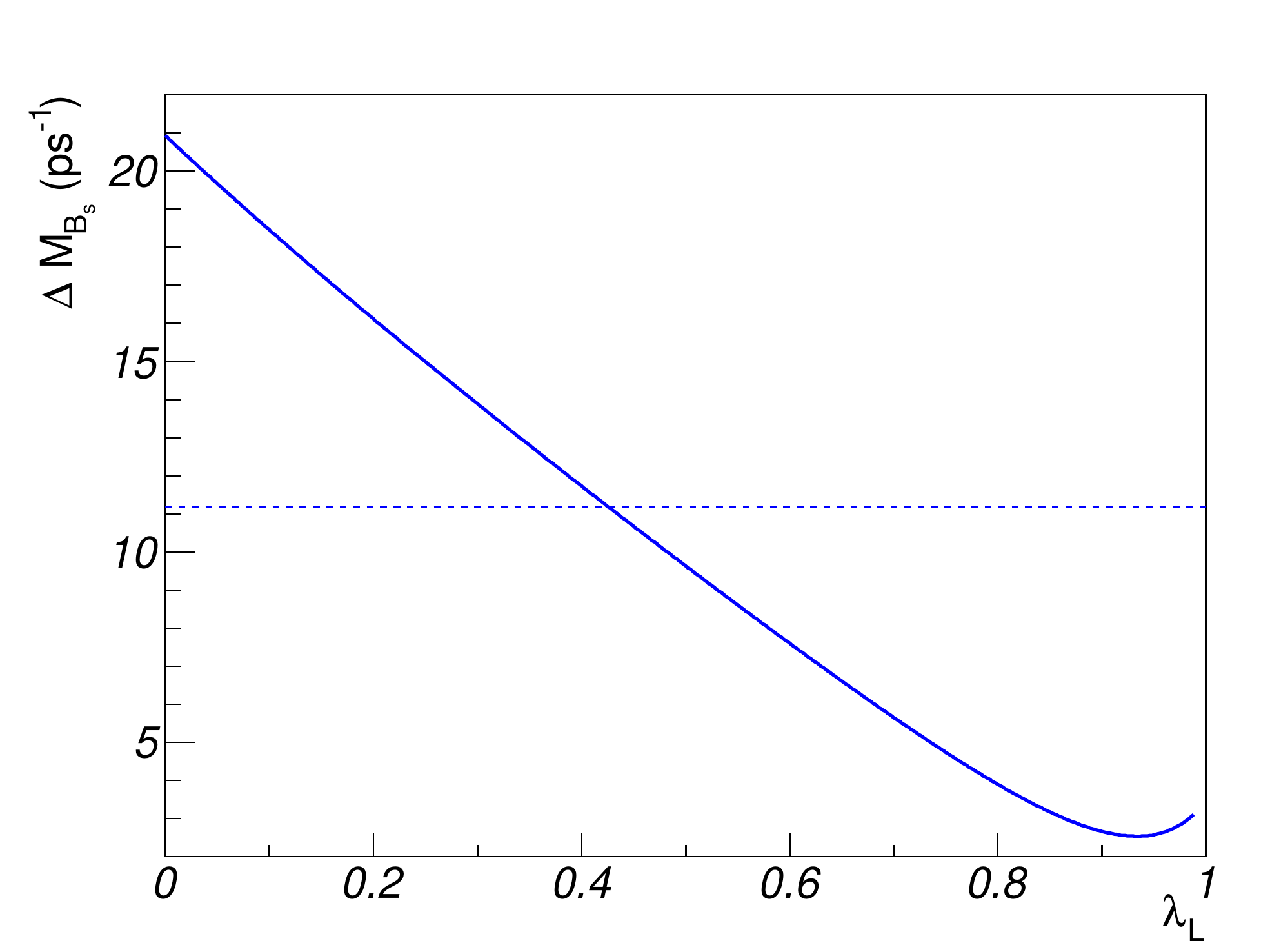} 
\end{center}
\caption{Same as Fig.\ \ref{fig:obs1} for the scenario II of Tab.\ \ref{tab:points}.}
\label{fig:obs2}
\end{figure}

\begin{figure}
\begin{center}
	\includegraphics[scale=0.35]{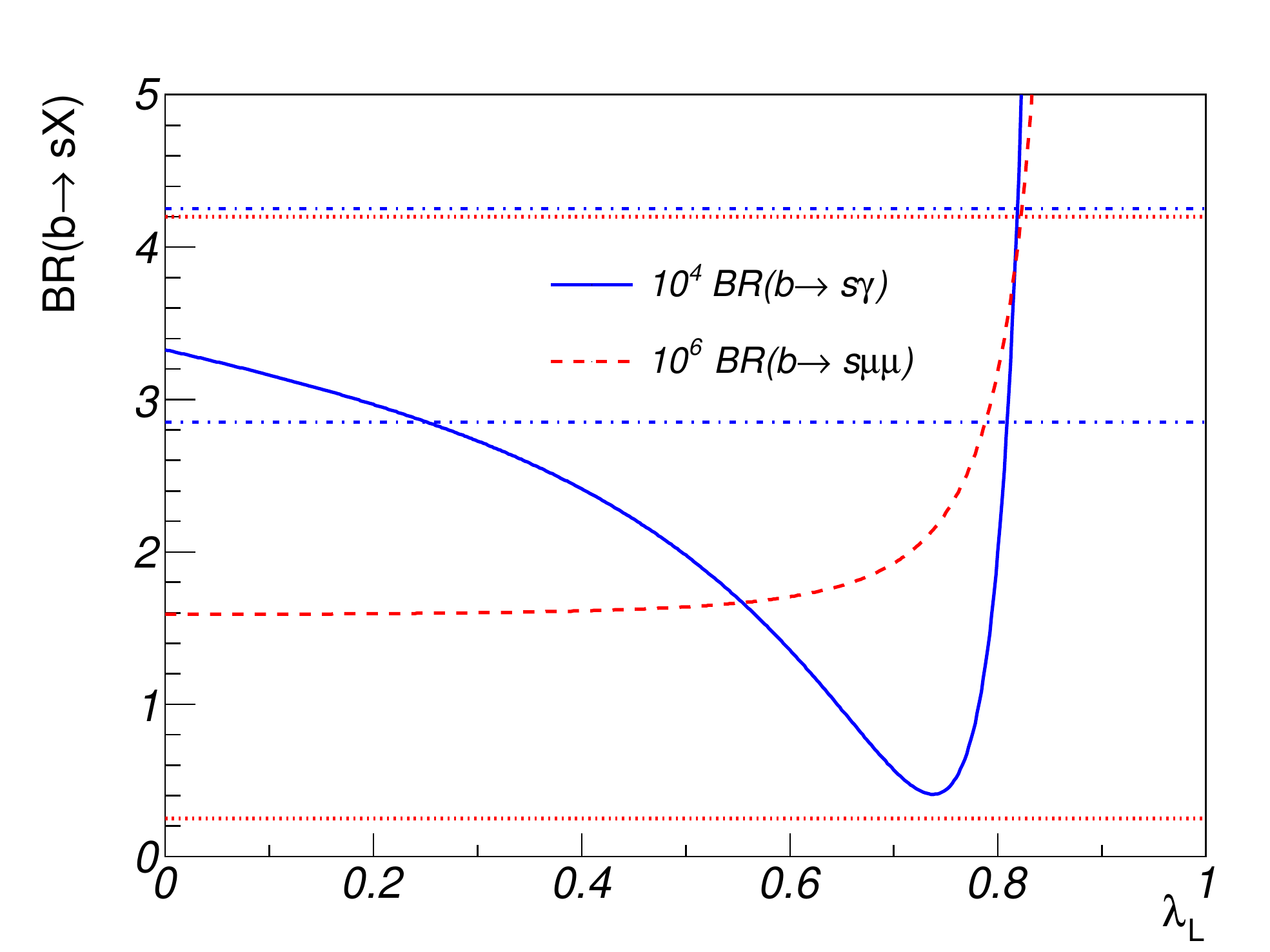} \qquad
	\includegraphics[scale=0.35]{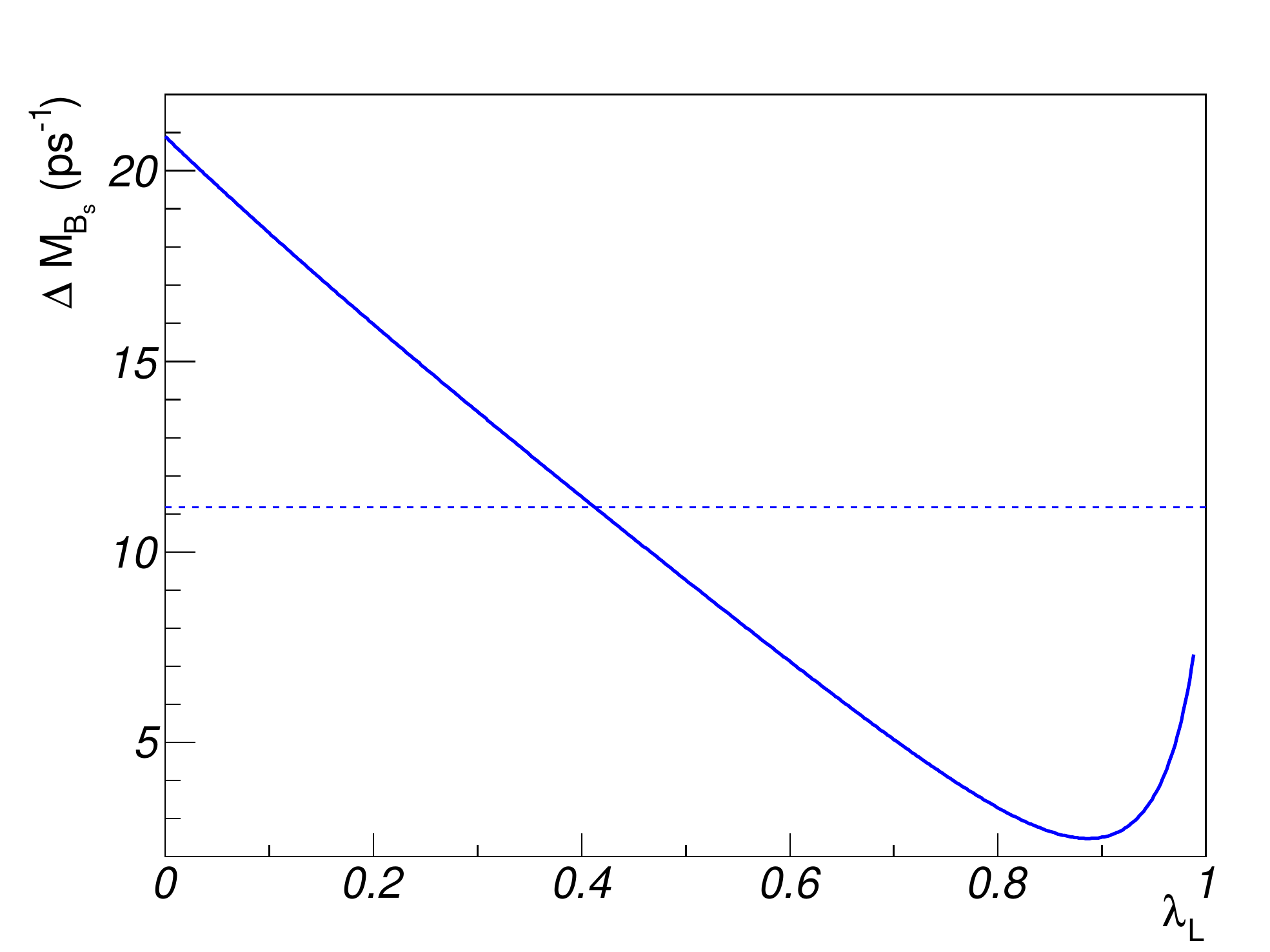} 
\end{center}
\caption{Same as Fig.\ \ref{fig:obs1} for the scenario III of Tab.\ \ref{tab:points}.}
\label{fig:obs3}
\end{figure}

In Figs.\ \ref{fig:obs1}, \ref{fig:obs2} and \ref{fig:obs3}, we show the
theoretical predictions for the
branching ratios of the rare decays $b\to s\gamma$ and $b\to s\mu\mu$ as well as
the meson-oscillation observable $\Delta M_{B_s^0}$ as a function of the
left-left mixing parameter $\lambda_L$. These three observables are rather 
sensitive to non-minimal flavour violation and can therefore be used to
constrain the possible values of the $\lambda$-parameters, contrary to 
the branching ratio $B_s^0\to
\mu^+\mu^-$, for which the predictions lie well below the observed upper bound of
Eq.\ \eqref{eq:bs0}, for any value for the $\lambda$-parameters, and are given
by $(5.0 \ldots 5.2) \times 10^{-9}$, $(5.2 \ldots 6.3) \times 10^{-9}$ and
$(5.0 \ldots 9.0) \times 10^{-9}$ for our scenarios I, II and III, respectively.
As already discussed above, the decay
$b\to s\gamma$ is very sensitive to additional flavour mixing, particularly in the
left-left sector. Together with its high experimental precision, this makes it
the most stringent constraint in this case. Hence, as shown in the figures, 
the lower limit of the interval given in Eq.\ (\ref{eq:bsg}) is
reached (at the $2\sigma$ confidence level) at $\lambda_L \sim 0.4$, according
to the corresponding values derived in Tab.\ \ref{tab:ranges}. 
The prediction of $\Delta
M_{B_s^0}$ also strongly depends on the mixing in the left-left chiral sector.
The lower bound of Eq.\
(\ref{eq:dMBs}) is reached at $\lambda_{\rm L} \sim 0.43$, making this constraint
competitive to the rare decay $b\to s\gamma$ despite the large theoretical
uncertainty. Finally, constraints from the $b\to s\mu\mu$ observable are weaker for
values of the mixing-parameter $\lambda_L \lesssim 0.6$, the theoretical
predictions showing no significant dependence on flavour mixing in that
$\lambda_{\rm L}$-region. 

\section{Non-minimally flavour violating gluino hadroproduction}
\label{sec:xsec}

\subsection{Generalized couplings in NMFV supersymmetry
  \label{sec:NMFVcouplings}}

In our previous publications, we have computed the cross sections for the
production of gaugino and squark pairs in the framework of non-minimally
flavour-violating supersymmetry  \cite{NMFV_mSUGRA, NMFV_GMSB}. In this work, we
add the missing channels related to gluino production.
We first introduce our conventions for the generalized strong and
electroweak couplings which will appear in our analytical calculations of the
cross sections. 
Non-minimal flavour violation in the strong sector can arise from interactions
between gluinos $\g$, quarks $q$ and left- (right-)handed squarks $\q_L$
($\q_R$) described by the (flavour-diagonal) Lagrangian \cite{Nilles:1983ge,
Haber:1984rc}
\beq
  \lag_{q\q\g} = \sqrt{2} g_s \Big[ 
       - \q_{Lf}^\dag T_a \big(\g^a P_L \bar q_f \big) 
       + \big(\bar q_f P_L \g^a \big) T_a \q_{Rf} \Big] + 
       \hc \ ,
\eeq
where $f$ stands for a flavour index, $T_a$ and $g_s$ are the fundamental
matrices and the coupling constant associated with the $SU(3)_c$ gauge group,
and $P_L$ denotes the left-chirality projection operator acting on
four-component spinors. Rotating to the mass-eigenstate basis, 
flavour violation is introduced through possible non-diagonal entries in the
matrices $R^q$. In this case, the left-handed and right-handed coupling strengths
become proportional to
\beq
  \Big\{ L_{\q_iq_j\g} ,  R_{\q_iq_j\g} \Big\} = 
  \Big\{ R^q_{ij},  -R^q_{i(j+3)} \Big\} \ .
\eeq 
Similarly, for the electroweak sector, non-minimally flavour-violating
interactions can arise in the chargino and neutralino couplings to quarks and
squarks. We start from the Lagrangian
\beq\bsp
   \lag_{q\q\X} = 
   &\ \sqrt{2} \gp \Big[ 
       - \frac16 \q_{Lf}^\dag \big(\bar\B P_L q_f \big) 
       + \frac23 \big(\bar u_f P_L \B \big) \u_{Rf}   
       - \frac13 \big(\bar d_f P_L \B \big) \sd_{Rf} \Big] -
     \frac{\sqrt{2}}{2} \gw \Q_f^\dag \sigma_k \big(\bar \W^k P_L Q_f \big)  \\
   &\ - (y_d)_{f\fp}\ \sd_f^\dag \bar\H_d P_L Q_\fp 
      + (y_d)_{f\fp}\ \Q_\fp \bar d_f P_L \H_d
      + (y_u)_{f\fp}\ \u_f^\dag \bar\H_u P_L Q_\fp 
      - (y_u)_{f\fp}\ \Q_\fp \bar u_f P_L \H_u + \hc \ , 
\esp\eeq
where $\B$, $\W$ and $\H_{\{u,d\}}$ are the four-component bino-, wino- and
higgsino-eigenstates and $Q$ ($\Q$), $u$ ($\u$) and $d$ ($\sd$) the doublet of
left-handed (s)quarks and the up- and down-type right-handed (s)quarks,
respectively.
In the Lagrangian above, we denote the hypercharge and
weak coupling constants $\gp$ and $\gw$, while the generators of
$SU(2)_L$ are given by $\sigma/2$, $\sigma$ being the Pauli matrices. Finally,
$y_u$ and $y_d$ are the up-type and down-type quark Yukawa
matrices, which once diagonalized are proportional to the quark masses,
\beq
  (\hat{y}_u)_{ij} = \frac{\sqrt{2} \gw m_{u_i}}{2 \mw \sin\beta} \delta_{ij} 
  \quad \text{and} \quad
  (\hat{y}_d)_{ij} = \frac{\sqrt{2} \gw m_{d_i}}{2\mw \cos\beta}  \delta_{ij}
  \ .
\eeq
Here, $\hat y_{\{u,d\}}$ denote the diagonalized Yukawa matrices. After rotating
to the mass-basis, one obtains the coupling strengths
\beq \bsp
  L_{\tilde{d}_j d_k \tilde{\chi}^0_i} =&\ \frac{(e_d - T^3_d) \sw N_{i1} + T^3_d
    \cw N_{i2}}{\sqrt{2} \cw} R^{d\ast}_{jk} + \frac{m_{d_k} N_{i3}}{2\sqrt{2}
    \mw \cos\beta} R^{d\ast}_{j(k+3)}\ ,\\
  R_{\tilde{d}_j d_k \tilde{\chi}_i^0} =&\ -\frac{e_d \sw N^\ast_{i1}}{\sqrt{2}
    \cw} R^{d\ast}_{j(k+3)} + \frac{m_{d_k} N^\ast_{i3}}{2\sqrt{2} \mw
    \cos\beta}R^{d\ast}_{jk}\ ,\\
  L_{\tilde{u}_j u_k \tilde{\chi}^0_i} =&\  \frac{(e_u - T^3_u) \sw N_{i1} +
    T^3_u \cw N_{i2}}{\sqrt{2} \cw} R^{u\ast}_{jk} + \frac{m_{u_k}
    N_{i4}}{2\sqrt{2} \mw \sin\beta}R^{u\ast}_{j(k+3)}\ ,\\
  R_{\tilde{u}_j u_k \tilde{\chi}_i^0} =&\ -\frac{e_u \sw N_{i1}^\ast}{\sqrt{2}
    \cw} R^{u\ast}_{j(k+3)} + \frac{m_{u_k} N_{i4}^\ast}{2\sqrt{2} \mw
    \sin\beta} R^{u\ast}_{jk} \ ,\\
  L_{\tilde{d}_j u_k \tilde{\chi}_i^\pm} =&\ \frac12 \sum_{l=1}^3 \bigg[ U_{i1}
    R^{d\ast}_{jl} - \frac{m_{d_l} U_{i2}}{\sqrt{2} \mw \cos\beta}
    R^{d\ast}_{j(l+3)} \bigg] V_{u_k d_l} \ ,\\
  R_{\tilde{d}_j u_k \tilde{\chi}_i^\pm} =&\ - \sum_{l=1}^3 \frac{m_{u_k}
    V_{i2}^\ast V_{u_k d_l}}{2\sqrt{2} \mw \sin\beta} R^{d\ast}_{jl} \ ,\\ 
  L_{\tilde{u}_j d_k \tilde{\chi}_i^\pm} =&\ \frac12 \sum_{l=1}^3 \bigg[ V_{i1}
    R^{u\ast}_{jl} - \frac{m_{u_l} V_{i2}}{\sqrt{2} \mw
    \sin\beta}R^{u\ast}_{j(l+3)}\bigg] V_{u_l d_k}^\ast \ ,\\ 
  R_{\tilde{u}_j d_k \tilde{\chi}_i^\pm} =&\ - \sum_{l=1}^3 \frac{m_{d_k}
    U^\ast_{i2} V^\ast_{u_l d_k}}{2\sqrt{2} \mw \cos\beta} R^{u\ast}_{jl}\ ,
\esp \label{eq:couplstr}\eeq
where we follow the notations of Sec.\ \ref{sec:model}. In addition, we introduce
the mass of the $W$-boson $\mw$, the cosine of the electroweak mixing angle
$\cw$, and the matrices $N$, $U$ and $V$ related to the gaugino/higgsino mixing.

\subsection{Analytical results \label{sec:analytics}} 

We compute the partonic cross sections related to gluino production,
\textit{i.e.},
for the processes
\beq
 a_{h_a}(p_a)\ b_{h_b}(p_b) \to 
     \g (p_1)\ \q^{(\ast)}_i (p_2) \ , \quad
     \g (p_1)\ \g (p_2) \quad \text{and} \quad
     \g (p_1)\ \tilde{\chi}^{\{\pm,0\}}_i(p_2)\ ,
\eeq
and present the results for definite helicities $h_{a,b}$ of the initial partons
$a,b=q,\bar{q},g$ in terms of the squark masses $m_{\q_j}$, the chargino and
neutralino masses $m_{\tilde\chi_j}$, the gluino mass $m_\g$, the
Mandelstam variables
\beq
 s=(p_a+p_b)^2 \ , \quad t=(p_a-p_1)^2 \ , \quad\text{and}\quad u=(p_a-p_2)^2 \ ,
\eeq
and the mass-subtracted Mandelstam variables
\beq
 t_\g = t - m^2_\g\ , \quad 
 u_\g = u - m^2_\g\ , \quad
 t_{\q_i} = t-m_{\q_i}^2 \ , \quad
 u_{\q_i} = u-m_{\q_i}^2 \ , \quad 
 t_{\tilde \chi_i} = t-m_{\tilde\chi_i}^2 \ , \quad
 u_{\tilde \chi_i} = u-m_{\tilde\chi_i}^2 \ . 
\eeq
Unpolarized partonic cross sections $\d \hat\sigma$ and single- and
double-polarized cross partonic sections $\d\hat\sigma_L$ and
$\d\hat\sigma_{LL}$, averaged over initial spins, can easily be derived from the
helicity-dependent result,
\beq\bsp
  \d\hat{\sigma} = &\ 
     \frac{\d\hat{\sigma}_{ 1, 1} + \d\hat{\sigma}_{1,-1} + 
     \d\hat{\sigma}_{-1, 1} + \d\hat{\sigma}_{-1,-1}}{4} \ , \\
  \d\Delta\hat{\sigma}_L =&\ 
     \frac{\d\hat{\sigma}_{ 1, 1} + \d\hat{\sigma}_{1,-1} - 
     \d\hat{\sigma}_{-1, 1} - \d\hat{\sigma}_{-1,-1}}{4} \ , \\
 \d\Delta\hat{\sigma}_{LL}=&\
     \frac{\d\hat{\sigma}_{ 1, 1} - \d\hat{\sigma}_{ 1,-1} - 
     \d\hat{\sigma}_{-1, 1} + \d\hat{\sigma}_{-1,-1}}{4} \ .
\esp\eeq

\begin{figure}
 \centering
 \includegraphics[width=.6\columnwidth]{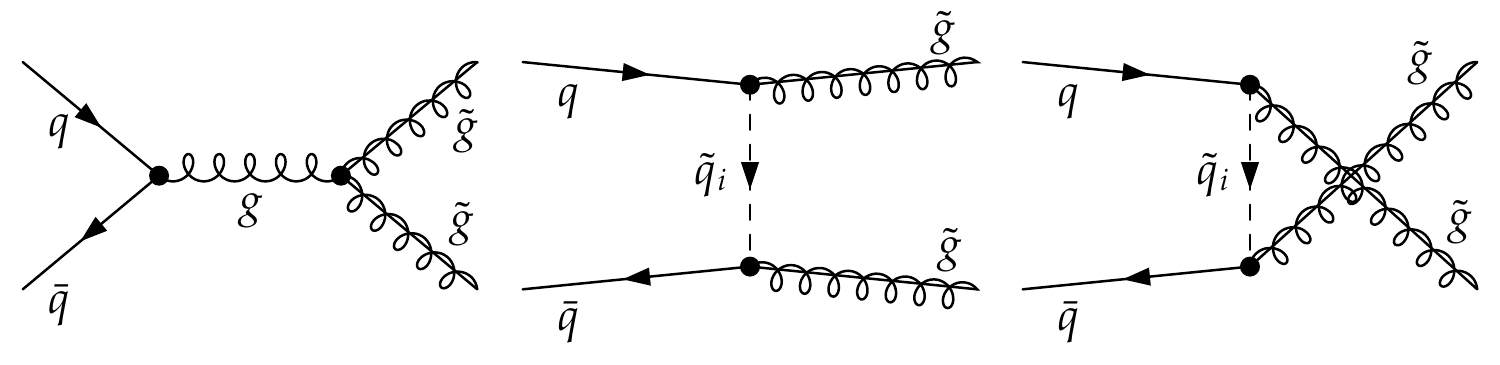}
 \includegraphics[width=.6\columnwidth]{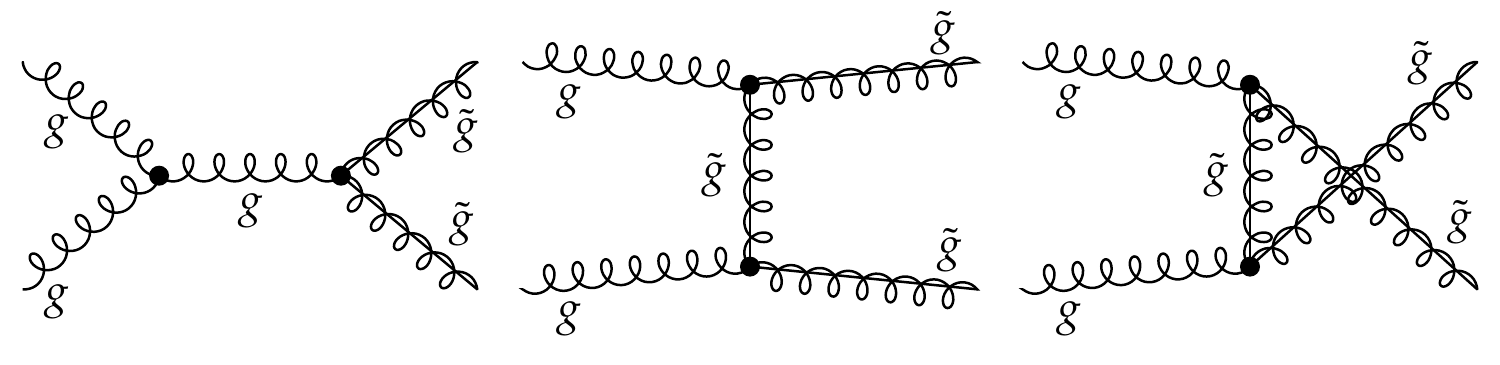}
 \caption{\label{fig:01}Tree-level Feynman diagrams for the production of a pair
of gluinos in quark-antiquark (top) and gluon-gluon collisions (bottom).}
\end{figure}

The strong production of a pair of gluinos proceeds either from the annihilation
of a quark-antiquark pair in the initial state, 
\beq
   q_{h_a}(p_a)\ \bar q^\prime_{h_b}(p_b) \to \g (p_1)\ \g (p_2) \ ,
\eeq
as shown in Fig.\ \ref{fig:01} (top), or from the fusion of two initial gluons,
\beq 
   g_{h_a}(p_a)\ g_{h_b}(p_b) \to \g (p_1)\ \g (p_2) \ , 
\eeq
as can be seen in the lower part of Fig.\ \ref{fig:01}. Since the latter is
independent of squark exchange, the corresponding differential partonic cross
section, averaged on the initial colour states and taking into account the
symmetry factor relative to the production of two identical particles, is
independent from any possible source of flavour violation in the squark sector. 
It is given by
\beq
  \frac{\d\hat\sigma}{\d t}(h_a,h_b) = 
  \frac{g_s^4 N_c^2}{16 \pi s^2 (N_c^2-1)} 
    \bigg(1-\frac{t_\g u_\g}{s^2}\bigg) \bigg[ (1-h_a h_b) 
      \bigg(-2 + \frac{s^2}{t_\g u_\g} + 4 \frac{s m_\g^2}{t_\g u_\g}\Big[1-
      \frac{s m_\g^2}{t_\g u_\g}\Big]\bigg) + 2  h_a h_b (s-2m_\g^2)
      \frac{m_\g^2 s^2}{t_\g^2 u_\g^2} \bigg]\ ,
\eeq
where $N_c$ denotes the number of colours.
This result agrees with those of Refs.\ \cite{Beenakker:1996ch,Gehrmann:2004xu}
after summing over the initial gluon polarizations.
Contrary, the quark-antiquark channel contains $t$-channel and $u$-channel
squark exchanges. However, even if a given squared diagram or interference term
depended on some combination of the element of the squark mixing matrices,
the sum over all possible squark exchanges is expected to considerably reduce
the flavour-violation dependence of the cross section. The latter can be
expressed as
\bea
 \frac{\d\hat\sigma}{\d t}(h_a,h_b) &=& 
   \frac{g_s^4}{128 \pi s^2 N_c^2} \Bigg[ 
     (1\!-\!h_a)(1\!+\!h_b)\bigg[ \frac{Q_{ss}}{s^2} + 
       \sum_{i=1}^6 \bigg\{ \frac{(Q_{st})_i^1}{s t_{\q_i}} \!+\!
       \frac{(Q_{su})_i^1}{s u_{\q_i}} \bigg\}  + 
       \sum_{i,j=1}^6 \bigg\{ \frac{(Q_{tt})_{ij}^{11}}{t_{\q_i} t_{\q_j}}\!+\!
       \frac{(Q_{uu})_{ij}^{11}}{u_{\q_i} u_{\q_j}}  \!+\!
       \frac{(Q_{tu})_{ij}^{11}}{t_{\q_i} u_{\q_j}} \bigg\} \bigg]  \nn \\ 
 & & \nn \quad + (1\!+\!h_a)(1\!-\!h_b)\bigg[ \frac{Q_{ss}}{s^2} + 
       \sum_{i=1}^6 \bigg\{ \frac{(Q_{st})_i^2}{s t_{\q_i}} \!+\!
       \frac{(Q_{su})_i^2}{s u_{\q_i}} \bigg\} + 
       \sum_{i,j=1}^6 \bigg\{ \frac{(Q_{tt})_{ij}^{22}}{t_{\q_i} t_{\q_j}}\!+\!
       \frac{(Q_{uu})_{ij}^{22}}{u_{\q_i} u_{\q_j}} \!+\!
       \frac{(Q_{tu})_{ij}^{22}}{t_{\q_i} u_{\q_j}} \bigg\} \bigg]\\
 & & \quad + (1\!-\!h_a)(1\!-\!h_b) \bigg[ 
       \sum_{i,j=1}^6 \bigg\{ \frac{(Q_{tt})_{ij}^{12}}{t_{\q_i} t_{\q_j}}\!+\!
       \frac{(Q_{uu})_{ij}^{12}}{u_{\q_i} u_{\q_j}} \!+\!
       \frac{(Q_{tu})_{ij}^{12}}{t_{\q_i} u_{\q_j}} \bigg\} \bigg]\\ 
 & & \nn \quad + (1\!+\!h_a)(1\!+\!h_b) \bigg[ 
       \sum_{i,j=1}^6 \bigg\{ \frac{(Q_{tt})_{ij}^{21}}{t_{\q_i} t_{\q_j}}\!+\!
       \frac{(Q_{uu})_{ij}^{21}}{u_{\q_i} u_{\q_j}} +
       \frac{(Q_{tu})_{ij}^{21}}{t_{\q_i} u_{\q_j}} \bigg\} \bigg] 
 \Bigg] 
\eea
with the generalized charges
\beq\bsp
    Q_{ss} =&\ 2 N_c (N_c^2 - 1)\Big( t_\g^2 + u_\g^2 + 2 m_\g^2 s\Big) \
      , \\
   (Q_{st})_i^m =&\ N_c (N_c^2 - 1)\
     {\rm Re} \Big[ {\cal C}^m_{\q_i q \g} {\cal C}^{m\ast}_{\q_i q^\prime \g}
     \Big] \Big( m_\g^2 s + t_\g^2\Big) \ , \\
   (Q_{su})_i^m =&\ N_c (N_c^2 - 1)\
     {\rm Re} \Big[ {\cal C}^m_{\q_i q \g} {\cal C}^{m\ast}_{\q_i q^\prime \g}
     \Big] \Big( m_\g^2 s + u_\g^2\Big) \ , \\
   (Q_{tt})_{ij}^{mn} =&\  \frac{(N_c^2 - 1)^2}{N_c}\
     {\cal C}^m_{\q_i q \g} {\cal C}^{n\ast}_{\q_i q^\prime \g}
     {\cal C}_{\q_j q \g}^{m\ast} {\cal C}^n_{\q_j q^\prime \g}\ t_\g^2 \ , \\ 
   (Q_{uu})_{ij}^{mn} =&\  \frac{(N_c^2 - 1)^2}{N_c}\
     {\cal C}^m_{\q_i q \g} {\cal C}^{n\ast}_{\q_i q^\prime \g}
     {\cal C}_{\q_j q \g}^{m\ast} {\cal C}^n_{\q_j q^\prime \g}\ u_\g^2 \ , \\ 
   (Q_{tu})_{ij}^{mn} =&\ 2 \frac{(N_c^2 - 1)}{N_c}\ {\rm Re} \Big[
     {\cal C}_{\q_i q \g}^m {\cal C}^{n\ast}_{\q_i q^\prime \g}
     {\cal C}_{\q_j q^\prime \g}^n {\cal C}^{m\ast}_{\q_j q \g}\Big]\ \Big(
     (1-\delta_{mn}) (m_\g^2 s - t_\g u_\g)  + \delta_{mn} m_\g^2 s \Big) \ , 
\esp \eeq
where for the sake of simplicity we have introduced the generic notation
\beq
 \big\{\mathcal{C}^1_{a b c}, \mathcal{C}^2_{a b c} \big\} = 
 \big\{ L_{a b c},R_{a b c} \big\} \ .
\eeq
This reproduces both the polarized and unpolarized results of Refs.\
\cite{Beenakker:1996ch,Gehrmann:2004xu} in the flavour-conserving MSSM limit.

\begin{figure}
 \centering
 \includegraphics[width=0.6\columnwidth]{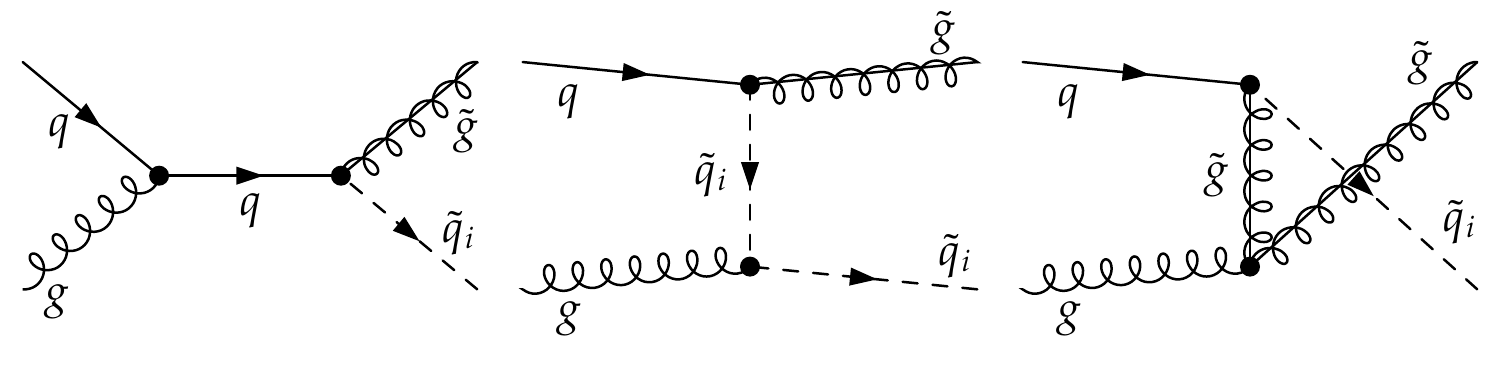}
 \caption{\label{fig:gosq}Tree-level Feynman diagrams for the associated
    production of gluinos and squarks.} 
\end{figure}
An associated pair of a gluino and a squark originates from quark-gluon initial
states,
\beq
  q_{h_a}(p_a)\ g_{h_b}(p_b) \to \g (p_1)\ \q_i (p_2) \ ,  
\eeq
and proceeds through an $s$-channel quark, $t$-channel squark, or $u$-channel
gluino exchange as it is illustrated in Fig.\ \ref{fig:gosq}. Since each
contribution involves a coupling between a quark, a squark and a gluino, this
process can in general violate flavour. The differential cross
section is given by
\beq\bsp
 \frac{\d\hat\sigma}{\d t}(h_a,h_b) =& \frac{g_s^4}{64 \pi s^2} \Bigg\{
    \bigg[ 
      (1\!-\!h_a)(1\!-\!h_b) \big|L_{\q_i q \g}\big|^2 + 
      (1\!+\!h_a)(1\!+\!h_b) \big|R_{\q_i q \g}\big|^2 \bigg]\\ 
    &\ \times \bigg[ 
      \frac{-(N_c^2-1) u_\g}{4 N_c^2 s} +
      \frac{s t \!+\! t_\g t_{\q_i}}{2 N_c^2 s t_{\q_i}} 
     + \frac{1}{2 u_\g} \bigg( 
      \frac{2 s u \!-\! s u_\g \!+\! 2 u_{\q_i} u_\g}{u_\g} + 
      \frac{m_\g^2 s \!+\! u_\g u_{\q_i}}{s} + 
      \frac{m_\g^2 m_{\q_i}^2 \!-\! t u}{t_{\q_i}} \bigg) 
     \bigg] \\
    +&\ \bigg[ 
     (1\!-\!h_a) \big|L_{\q_i q \g}\big|^2 + 
     (1\!+\!h_a) \big|R_{\q_i q \g}\big|^2 \bigg] \\
  &\ \times \bigg[ 
     \frac{(N_c^2-1) m_{\q_i}^2 t_\g}{2 N_c^2 t_{\q_i}^2}-
     \frac{t_\g (s \!+\! t_\g)}{2 N_c^2 s t_{\q_i}}
   \frac{1}{2 u_\g} \bigg(
      \frac{2 u t_\g}{u_\g} - 
      \frac{u_{\q_i} (s \!+\!u_{\q_i}) }{s} +
      \frac{2 u \!-\! u_{\q_i}}{t_{\q_i}} \bigg)
     \bigg] \Bigg\} \ , 
\esp\eeq 
which agrees again with the polarized and unpolarized results of Refs.\
\cite{Beenakker:1996ch,Gehrmann:2004xu} in the flavour-conserving MSSM limit and 
after summing over mass-degenerate squarks. The cross section for the
charge-conjugate process can be easily derived by replacing $h_a\to-h_a$.

\begin{figure}
 \centering
 \includegraphics[width=0.40\columnwidth]{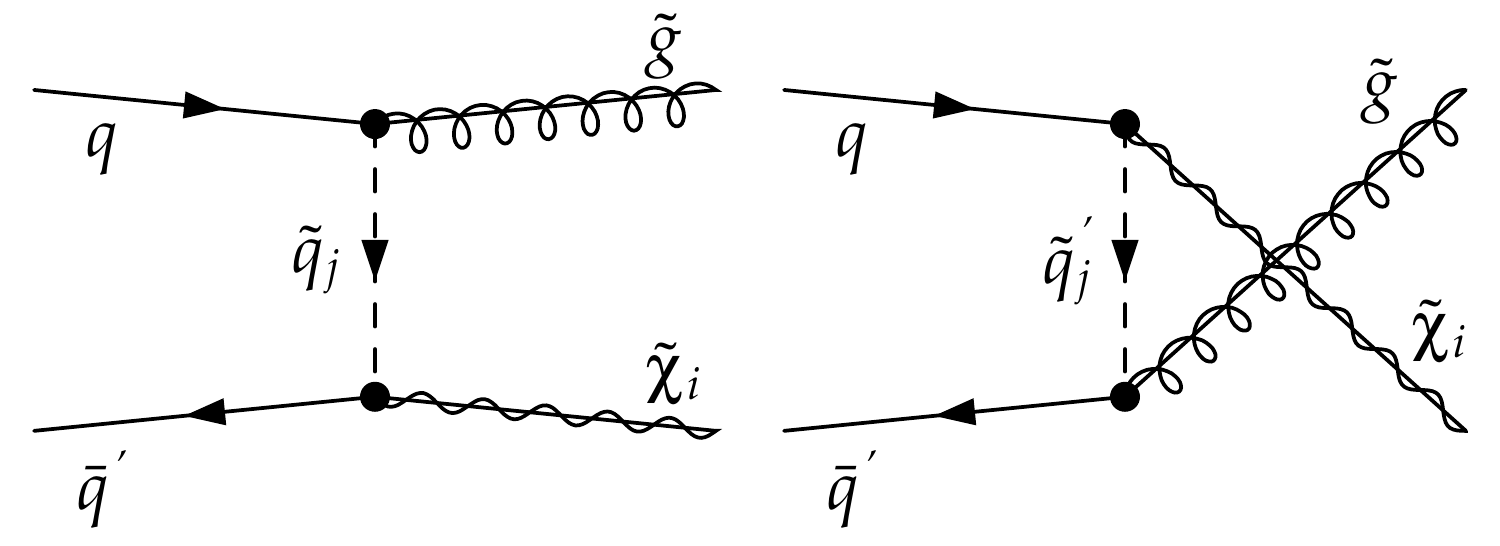}
 \caption{\label{fig:03}Tree-level Feynman diagrams for the associated
production of gluinos and gauginos.} 
\end{figure}
Finally, the associated production of a gluino and a chargino or of a gluino and
a neutralino,
\beq
  q_{h_a}(p_a)\ \bar q_{h_b}(p_b) \to \g (p_1)\ \tilde\chi^{\pm,0}_i (p_2) \ ,  
\eeq
can be mediated through a $t$-channel or $u$-channel squark exchange, as shown
in Fig.\ \ref{fig:03}, and can thus involve flavour-violating interactions. The
differential cross section is given by
\beq\bsp
 \frac{\d\hat\sigma}{\d t}(h_a,h_b) =& \frac{g_s^2 g_W^2 (N_c^2-1)}{8 \pi s^2
N_c^2} \sum_{i,j=1}^6 \bigg\{ (1\!-\!h_a)(1\!+\!h_b)\bigg[\frac{(Q_{tt})_{ij}^{11}}{t_{\q_i} t_{\q_j}}\!+\!
       \frac{(Q_{uu})_{ij}^{11}}{u_{\q_i} u_{\q_j}}  \!+\!
       \frac{(Q_{tu})_{ij}^{11}}{t_{\q_i} u_{\q_j}} \bigg] \\ 
 &\ + (1\!+\!h_a)(1\!-\!h_b)\bigg[\frac{(Q_{tt})_{ij}^{22}}{t_{\q_i} t_{\q_j}}\!+\!
       \frac{(Q_{uu})_{ij}^{22}}{u_{\q_i} u_{\q_j}}  \!+\!
       \frac{(Q_{tu})_{ij}^{22}}{t_{\q_i} u_{\q_j}} \bigg]\\
 &\ + (1\!-\!h_a)(1\!-\!h_b) \bigg[ 
       \frac{(Q_{tt})_{ij}^{12}}{t_{\q_i} t_{\q_j}}\!+\!
       \frac{(Q_{uu})_{ij}^{12}}{u_{\q_i} u_{\q_j}} \!+\!
       \frac{(Q_{tu})_{ij}^{12}}{t_{\q_i} u_{\q_j}}\bigg]\\ 
 &\ + (1\!+\!h_a)(1\!+\!h_b) \bigg[ 
       \frac{(Q_{tt})_{ij}^{21}}{t_{\q_i} t_{\q_j}}\!+\!
       \frac{(Q_{uu})_{ij}^{21}}{u_{\q_i} u_{\q_j}} +
       \frac{(Q_{tu})_{ij}^{21}}{t_{\q_i} u_{\q_j}}\bigg]  \bigg\} \ , 
\esp\eeq 
where we have introduced the generalized charges,
\beq\bsp
  (Q_{tt})_{ij}^{mn} =&\  {\cal C}^{m\ast}_{\q_i q \g} {\cal C}^{n\ast}_{\q_i
    q^\prime \tilde\chi_i} {\cal C}_{\q_j q \g}^m {\cal C}^n_{\q_j q^\prime \tilde\chi_i} 
    t_\g t_{\tilde \chi_i} \ , \\
  (Q_{uu})_{ij}^{mn} =&\  {\cal C}^{n\ast}_{\q_i q^\prime \g} {\cal
    C}^{m\ast}_{\q_i q \tilde\chi_i} {\cal C}_{\q_j q^\prime \g}^n {\cal
    C}^m_{\q_j q \tilde\chi_i} u_\g u_{\tilde \chi_i} \ , \\
(Q_{tu})_{ij}^{mn} =&\ 2 {\rm Re} \Big[ 
   {\cal C}^m_{\q_i q \g} {\cal C}^n_{\q_i q^\prime \tilde\chi_i}
   {\cal C}_{\q_j q^\prime \g}^n {\cal C}^m_{\q_j q \tilde\chi_i} u_\g u_{\tilde
   \chi_i} \Big] \Big(
     (1-\delta_{mn}) (u t - m_\g^2 m_{\tilde \chi_i}^2)  - \delta_{mn} m_\g
     m_{\tilde\chi_i} s \Big) \ .
\esp\eeq
\subsection{Numerical predictions for NMFV gluino production at the LHC}

In this section, we present numerical predictions in the context of
non-minimally flavour violating supersymmetry at the LHC for cross
sections related to the production of gluino pairs as well as to the one of 
associated pairs of a gluino and a squark, an
antisquark, a chargino or a neutralino. Squarks and gluinos are
expected to be copiously produced at the LHC, due to their strong couplings to
quarks and gluons. However, in the case of the benchmark scenarios presented in 
Tab.\ \ref{tab:points}, the high mass of the coloured superpartners drastically
reduces the LHC sensitivity, most of the channels being 
hence largely phase-space suppressed.
Therefore, we focus on $pp$-collisions at the LHC design centre-of-mass energy
of $\sqrt{S}=14$ TeV, supposed to be reached in the second phase of the running
of the LHC, after the shutdown of 2013. 

Thanks
to the QCD factorization theorem, total hadronic production cross sections can
be computed by convolving the partonic cross sections derived in Sec.\
\ref{sec:analytics}, summed and averaged over final and initial spins,
respectively,
with the universal parton densities $f_{a/p}$ and $f_{b/p}$ of partons
$a,b$ in the proton, which depend on the longitudinal momentum fractions of the
two partons $x_{a,b} = \sqrt{\tau}e^{\pm y}$ and on the unphysical factorization
scale $\mu_F$, 
\beq
 \sigma =
 \int_{4m^2/S}^1\!\d\tau\!\!
 \int_{-1/2\ln\tau}^{1/2\ln\tau}\!\!\d y
 \int_{t_{\min}}^{t_{\max}} \d t \
 f_{a/p}(x_a,\mu_F) \ f_{b/p}(x_b,\mu_F) \ {\d\hat{\sigma}\over\d t} \ .
\eeq
Neglecting all quark masses but the top mass, we employ the leading order (LO) set of
the CTEQ6 parton density fit \cite{Pumplin:2002vw}, which includes $n_f=5$ light
quark flavours and the gluon, but no top-quark density. Consistently, the strong
coupling constant $g_s$ is evaluated with the corresponding LO value of the QCD 
scale $\Lambda_{\rm LO}^{n_f=5}=165$ MeV. For all our results, we 
identify the renormalization scale
$\mu_R$ with the factorization scale $\mu_F$ and set the scales to the
average mass of the final state supersymmetric particles $m$.

\begin{table}
\caption{Cross sections, in fb, for the production of pairs of gluinos
($pp\to\tilde{g}\tilde{g}$), for the one of associated pairs of gluinos and
charginos ($pp\to\tilde{g}\tilde{\chi}^+_i+\tilde{g}\tilde{\chi}^-_i$ for
$i=1,2$) and for the one of associated pairs of gluinos and neutralinos
($pp\to\tilde{g}\tilde{\chi}^0_i$ for $i=1,2,3,4$), 
for the reference scenarios presented in Tab.\ \ref{tab:points}.}
\label{tab:xsec}
\begin{tabular}{|c|c|cc|cccc|}
	\hline
	  & $\quad\tilde{g}\tilde{g}\quad$ & $\quad\tilde{g}\tilde{\chi}^{\pm}_1\quad$ & $\quad\tilde{g}\tilde{\chi}^{\pm}_2\quad$ & $\quad\tilde{g}\tilde{\chi}^0_1\quad$ & $\quad\tilde{g}\tilde{\chi}^0_2\quad$ & $\quad\tilde{g}\tilde{\chi}^0_3\quad$ & $\quad\tilde{g}\tilde{\chi}^0_4\quad$ \\
	\hline
	SPS9 & 93.8 & 43.8 & 0.055 & 5.1 & 0.46 & 0.006 & 0.001  \\
	\hline
	   I & 76.5 & 29.0 & 0.040 & 3.3 & 0.38 & 0.004 & 0.001  \\
	  II & 58.9 & 10.8 & 0.022 & 1.2 & 0.19 & 0.002 & 0.001  \\
	 III & 52.5 & 4.2 & 0.014 & 0.47 & 0.085 & 0.001 & 0.001  \\
	\hline
\end{tabular}
\end{table}

For gluino pair production as well as for the associated production of a
chargino or a neutralino with a gluino, only the $t$- and $u$-channel diagrams 
depend on the flavour-violating parameters in the squark sector since they
contain a squark propagator  (see the Feynman diagrams shown in 
Figs.\ \ref{fig:01} and \ref{fig:03}).
However, all squark eigenstates contribute to the total cross section and the
corresponding diagrams must be summed over, leading subsequently to 
production cross sections insensitive to non-minimal flavour-violation. 
We present these results, therefore independent of the $\lambda$-parameters, 
in Tab.\ \ref{tab:xsec}, both
for our scenarios I, II and III as well as for the SPS 9 benchmark point as a
reference. Strong gluino pair production is clearly dominant, and the 
luminosity required to observe possible signal events is not so high. In
contrast, the cross sections
related to the semi-weak production of a gluino and a chargino/neutralino vary
from ${\cal O}(10)$ fb for the lightest chargino and neutralino case to
the barely visible level of ${\cal O}(10^{-3})$ fb for the heavier
superpartners.

\begin{figure}
\begin{center}
	\includegraphics[scale=0.45]{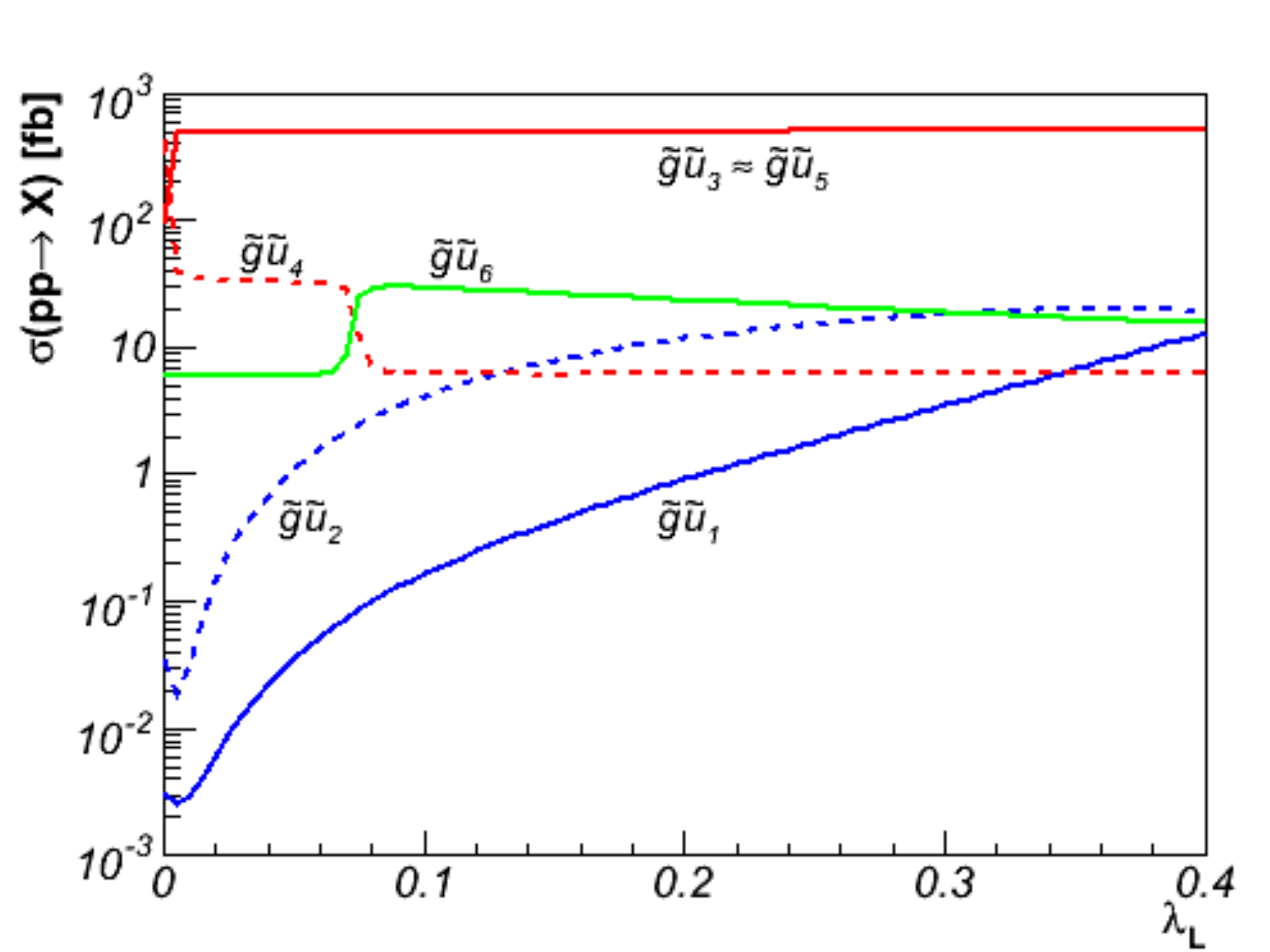}
	\includegraphics[scale=0.45]{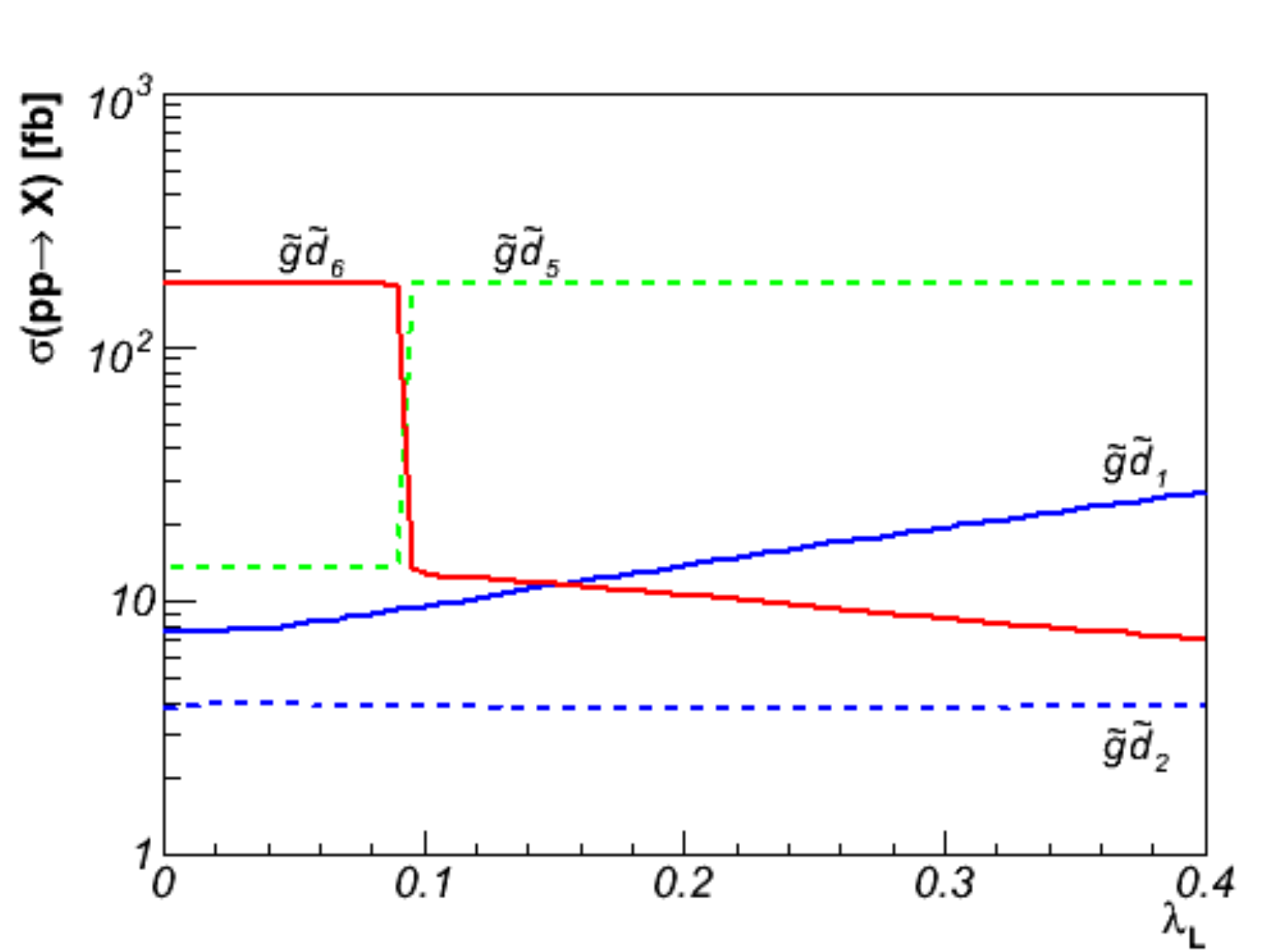} \\
	\includegraphics[scale=0.45]{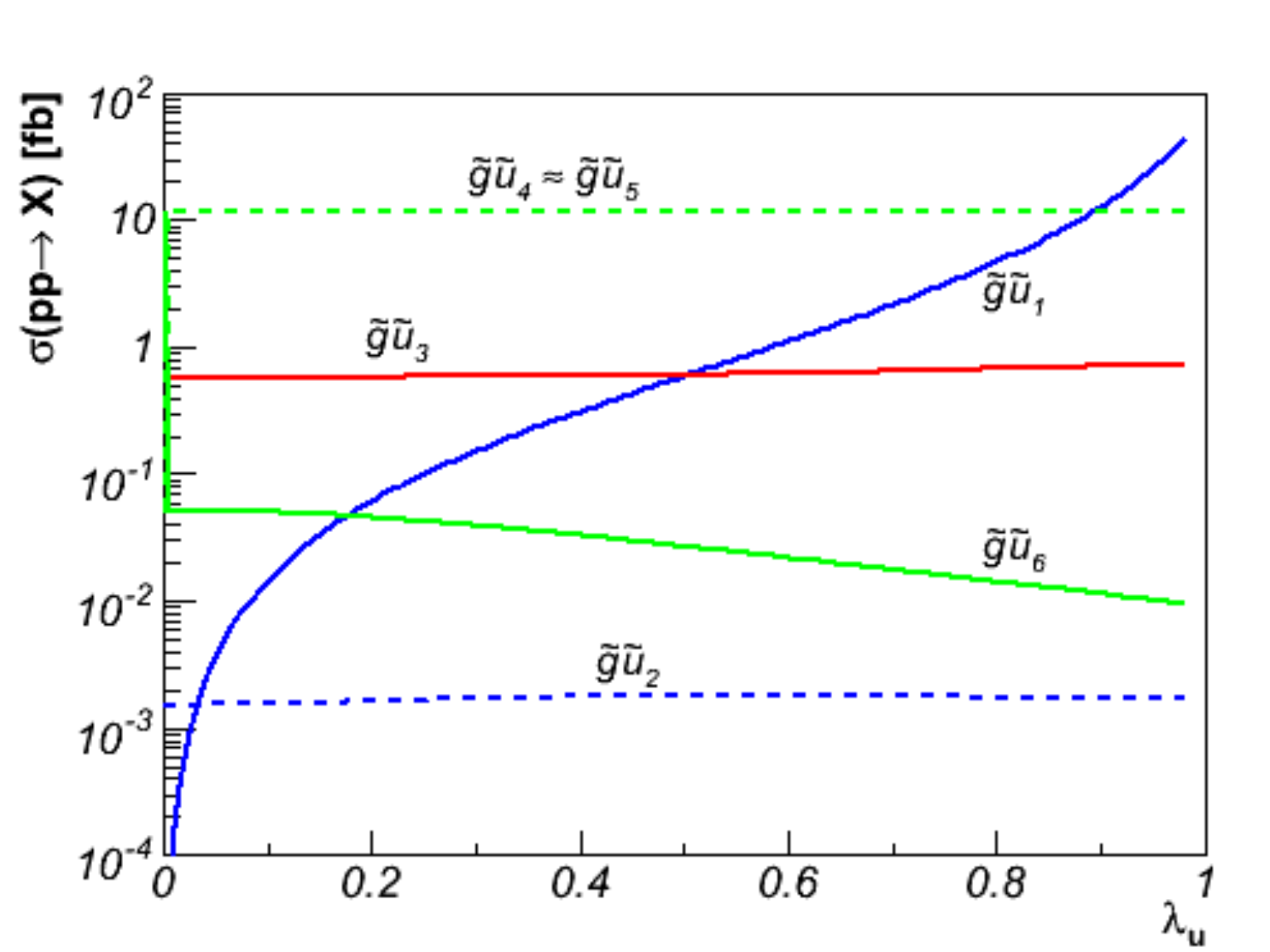}
	\includegraphics[scale=0.45]{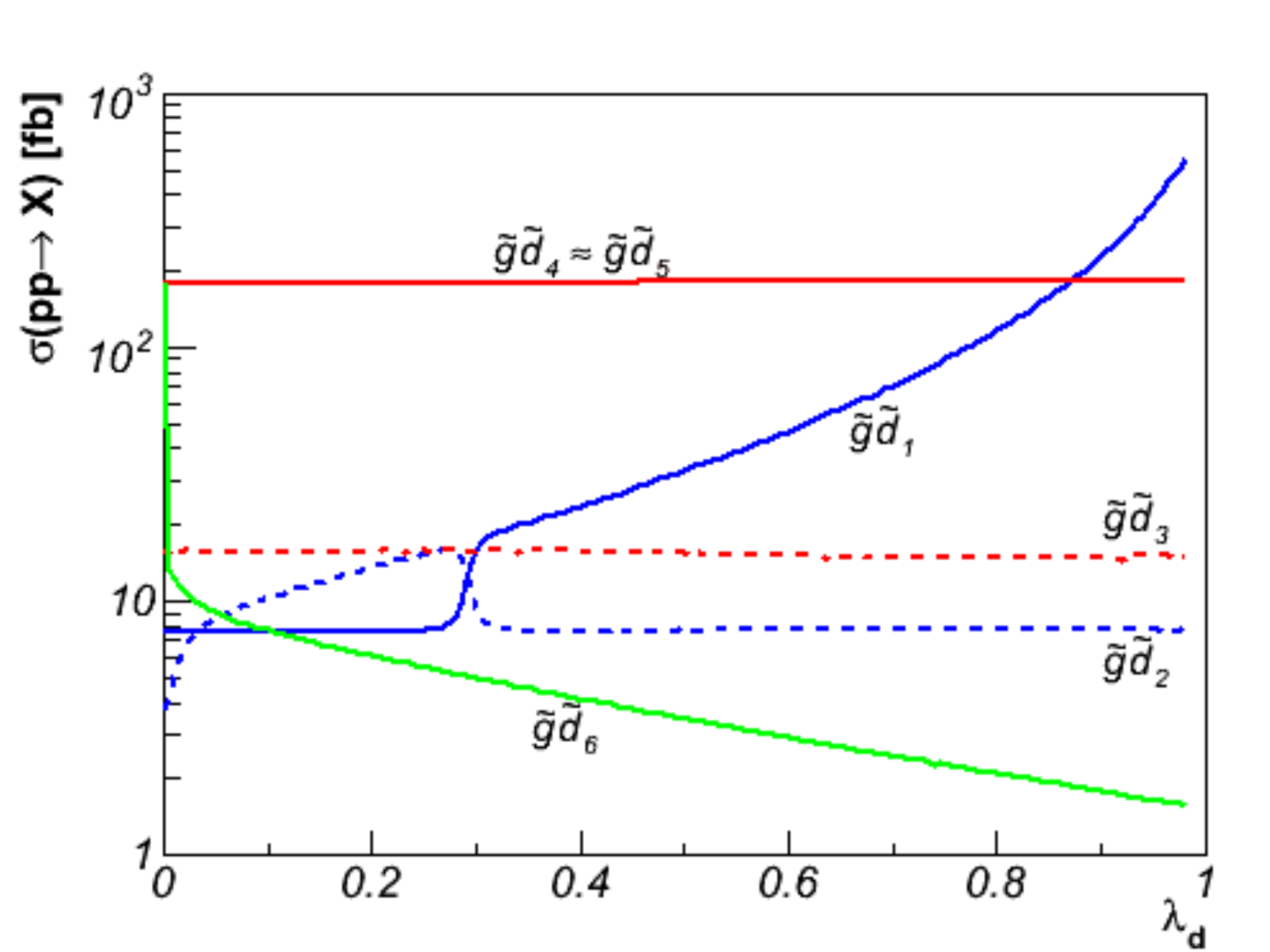}
\end{center}
\caption{Cross-sections of gluino production in association with up- and
down-type squarks for various ranges of the flavour-mixing parameters $\lambda_L$, $\lambda_u$, and $\lambda_d$ for the reference scenario I of Tab.\ \ref{tab:points}.}
\label{fig:xsec1}
\end{figure}

\begin{figure}
\begin{center}
	\includegraphics[scale=0.45]{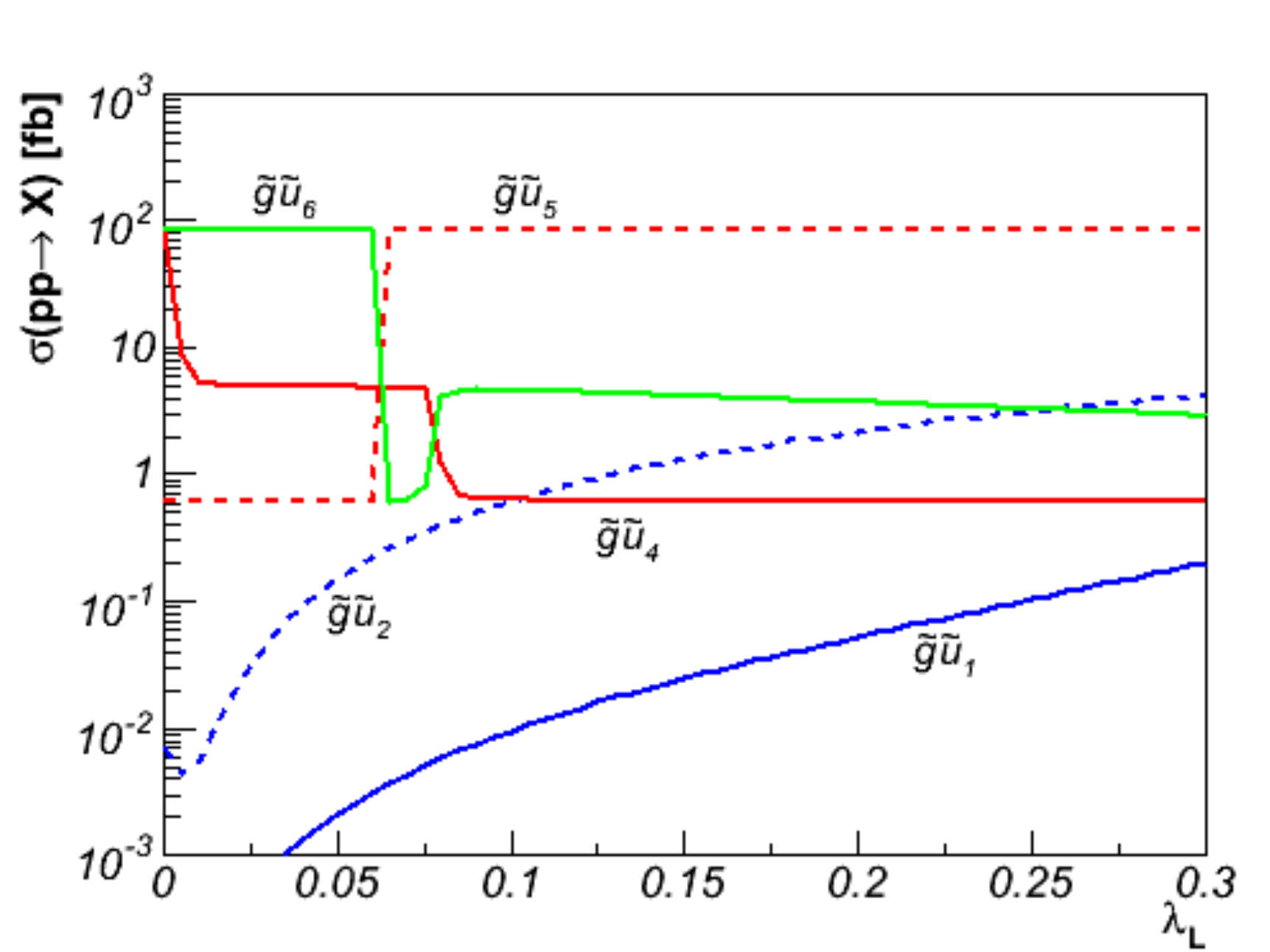}
	\includegraphics[scale=0.45]{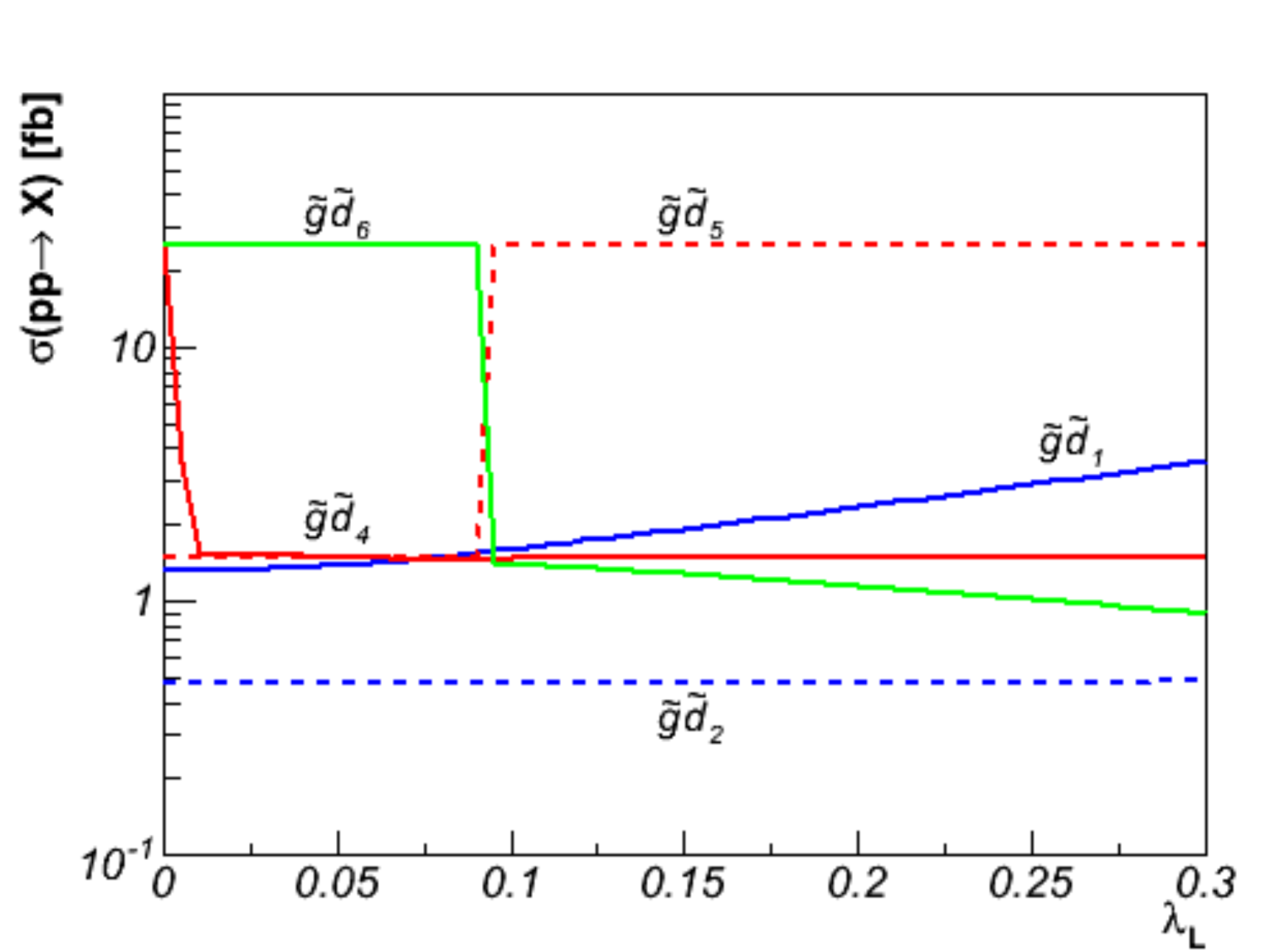} \\
	\includegraphics[scale=0.45]{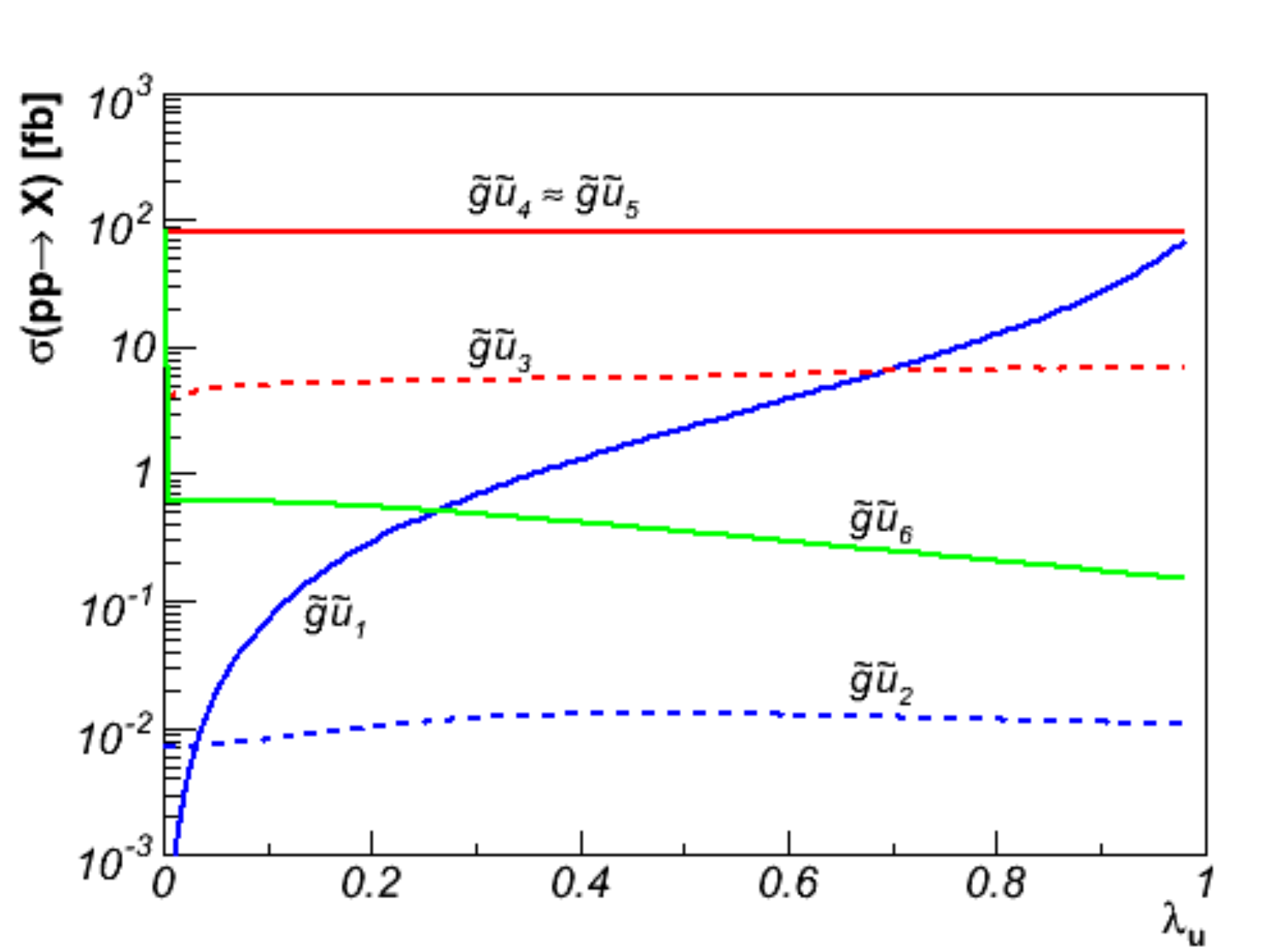}
	\includegraphics[scale=0.45]{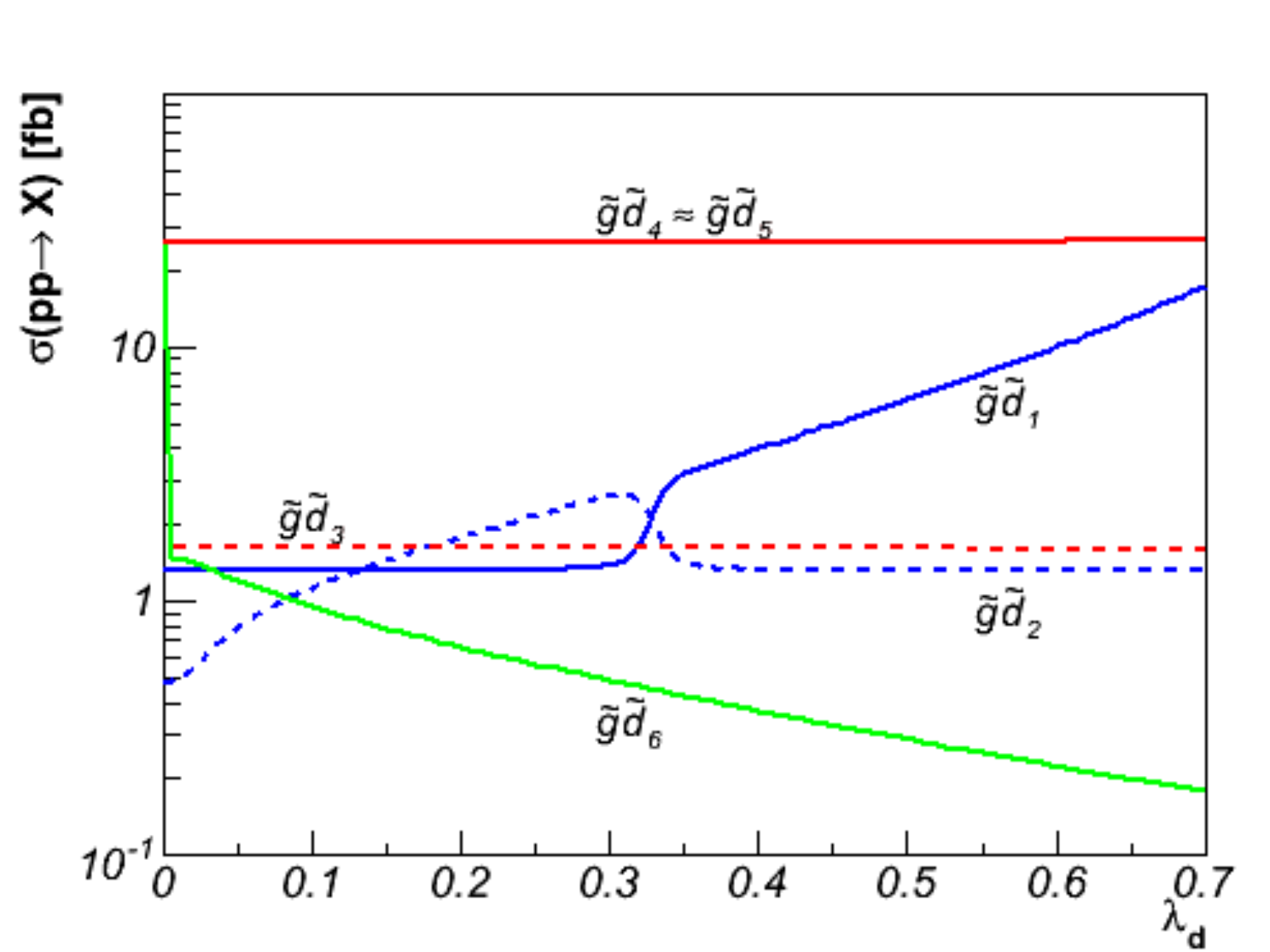}
\end{center}
\caption{Same as Fig.\ \ref{fig:xsec1} for reference scenario II.}
\label{fig:xsec2}
\end{figure}

\begin{figure}
\begin{center}
	\includegraphics[scale=0.45]{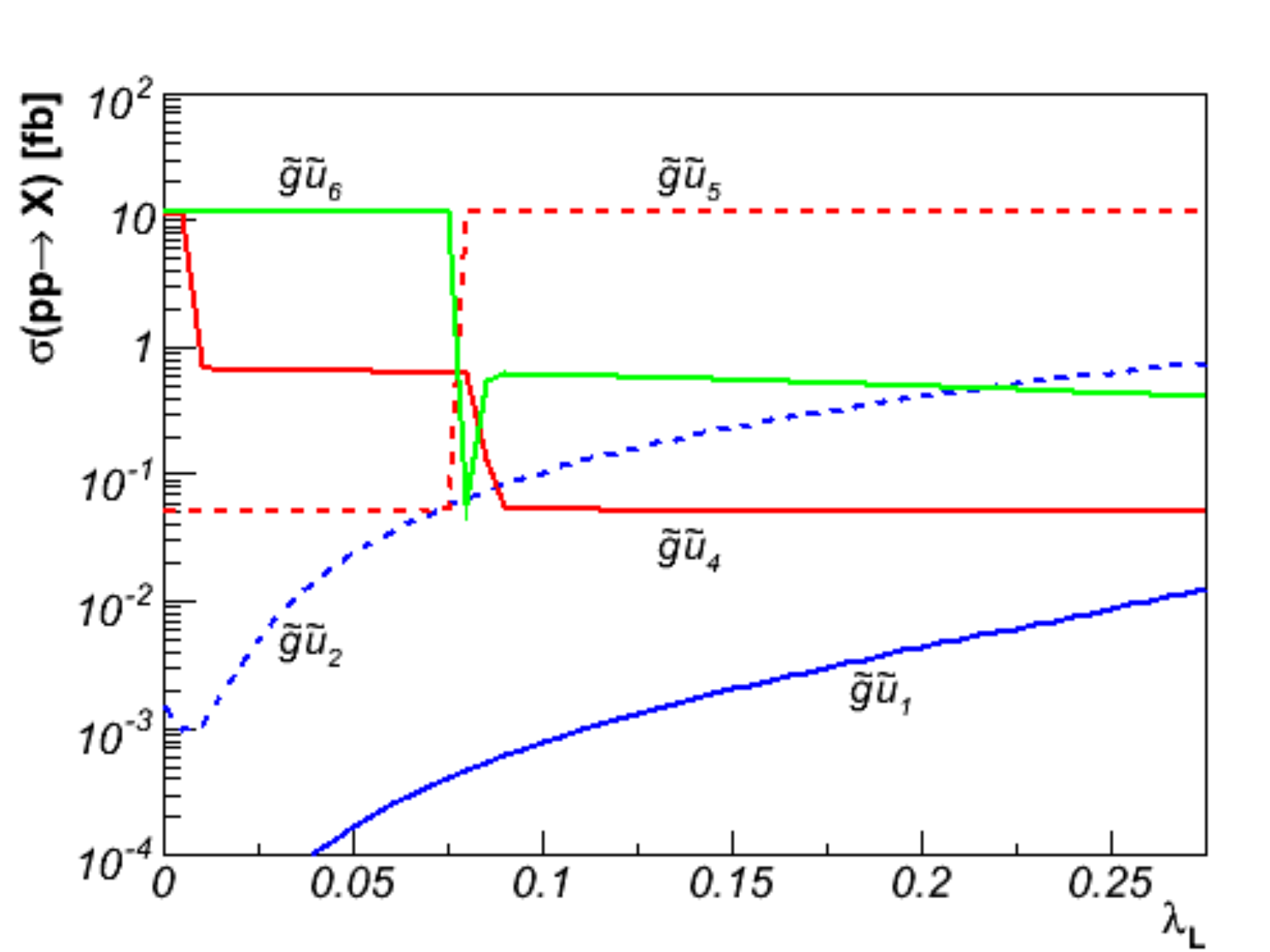}
	\includegraphics[scale=0.45]{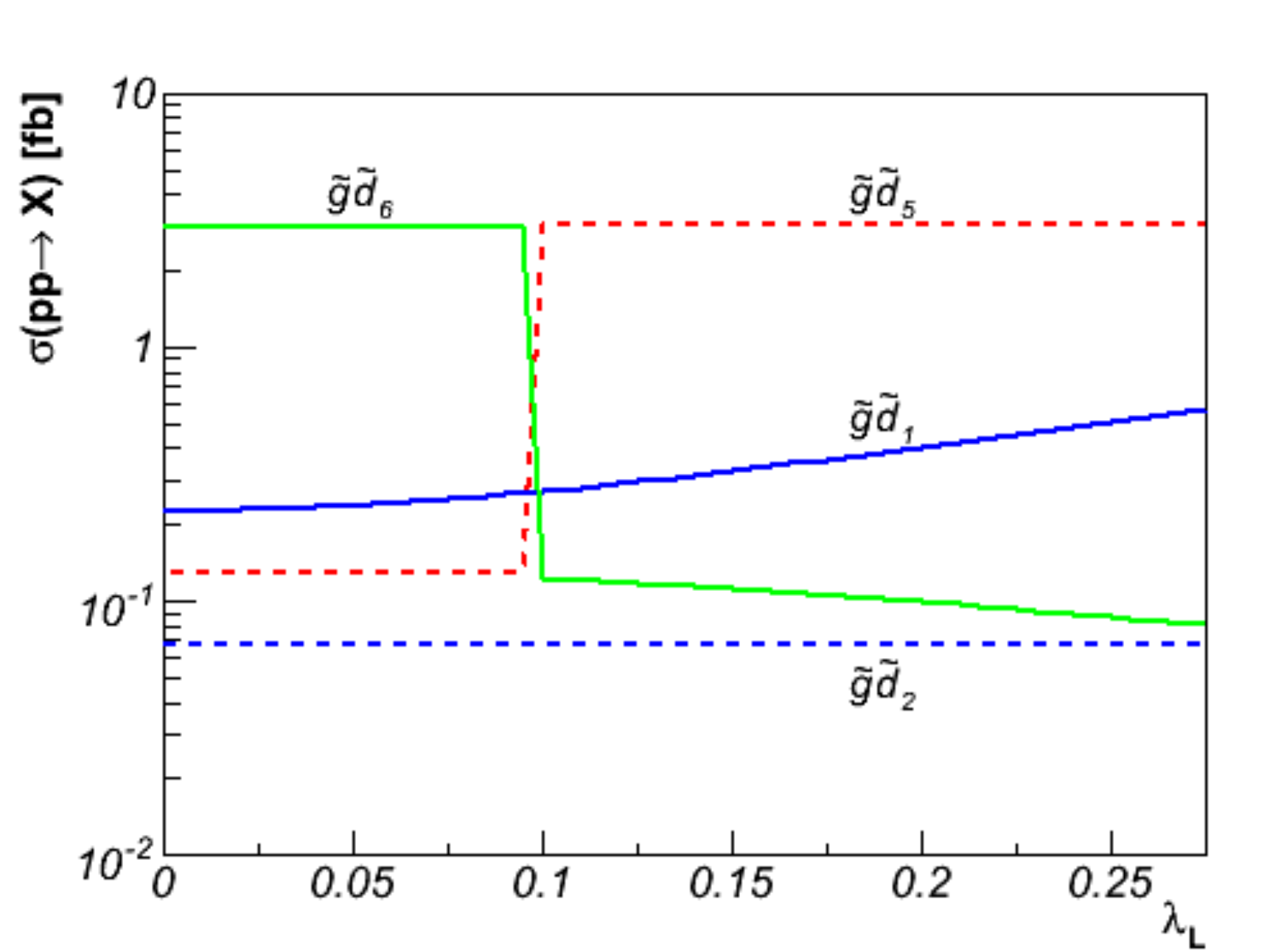} \\
	\includegraphics[scale=0.45]{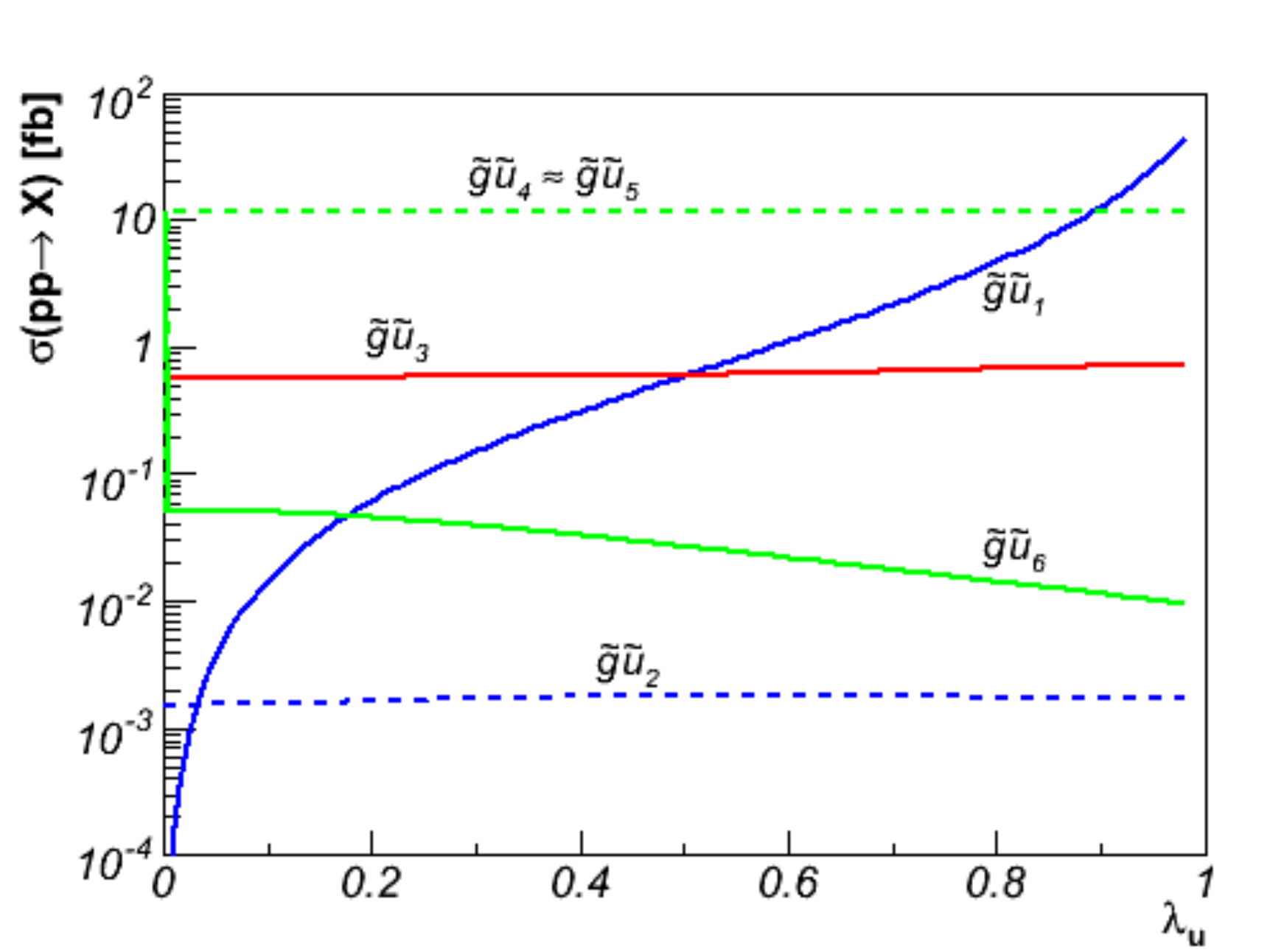}
	\includegraphics[scale=0.45]{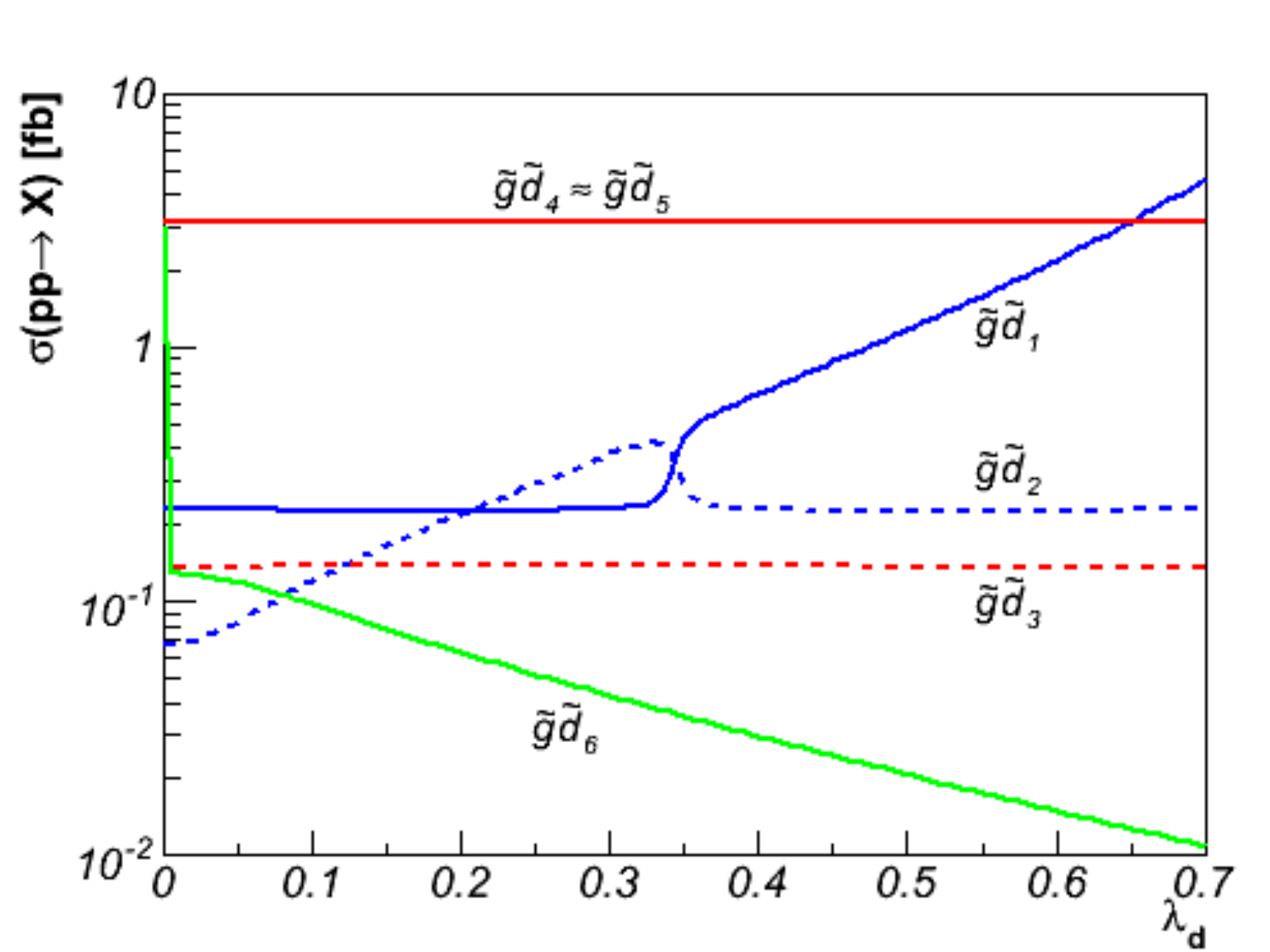}
\end{center}
\caption{Same as Fig.\ \ref{fig:xsec1} for reference scenario III.}
\label{fig:xsec3}
\end{figure}

Contrary to the previous cases, associated squark and gluino production shows an
interesting dependence on the non-minimal flavour-violation $\lambda$-parameters,
as illustrated in Figs.\ \ref{fig:xsec1}, \ref{fig:xsec2} and \ref{fig:xsec3}
for our scenarios I, II and III, respectively. Moreover, the cross sections are
fairly high, reaching the level of several hundereds of fb for many channels,
which makes them nice candidates to study non-minimal flavour violation in
supersymmetry at the LHC. 
Analyzing the dependence of the cross section on the $\lambda$-parameters, 
sharp and smooth transitions can
be observed, the first ones being related to the presence of an avoided crossing
and the second ones to a smooth change in the flavour content of the relevant
eigenstate (see Section \ref{sec:benchmark}). Indeed, at the point where the
mass of two eigenstates should cross, \eg, at $\lambda_{\rm L} \sim 0.07$ on the
top-left panel of Fig.\ \ref{fig:xsec1} or at $\lambda_d \sim 0.32$ on the
lower-right panel of Fig.\ \ref{fig:xsec2}, the flavour content of both
squarks is exchanged
and the same sharp transition is observed at the level of the cross
section. Smooth increases and decreases in the cross section with the values of
the $\lambda$-parameters have two sources. First, second and third generation
squark mixing induces larger mass
splitting, as illustrated in Figs.\ \ref{fig:squarks1}, \ref{fig:squarks2} and
\ref{fig:squarks3}. This renders certain channels 
phase-space favoured and other
channels phase-space suppressed. Secondly, the magnitude of the cross
section is connected to the flavour content of the
squark produced in association with the gluino, since producing a given flavour
of squark requires an initial quark of the same flavour, as shown in the Feynman
diagrams of Fig.\ \ref{fig:gosq}. As an example, the lightest 
down-type squark mass eigenstate is a pure sbottom state for small values
of $\lambda_d \lesssim 0.3$, whilst it becomes a mixed state for $\lambda_d
\gtrsim 0.3$, with a larger and larger strange squark component with increasing
values of $\lambda_d$, as it can be seen from the upper-right panel of Fig.\
\ref{fig:squarks1}. Consequently, as presented in lower-right panel of 
Fig.\ \ref{fig:xsec1}, the cross section related to the process
$p p \to \g \tilde d_1$ is small and constant for $\lambda_d \lesssim 0.3$
($\sigma \approx 8$ fb) and gets larger and larger for increasing $\lambda_d$,
even reaching a couple of hundreds of fb for $\lambda_d \gtrsim 0.8$. The
opposite effects can be observed for the process $p p \to \g \tilde u_6$ in the
context of the scenario III, as shown in the lower-left panel of Fig.\
\ref{fig:xsec3}.

\section{Cosmological aspects \label{sec:cosmo}}

Among the most compelling evidences for physics beyond the Standard Model is the presence of cold dark matter (CDM) in our Universe. Its relic density is today  constrained to be
\beq
	\Omega_{\rm CDM}h^2 ~=~ 0.1123 \pm 0.0035
\label{eq:omh2}
\eeq
from recent WMAP data combined with measurements related to supernov\ae\ and baryonic acoustic oscillations \cite{WMAP}. Here, $h$ is the present Hubble expansion rate in units of $100~{\rm km\,s^{-1}\,Mpc^{-1}}$. New physics models should therefore include a viable dark matter candidate that can account for the above amount of dark matter.

In AMSB models, the lightest of the four neutralinos is the lightest
superpartner and therefore the dark matter candidate, if R-parity is assumed to
be conserved. After the renormalization group evolution from the high-scale
parameters to the weak scale, the wino mass parameter $M_2$ turns out to be
smaller than the bino and gluino masses $M_1$ and $M_3$. In consequence, in AMSB
scenarios the lightest neutralino is wino-like. This is in contrast to, e.g.,
models based on minimal supergravity where usually $M_1 < M_2$ leading to a
bino-like LSP. Since also the chargino mass is governed by $M_2$, the mass
difference between the lightest neutralino and the lightest chargino is rather
small. For our reference scenarios, the mass difference is less than a GeV, as
can be seen in Tab.\ \ref{tab:points}. Due to the larger pair annihilation
cross section as compared to the bino and due to efficient co-annihilations with
the chargino, the resulting relic density of the thermally produced neutralino
is usually one or two orders of magnitude below the range given in Eq.\
(\ref{eq:omh2}) \cite{ChenDreesGunion, MoroiRandall}. Using the public programme
{\tt DarkSUSY} \cite{DarkSUSY}, we obtain the values
$\Omega_{\tilde{\chi}_1^0}h^2 = 8.57\cdot 10^{-4},~7.87\cdot 10^{-4},~7.83\cdot
10^{-4}$, and $7.83\cdot 10^{-4}$ for the scenarios of Tab.\ \ref{tab:points},
respectively.

\begin{figure}
\begin{center}
	\includegraphics[scale=0.4]{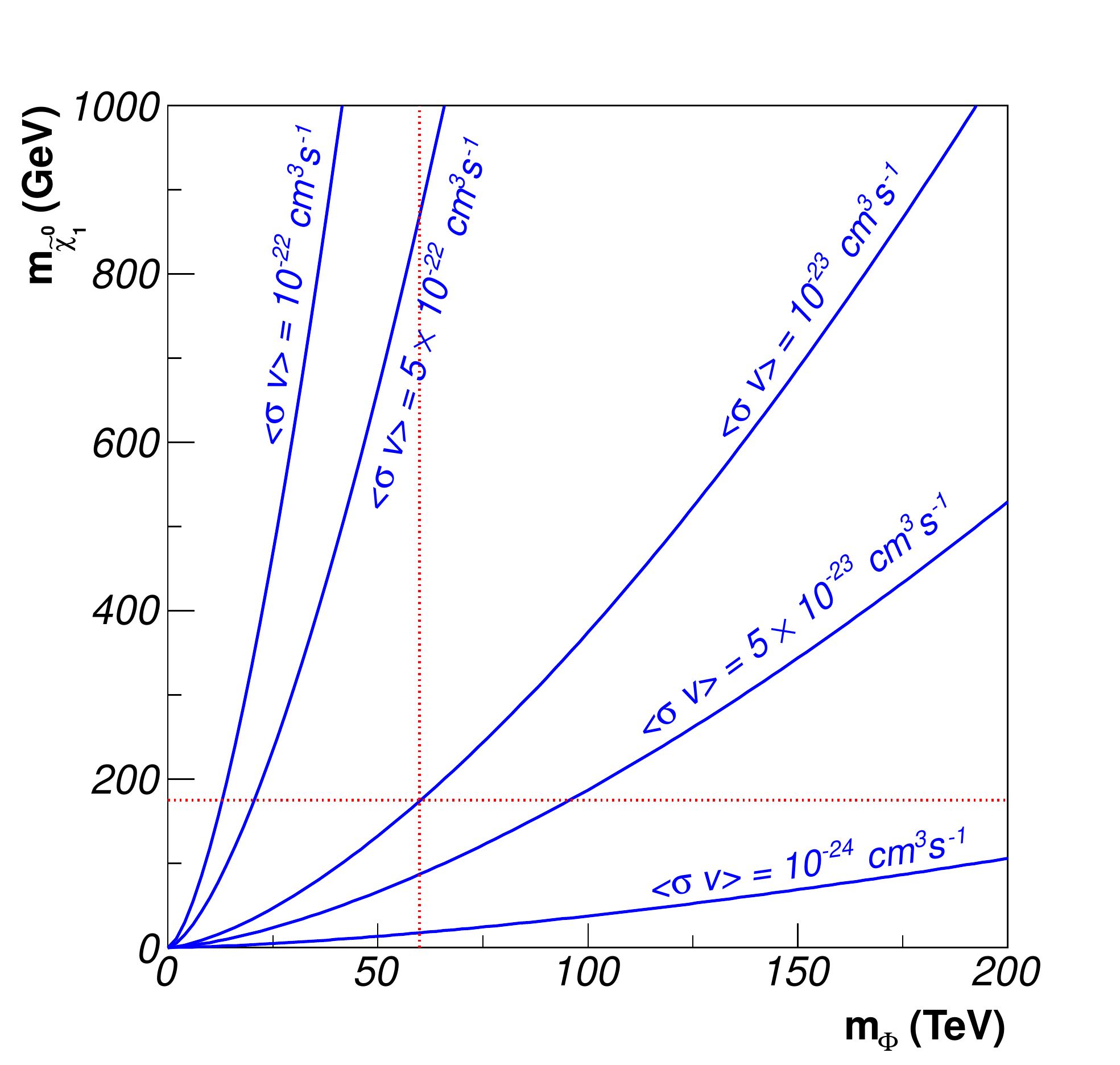}
	\includegraphics[scale=0.4]{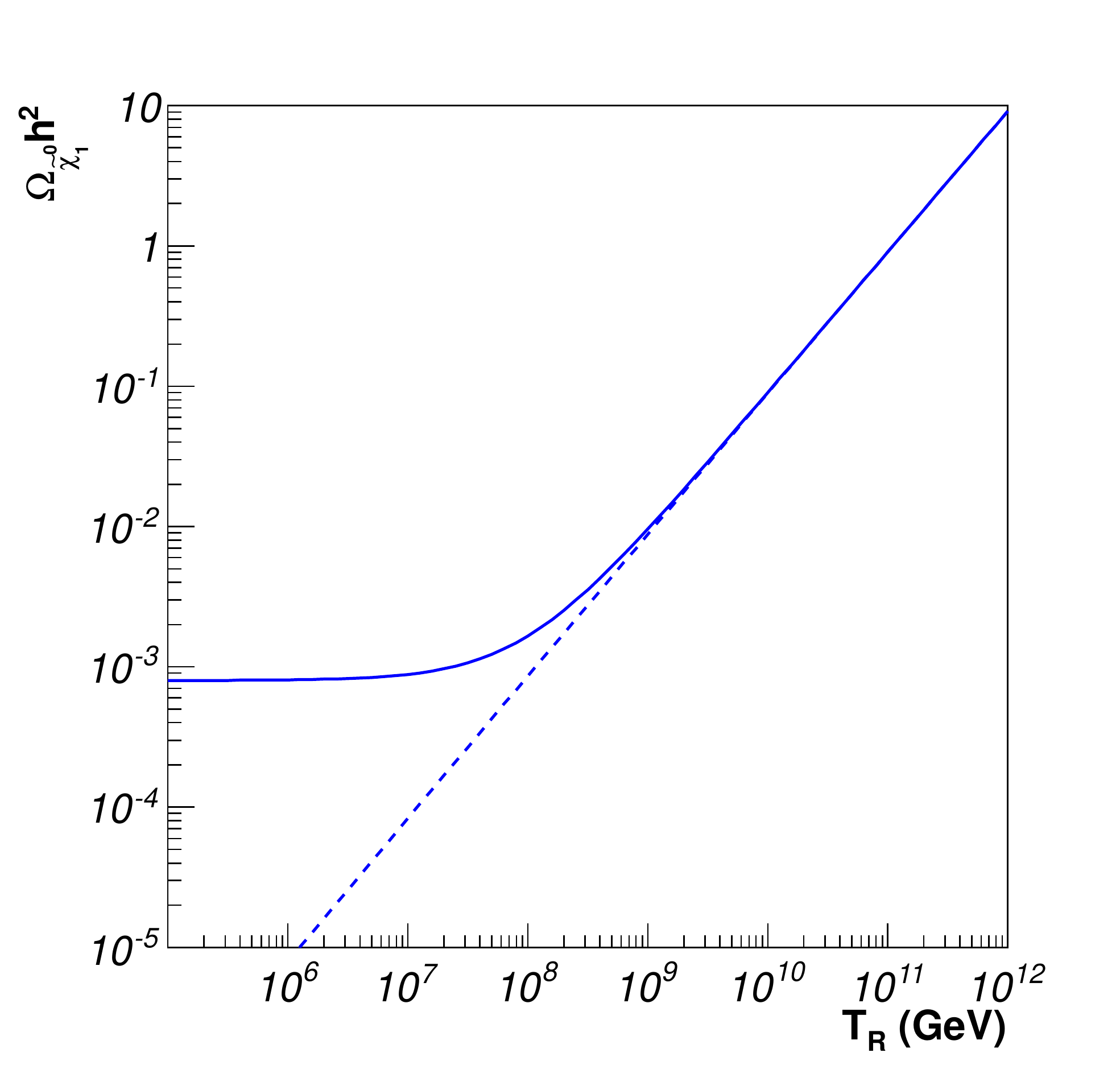}
\end{center}
\caption{Left: Isolines corresponding to $\Omega_{\tilde{\chi}_1^0}h^2=0.1123$ in the $m_{\Phi}$--$m_{\tilde{\chi}_1^0}$ planes for different values of $\langle\sigma v\rangle$. The dotted lines correspond to $m_{\tilde{\chi}^0_1}=175$ GeV and $m_{\Phi}=m_{3/2}=60$ TeV. Right: The neutralino relic density $\Omega_{\tilde{\chi}^0_1}h^2$ from gravitino decay as a function of the reheating temperature $T_R$ for a typical AMSB scenarios with $m_{\tilde{\chi}_1^0}=175$ GeV, $m_{3/2}=60$ TeV, as it is the case for the scenarios of Tab.\ \ref{tab:points}. The solid line includes a contribution of $\Omega_{\tilde{\chi}}h^2=8\cdot 10^{-4}$ from thermal neutralinos, while the dashed line indicates the contribution from gravitino decay only.}
\label{fig:cosmo}
\end{figure}

However, thermal production of neutralinos is not the only mechanism to be considered. Possible non-thermal production modes include the decay of heavy fields such as moduli or gravitinos in the early universe \cite{MoroiRandall, PhDShibi}. Moreover, axions and axinos can contribute to the dark matter relic abundance \cite{Covi1999, Covi2001}. The contribution to the neutralino relic density from moduli decay can be estimated as \cite{Acharya2009}
\beq
	\Omega_{\tilde{\chi}^0_1}h^2 ~\simeq~ 0.1 \left( \frac{m_{\tilde{\chi}^0_1}}{100~{\rm GeV}} \right) \left( \frac{10.75}{g_{*}} \right)^{1/4} 
			\left( \frac{3\cdot 10^{-24}~{\rm cm}^3/{\rm s}}{\langle \sigma v \rangle} \right) \left( \frac{100~{\rm TeV}}{m_{\Phi}} \right)^{3/2} .
\eeq
It depends on the neutralino mass $m_{\tilde{\chi}_1^0}$, the mass of the moduli
fields $m_{\Phi}$, the effective number of degrees of freedom $g_{*}$, and the
thermally averaged annihilation cross-section $\langle \sigma v \rangle$. In
Fig.\ \ref{fig:cosmo}, we depict isolines of the neutralino relic density in the
$m_{\tilde{\chi}_1^0}$--$m_{\Phi}$ plane, assuming $g_* \sim 10.75$
\cite{MoroiRandall} and different values of the annihilation cross-section
$\langle \sigma v \rangle$. Our typical scenarios lead to a neutralino relic
density of about $\Omega_{\tilde{\chi}_1^0}h^2 = 8\cdot 10^{-4}$, which is
consistent with a neutralino annihilation cross-section of $\langle \sigma v
\rangle \sim 10^{-23}~{\rm cm}^3{\rm s}^{-1}$. For such cross-sections, a
neutralino mass of $m_{\tilde{\chi}_1^0} \sim 175$ GeV and moduli masses
comparable with the gravitino mass, $m_{\Phi} \sim m_{3/2} \sim 60$ TeV, which is
consistent with gravity and anomaly mediation \cite{Acharya2009}, this yields
nearly the measured abundance of Eq.\ (\ref{eq:omh2}).

The neutralino relic abundance from gravitino decay is directly related to the thermal abundance $\Omega_{3/2}h^2$, that the gravitino would have if it did not decay. The latter is obtained from computing the thermal production of gravitinos in the early universe. More precisely, the resulting neutralino relic density can then be evaluated according to \cite{PradlerSteffen}
\beq
	\Omega_{\tilde{\chi}^0_1}h^2 ~=~ \frac{m_{\tilde{\chi}^0_1}}{m_{3/2}} \Omega_{3/2}h^2 ~\simeq~ 
		\frac{m_{\tilde{\chi}^0_1}}{m_{3/2}} 
		\biggr( \frac{m_{3/2}}{100~{\rm GeV}} \biggr) \left( \frac{T_R}{10^{10}~{\rm GeV}} \right)
		\sum_{i=1}^3 \omega_i g_i^2 \left( 1 + \frac{M^2_i}{3 m^2_{3/2}} \right) \log\frac{k_i}{g_i} ,
\label{eq:omh2grav}
\eeq
where the ratio of neutralino and gravitino mass expresses the fact that each gravitino decays into one stable neutralino. The thermal gravitino production depends linearly on the gravitino mass $m_{3/2}$ and the reheating temperature $T_R$ of the universe after inflation. The sum runs over the three gauge groups $U(1)$, $SU(2)$, and $SU(3)$, $g_i$ are the coupling constants of the three gauge groups, and $M_i$ the mass parameters of the associated gauginos. Note that in Eq.\ (\ref{eq:omh2grav}), $g_i$ and $M_i$ are evaluated at the reheating scale $T_R$. The constants $\omega_i$ and $k_i$ are given by $\omega_i = 0.018,~ 0.044,~ 0.117$ and $k_i = 1.266,~ 1.312,~ 1.271$ for $i=1,2,3$, respectively \cite{PradlerSteffen}.

The right panel of Fig.\ \ref{fig:cosmo} shows the resulting neutralino relic density as a function of the reheating temperature for the situation corresponding to the scenarios of Tab.\ \ref{tab:points}. For low values of $T_R$, the thermal neutralino production dominates, leading to the value of $\Omega_{\tilde{\chi}_1^0}h^2 = 8.57\cdot 10^{-4}$ already mentioned above. For $T_R \gtrsim 10^7$~GeV, the contribution from gravitino decay becomes dominant and $\Omega_{\tilde{\chi}_1^0}h^2$ grows linearly with $T_R$. As can be seen, the observed relic density of $\Omega_{\tilde{\chi}_1^0}h^2 \sim 0.11$ is obtained for a reheating temperature of $T_R \sim 10^{10}$~GeV, which is well compatible with thermal leptogenesis \cite{Buchmuller:2004nz}.

The relic abundance of the neutralino may also depend on flavour violating entries of the squark (or slepton) mass matrices. In Ref.\ \cite{QFV_DarkMatter} this has been studied for the case of minimal supergravity scenarios, where flavour violating couplings can modify the annihilation and coannihilation modes that enter the Boltzmann equation in the typical scenario with thermal production of neutralinos. Similar conclusions have been found for flavour non-diagonal entries in the leptonic soft matrices \cite{LFV_DarkMatter}. Flavour-violating effects are, of course, also present in the discussed cases of moduli or gravitino decay. A full study of their impact within this context is, however, clearly beyond the scope of this work.
\section{Conclusion \label{sec:conclusion}}

In this paper, we have studied the consequences of non-minimal
flavour violation in minimal anomaly-mediated supersymmetry
breaking models, where tachyonic sleptons are avoided
by introducing a common scalar mass similar to the one introduced
in minimal supergravity. In these scenarios, new sources of 
flavour violation are in general introduced at high scales,
leading to different flavour mixings for SM particles and their
superpartners at the weak scale.

Using the conventional parameterization of squark mixing at the
weak scale, we analyzed the viable AMSB parameter space in the
light of the latest limits on low-energy observables and of the
latest results from direct searches for Higgs and SUSY particles
at the LHC.
We found that intermediate values of $\tan\beta=10...30$ and
relatively large scalar masses of $m_0=1...3$ TeV, increasing
with $\tan\beta$, were preferred and allowed for sizeable flavour
violation in the left-left and essentially unconstrained
flavour violation in the right-right squark sectors. 

We completed our analytical calculations of flavour-violating
supersymmetric particle production at hadron colliders with those
related to gluino pair production and to the associated
production of gluinos with charginos and neutralinos 
as well as with squarks.
The corresponding cross sections were expected to be large due to
the strong coupling of gluinos to the initial quarks and gluons.
Flavour violation effects were expected to be only significant
for the associated production of gluinos and squarks, since the
other processes involved (almost) complete sums over internal
squark exchanges.

This was confirmed in our numerical analysis for the high-energy
phase of the LHC, where phenomena such as avoided crossings or
smooth flavour dependences known from our previous studies could
again be observed. For the experimental analysis of the ensuing
cascade decays, leading to final states with second and third
generation quarks and missing transverse energy, we referred the
reader to previously published dedicated studies performed, \eg,
in supergravity models. The corresponding analysis in AMSB models
was beyond the scope of this paper and is left for future work.

Finally, we briefly addressed the related cosmological aspects,
showing that the well-known problem of dark matter underabundance
in minimal AMSB models could be solved with moduli or gravitino decays
also in the presence of flavour violation.

\acknowledgments
The authors would like to thank W.~Porod for helpful discussions. 
This work has been supported by the Theory-LHC France-initiative of the
CNRS/IN2P3 and  by the ``Helmholtz Alliance for Astroparticle Phyics HAP''
funded by the Initiative and Networking Fund of the Helmholtz Association.


\end{document}